\documentclass[a4paper,11pt]{article}
\pdfoutput=1 

\usepackage{jinstpub} 

\usepackage{graphicx}
\usepackage{siunitx}
\usepackage{hyperref}
\usepackage{enumitem}
\usepackage{makecell}

\title{The Liquid Argon In A Testbeam (LArIAT) Experiment}

\collaboration{%
\includegraphics[height=17mm]{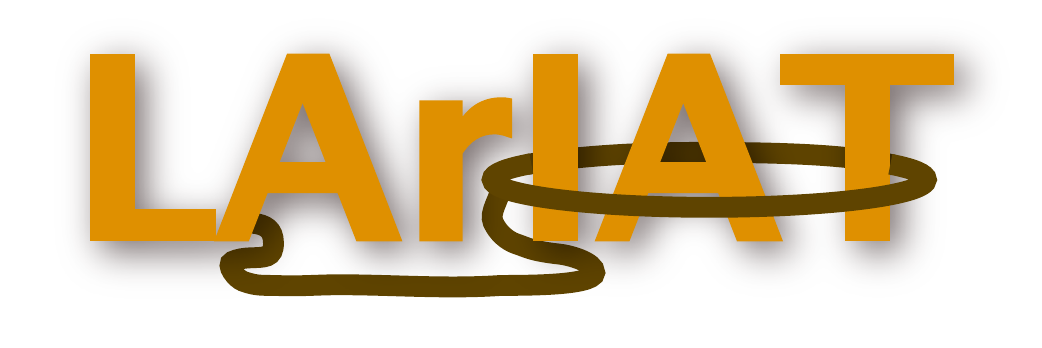}\\[6pt]
LArIAT Collaboration}

\author[8]{R.~Acciarri}
\author[21]{C.~Adams}
\author[17]{J.~A.~Asaadi}
\author[8]{M.~Backfish}
\author[8]{W.~Badgett}
\author[8]{B.~Baller}
\author[9,a]{O.~Benevides~Rodrigues\note{Present address: Syracuse University, Syracuse, NY 13244, USA}}
\author[3]{F.~d.~M.~Blaszczyk}
\author[6]{R.~Bouabid}
\author[13]{C.~Bromberg}
% ...

% ...
\affiliation[3]{Boston University, Boston, MA 02215, USA}
% ...
\affiliation[6]{University of Chicago, Chicago, IL 60637, USA}
% ...
\affiliation[8]{Fermi National Accelerator Laboratory, Batavia, IL 60510, USA}
\affiliation[9]{Universidade Federal de Goi\'{a}s, Goi\'{a}s, CEP 74690-900, Brasil}
% ...
\affiliation[13]{Michigan State University, East Lansing, MI 48824, USA}
% ...
\affiliation[17]{University of Texas at Arlington, Arlington, TX 76019, USA}
% ...
\affiliation[21]{Yale University, New Haven, CT 06520, USA}

\emailAdd{lariat\_authors@fnal.gov}

\abstract{The LArIAT liquid argon time projection chamber, placed in a tertiary beam of charged particles at the Fermilab Test Beam Facility, has collected large samples of pions, muons, electrons, protons, and kaons in the momentum range $\sim300$-$1400$~MeV/c. This paper describes the main aspects of the detector and beamline, and also reports on calibrations performed for the detector and beamline components.}

\begin{document}
\maketitle
\flushbottom

\section{Introduction}\label{sec:Introduction}
Liquid Argon Time Projection Chambers (LArTPCs) offer fine-grained
tracking as well as precise calorimetry and particle identification
capabilities, making them well-suited to the study of neutrino-nucleus
interactions. LArTPCs have been chosen for the study of neutrino oscillations over both short baselines ($<1$~km) and long baselines ($>1000$~km) by experiments such as SBND, MicroBooNE, ICARUS, and DUNE~\cite{SBN-proposal,MicroBooNE-det,ICARUS-T600,DUNE-CDRvol1,DUNE-CDRvol2,DUNE-CDRvol3,DUNE-CDRvol4}.

\begin{figure}[htb]
\begin{centering}
\includegraphics[width=0.6\textwidth]{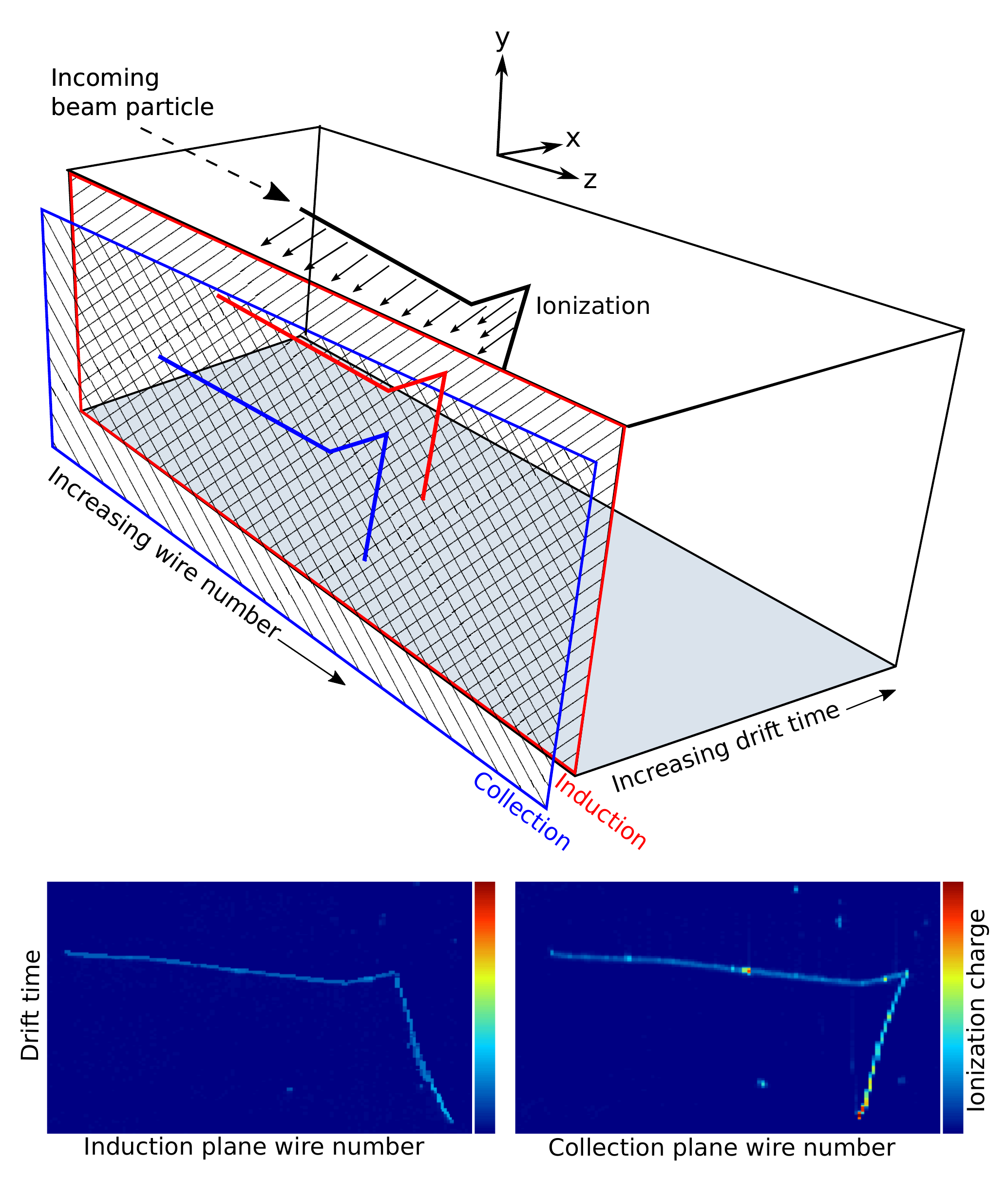}
\caption{The development of a LArTPC signal, produced by a charge particle which scatters within the medium, causing a kinked track. \emph{Top}: Ionization electrons drift toward the wire planes. Signals are formed first on the wires of the induction plane and then on the wires of the collection plane. \emph{Bottom}: LArIAT data induction and collection plane views of the particle's trajectory, shown in wire number vs. time. The amplitude of the wire signals determines the magnitude of the ionization charge. The drift time of the signals is used to measure each track's coordinates along the normal to the two planes.  The two planar views can be combined to form a full 3D track.}
\label{fig:lartpc_schematic}
\end{centering}
\end{figure}

When charged particles traverse liquid argon they leave a trail of
ionization electrons in their wake. In a simplified picture, the ionization electrons in a LArTPC
drift at constant speed along the electric field lines, which are directed
perpendicular to the beam. Planes of pads or parallel wires are located far
to one side of the drift volume, and the drifting electrons induce an electrical signal, or are collected on the pads/wires.  These signals are read out by the detector electronics. The sense wires used in the LArIAT LArTPC  are arranged in planes of parallel wires, oriented perpendicular to the E-field.  The wires of each plane are directed at an angle relative to the wires in the adjacent plane, as shown schematically in the upper panel of figure~\ref{fig:lartpc_schematic}. The bottom panels of the same figure show the projection of the 3-dimensional track onto the 2-dimensional wire planes, in wire number vs. drift time (not Cartesian coordinates as in the 3D view). 

Using the known E-field (and therefore the drift speed of the
electrons), the timing of the initial particle interaction, and the magnitude and duration of the signals on the sense wires,
a three-dimensional image of a particle interaction can be reconstructed.
Both the magnitude of the deposited charge and the topology of the
ionization are used for particle identification and calorimetry.

In addition to the ionization electrons, the liquid argon produces
scintillation light.
The light consists of two components: a prompt component with a lifetime of several nanoseconds and a delayed component with a lifetime in excess of 1~$\mu$s. Both arise from the decay of Ar$_2^*$ excimers, either from self-trapping excitons or by the recombination luminescence
of an ionization electron, argon ion and neutral atom. The prompt component is produced from the excimer spin-singlet state and the delayed component, from its triplet state. 
The yields of ionization electrons and scintillation light are largely complementary: the
amount of scintillation light decreases as the external electric field
increases. The yield of both ionization electrons and
scintillation light can be reduced by the presence of impurities. The
scintillation light is produced in a narrow spectrum peaked at 128~nm.
To enable the use of traditional light detection devices, which operate 
in visible wavelengths, a wavelength-shifting material must be introduced into the sensitive volume. The scintillation light signal, prompt in relation to the slowly-forming signal from the ionization, can be used as a trigger, indicating activity within the LArTPC. The scintillation light can also be used to enhance calorimetric measurements made with the ionization signal.

LArTPC technology~\cite{Willis1974,rubbia1977} was pioneered by the ICARUS collaboration which, through its run at the Laboratori Nazionali Gran Sasso~\cite{ICARUS-LNGS}, demonstrated the feasibility of using LArTPCs in long-baseline underground neutrino experiments. Following this effort, and marking the start of the US-LArTPC program, the Argon Neutrino Teststand (ArgoNeuT) experiment~\cite{ArgoNeuT-det} was deployed in the NuMI 
beamline~\cite{NuMI} at the Fermi National Accelerator Laboratory (Fermilab). The ArgoNeuT collaboration has published some of the first neutrino-argon cross section measurements, along with a number of calibration and detector studies, e.g., references~\cite{ArgoNeuT-CCincl,ArgoNeuT-b2b,ArgoNeuT-muoncalib}. Since LArTPCs, unlike Cherenkov detectors, can distinguish between electron- and photon-induced showers via the difference in ionization charge deposited in the detector, the experiments investigating the LSND/MiniBooNE electron neutrino appearance anomaly~\cite{LSND-appearance,MiniBooNE-appearance} have chosen LArTPC technology. The first of these is MicroBooNE~\cite{MicroBooNE-det}, an 87-ton active mass LArTPC located 470~m
downstream of the Booster Neutrino Beam (BNB) target, just upstream of the
MiniBooNE detector. The next phase, the  Short-Baseline Neutrino (SBN)
program, will see the addition of two more functionally identical LArTPCs
located on-axis in the BNB. The Short-Baseline Near Detector (SBND) will
be a new, 112-ton active mass LArTPC, situated 110~m downstream of the
BNB target. SBND will measure the unoscillated neutrino flux in the BNB, enabling searches in both the neutrino appearance and disappearance channels. The newly-upgraded ICARUS-T600 detector (476-ton active mass), formerly installed at the Laboratori Nazionali Gran Sasso and now deployed some 600~m downstream from the BNB target, will serve as the far detector. The large fiducial mass of ICARUS provides the SBN program with the experimental sensitivity to determine the nature of the $\nu_e$ appearance anomalies.

After its commissioning in the next decade, the Deep
Underground Neutrino Experiment (DUNE)~\cite{DUNE-CDRvol1,DUNE-CDRvol2,DUNE-CDRvol3,DUNE-CDRvol4} will be the flagship neutrino experiment in the US for many years.  DUNE will be located 1300~km from Fermilab, at the
Sanford Underground Research Facility (SURF) in Lead, South Dakota. In
DUNE, a high-power, wide-band beam of muon neutrinos and antineutrinos will be directed from Fermilab to the four, 10-kiloton fiducial mass LArTPCs which comprise its far detector.  By measuring the difference between the rate of appearance of electron neutrinos from a beam of muon neutrinos compared to the rate of appearance of electron antineutrinos from a beam of muon antineutrinos, DUNE will make measurements of both the CP-violating phase and of the relative ordering of the neutrino mass states.

In all of these experiments, the calibration and characterization of the
LArTPC response is a critical first step towards physics measurements.
However, the {\it in situ} calibration data available to the LArTPCs at
Fermilab is limited largely to cosmic rays and UV lasers. For DUNE, which is located deep underground, cosmic calibration data will be 
even more scarce. The goal of the LArIAT experiment, located in the
Fermilab Test Beam Facility (FTBF)~\cite{FTBF}, is to study the interactions of charged hadrons and leptons in liquid argon, free from the additional uncertainties that arise from nuclear effects in neutrino interactions~\cite{LArIAT-whitepaper}.  The charged particles produced in the LArIAT test beam are of the same type and energy range as the charged particles that are produced by neutrino-argon ($\nu$-${\textrm{Ar}}$) scattering and are therefore relevant to both the short- and long-baseline neutrino experiments. Using particles of measured momentum and type, reconstruction tools, particle identification (PID) algorithms, and calorimetry models can all be tested without excessive reliance on simulation.  The very flexible LArIAT experimental setup is also used in the development of new technologies that will enhance the capabilities of future LArTPCs.

\subsection{Overview}\label{sec:ExpermentOverview}



Details of the LArIAT experiment are presented in the following sections.
LArIAT's scientific goals are outlined in the next subsection. The LArIAT beamline and associated instrumentation are described in section~\ref{sec:Beamline}.  Section~\ref{sec:LArIATDetector} is devoted to the LArIAT TPC, including the light collection system, the front end electronics and data acquisition system. Experiment monitoring systems and data quality monitoring are described in section~\ref{sec:Monitoring}.  This is followed by a summary of data-taking periods and associated changes to detector configuration in section~\ref{sec:DataCampaigns}.  Beamline and detector performance are the subject of section~\ref{sec:DetectorPerformance}.

\subsection{Science Goals}\label{sec:ScienceGoals}

\begin{figure}
\begin{centering}
\includegraphics[width=\textwidth]{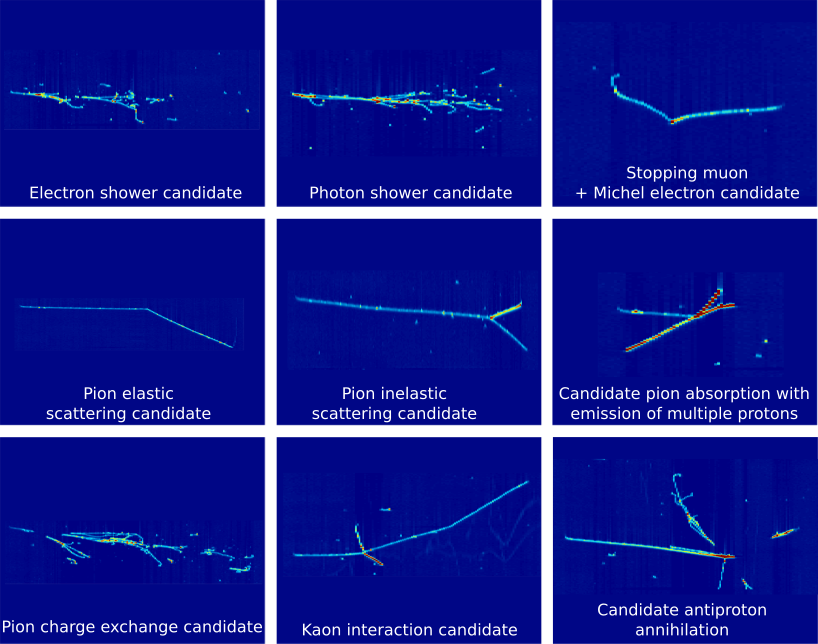}
\caption{Various candidate events from LArIAT Run II data, with tracks entering from the left. Each graph shows the collection plane view of time versus wire number (not to same scale in each graph).  The density of ionization charge is indicated by the color, with red the most dense. }
\label{fig:particle-topologies}
\end{centering}
\end{figure}

The broad scientific goals of the Fermilab neutrino program make many
demands on the LArTPC technology as well as on event generation and
detector simulation tools. This section reviews some of the most critical
questions and the role that LArIAT plays in resolving them. The detector used in a successful resolution of the short-baseline neutrino
anomaly, or in a measurement of CP violation, must have the ability
to distinguish between electromagnetic showers originating from photons and
those originating from electrons. In previous experiments that have relied on
Cherenkov radiation and/or scintillation light to identify electromagnetic
showers, photons and electrons leave very similar signals in the detector.
In particular, the largest background
to $\nu_e$ charged current interactions arises from neutral current
$\pi^0$ production in which only one photon from the prompt $\pi^0$ decay
is detected: either the two photon showers overlap too much to be separated,
or one of the photons escapes the detector before it converts to an $e^+e^-$ pair that can be detected. In a LArTPC, fully active calorimetric sampling of the
developing shower, as well as detailed visualization of the event topology
can help separate electron- and gamma-induced showers. Because a high
energy gamma remains invisible until it pair produces, the beginning of
a photon-induced shower should be physically separated from the
neutrino interaction vertex. Moreover, the
$e^+e^-$ pair will deposit twice as much ionization energy per unit length
as the single electron from a charged current neutrino interaction.
The ArgoNeuT collaboration has produced the only experimental investigation of
$e/\gamma$ separation in a LArTPC~\cite{ArgoNeuT-nue}, a study based on a small sample of electron- and photon-induced showers. LArIAT has collected a large sample of electron showers produced by the FTBF beam. LArIAT has also collected
a large sample of photon-induced showers: from electron bremsstrahlung or
from $\pi^0$s produced in pion charge-exchange reactions. Examples of electron-induced and photon-induced showers are shown in figure~\ref{fig:particle-topologies}.

At energies of a few GeV, neutrino interactions with nuclei produce
significant numbers of charged pions. While pion-nucleon interactions
are of fundamental theoretical interest in their own right, they are also
a critical source of systematic uncertainty in LArTPC neutrino
experiments which involve moderately heavy nuclei.
There is uncertainty in the primary neutrino interaction with the nucleus,
particularly in the interactions of the scattered particles as they
emerge from the recoiling nucleus. LArIAT measurements will enable data-driven improvements to the simulation of neutrino-nucleon scattering in event generators
such as GENIE~\cite{GENIE}. There is also uncertainty
in the subsequent interactions of the final state particles with other
liquid argon nuclei.
The reconstruction of the incoming neutrino's energy, which is
determined by the energies and types of the final state particles, would be
adversely affected by undetected or incorrectly identified pions.

Although extensive pion scattering experiments over the past
40 years have provided detailed measurements of differential cross sections
for a variety of final state kinematic variables, the uncertainties on the
total hadronic cross sections range from 10\% to 30\% for light nuclei and
are even larger for heavy nuclei~\cite{Wilkin:1973xd,BINON1970168,Ashery1981,Allardyce1973,Compilation1,Compilation2}. In particular, experimental data for
pion-Ar interactions are sparse and current Monte Carlo simulations, such as Geant4, determine argon cross sections by interpolating data from lighter and heavier nuclei. The initial goal of the LArIAT pion interaction analysis is to measure the total hadronic interaction cross section as a function of pion kinetic energy in the range between 400 MeV and 1.2 GeV. Subsequent studies will focus on identification algorithms for each of the pion's individual interaction modes with argon nuclei, including pion absorption, charge exchange, and elastic and inelastic scattering, and measurements of the cross sections of these exclusive interaction channels. In addition to hadron-argon cross section measurements, the LArTPC technology provides a unique opportunity to tune Geant4 hadron models. By reconstructing the kinematics of final state particles in hadron interactions, several nuclear cascade models can be tested.

The LArIAT collaboration will also study kaon identification and
reconstruction efficiencies and measure the $K^{\pm}$-Ar interaction cross
sections. These studies are of particular relevance to future proton decay
searches with the DUNE detector, which will be highly sensitive to
decay modes involving kaons, such as $p\rightarrow \bar{\nu} K^+$; separating the kaons from pions and protons will require a detailed understanding of kaon
interactions in argon. As with pions, simulation packages determine 
kaon interaction cross sections on liquid argon by interpolating
data from lighter and heavier nuclei. The small but significant
fraction of kaons in the LArIAT beamline permits direct measurements
of those crucial cross sections.

In addition to these more traditional hadronic cross sections,
LArIAT has the opportunity to make the first-ever studies of both
hadron multiplicity and final state topologies in proton-antiproton annihilation in a liquid argon detector. Again, the results will be important for tuning simulations. They will also help constrain the nuclear modeling used in searches for neutron-antineutron oscillations, a baryon number violating phenomenon predicted by many extensions to the Standard Model.
Figure~\ref{fig:particle-topologies} shows an example of a candidate proton-antiproton annihilation at rest, with multiple hadrons exiting the interaction vertex.

Although it has been used to form a prompt trigger,
the full exploitation of scintillation light, for particle identification
and calorimetry, has never been demonstrated in a liquid argon detector.
The LArIAT collaboration has implemented a highly efficient, light collection
system, adapted from similar designs found in dark matter liquid
argon detectors~\cite{WArp-Acciarri,ArDM,DarkSide}. The
goal of the system is to improve the calorimetric reconstruction of events
by combining the collected light with ionization measured on the wires, without
the need to rely on recombination models to account for the energy lost
to light production. Results from LArIAT on this topic are presented in reference~\cite{LArIATMichel}.  There is also the possibility of using the relative
fraction of "early" and "late" scintillation
light to enhance particle identification techniques.
The LArIAT light collection system uses wavelength-shifter-coated reflector
foils lining the interior of the LArTPC to shift the vacuum ultraviolet (VUV) scintillation
to a longer wavelength, at which standard light detectors are sensitive.
The LArIAT detector also provides a testing ground for the exploration
of new, light-detection techniques, various coating methods, as well as
side-by-side comparisons of new devices (e.g., VUV-sensitive silicon photomultipliers) with
more traditional PMTs. The design and performance of the LArIAT light
collection system,
in various configurations, is described in section~\ref{sec:PhotonSystem}.

%
The LArIAT experiment was also designed for comprehensive studies of particle
identification in a LArTPC. For example, when charged hadrons propagate through
liquid argon and come to a stop within the TPC (and without an inelastic
interaction), the pattern of energy deposition along the track can be used
to determine the particle type. Using curves of particle $dE/dx$ vs.
{\it residual range}, the remaining distance before a track's stopping point, ArgoNeuT demonstrated that protons could be separated from pions and muons. LArIAT will enhance and extend the ArgoNeuT results with
high-statistics, beam-test data, collected with known particle species. The beam composition includes a reasonable fraction of kaons for the higher-energy beam settings, which will allow studies of the separation of protons from kaons and kaons from pions and muons.

A high-statistics sample of positive and negative muons, produced and
tagged in the LArIAT beamline, makes possible a different kind of PID
study: determining the sign of a particle's charge
without a magnetic field.  A $\mu^+$ decay results in the emission of an $e^+$
with a well-known energy spectrum while a stopping $\mu^-$ will either decay or
capture with a known probability. The capture of the $\mu^-$ is accompanied by
the emission of a neutron and a photon. Thus topological and calorimetric
differences should distinguish $\mu^-$ capture from $\mu^+$ decay.

\section{Beamline instrumentation and performance}\label{sec:Beamline}
\begin{figure}[htb]
\begin{centering}
\includegraphics[width=\textwidth]{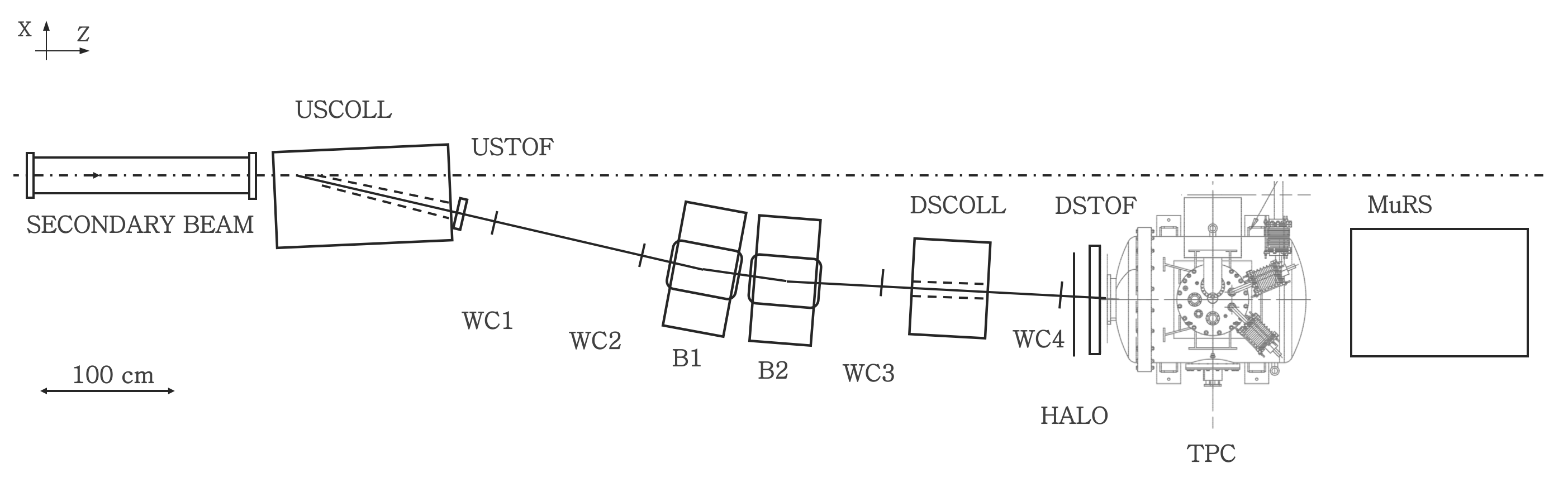}
\caption{Bird's eye view of the tertiary beam. }
\label{fig:tert-layout}
\end{centering}
\end{figure}
\noindent

The goal of the LArIAT beam line systems is to provide a well-understood, low intensity, charged particle beam with a momentum range between 0.3 and 1.4 GeV/c. The tertiary beamline which serves the LArIAT experiment is located in Fermilab's MC7 experimental hall. To produce the tertiary beam, the primary proton beam from the Main Injector is first focused on
a tungsten target, located further upstream at MC6. A secondary beam is momentum-selected
in the range between 8 and 80~GeV/c and then transported to MC7, where it
is focused onto a copper target. The copper target, and the steel collimator in which
it is mounted, mark the beginning of the tertiary particle beam and the LArIAT apparatus. 

A top-down view of the tertiary beam line and LArIAT apparatus is shown in figure~\ref{fig:tert-layout}; the secondary pion beam is incident from the left. The tertiary beam is formed from particles that are produced in the copper target and exit the upstream collimator (USCOLL) at an angle of 13 degrees relative to the secondary beam.  They first enter the upstream time-of-flight (USTOF) detector. The particles then pass through the magnetic spectrometer: two multi-wire proportional chambers (WC1 and WC2), followed by two dipole bending magnets (B1 and B2), and then a second set of wire chambers (WC3 and WC4). Particles which exit WC3 also pass through the downstream collimator (DSCOLL) and then through a pair of aerogel Cherenkov detectors (not pictured) before reaching WC4. A beam halo veto counter (HALO), with a 13~cm aperture, tags errant particles emerging from the DSCOLL. The time of flight measurement is completed by the downstream time-of-flight (DSTOF) counter, which sits just downstream of the HALO detector and just upstream of the TPC cryostat. Particles which pass all the way through the TPC, mostly pions and muons, are tagged by a set of scintillator paddles (not pictured) and measured in a small range stack (MuRS).

\begin{figure}[htb]
\begin{centering}
\includegraphics[width=0.49\textwidth]{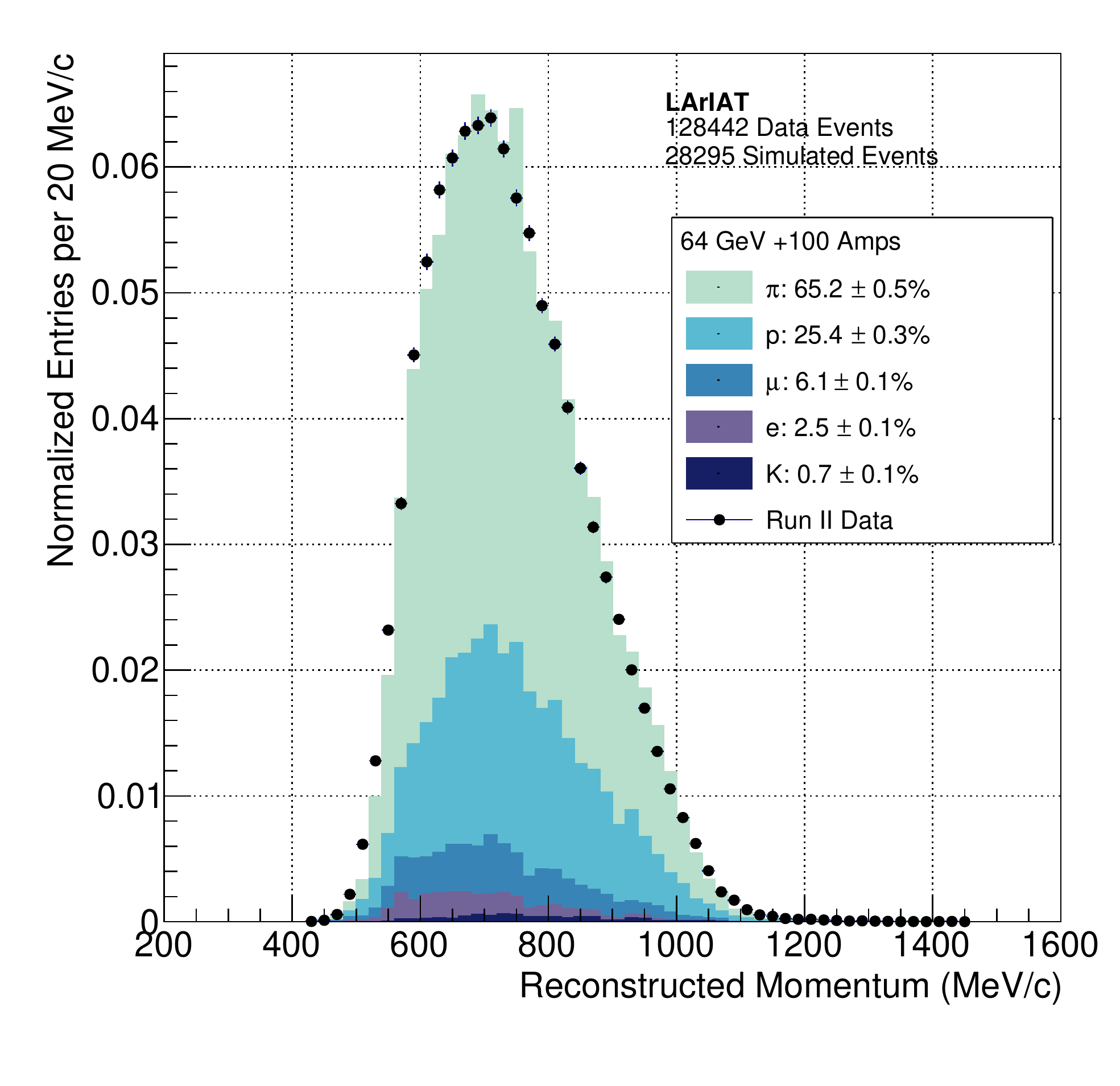}
\includegraphics[width=0.49\textwidth]{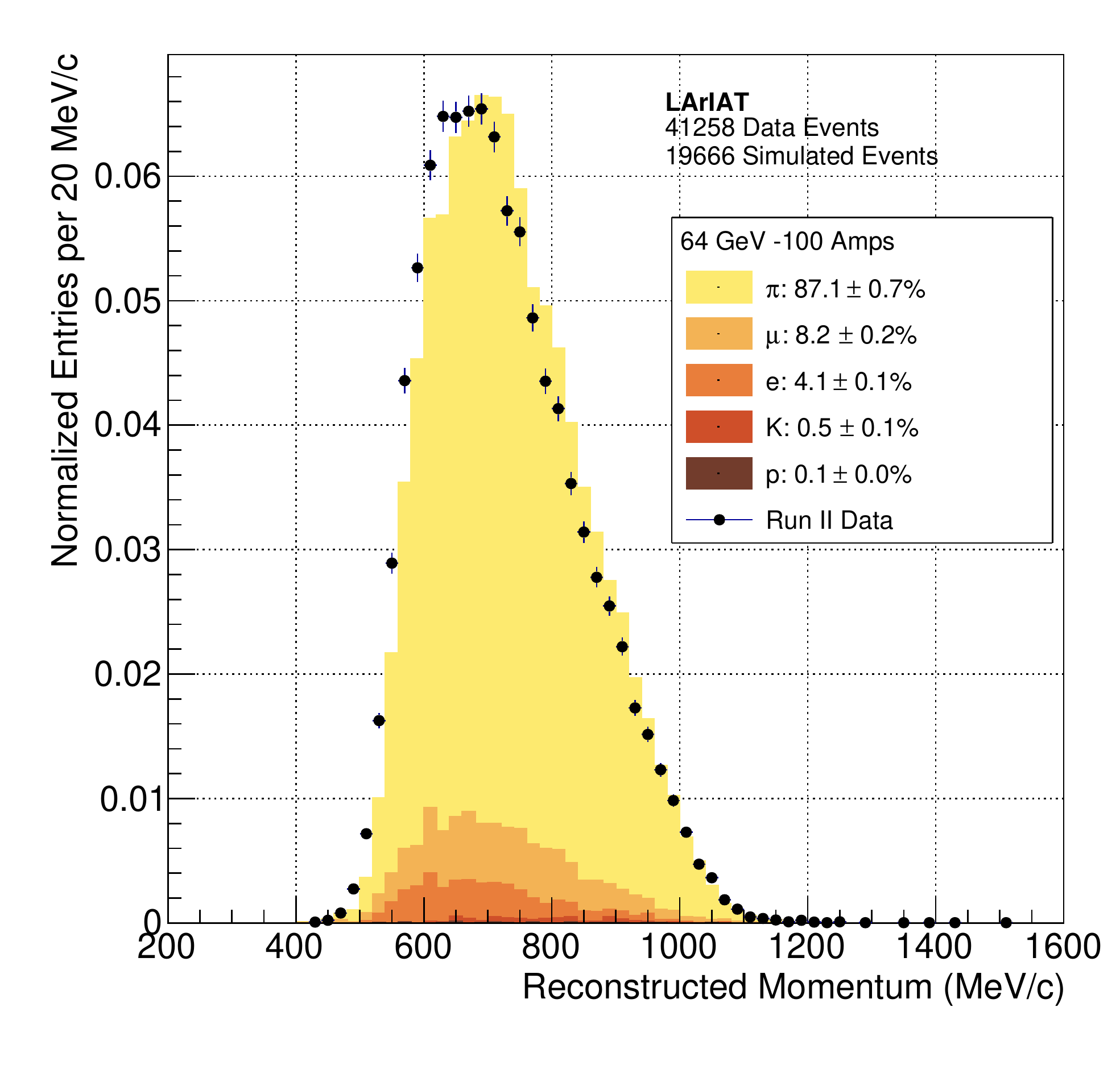}
\includegraphics[width=0.49\textwidth]{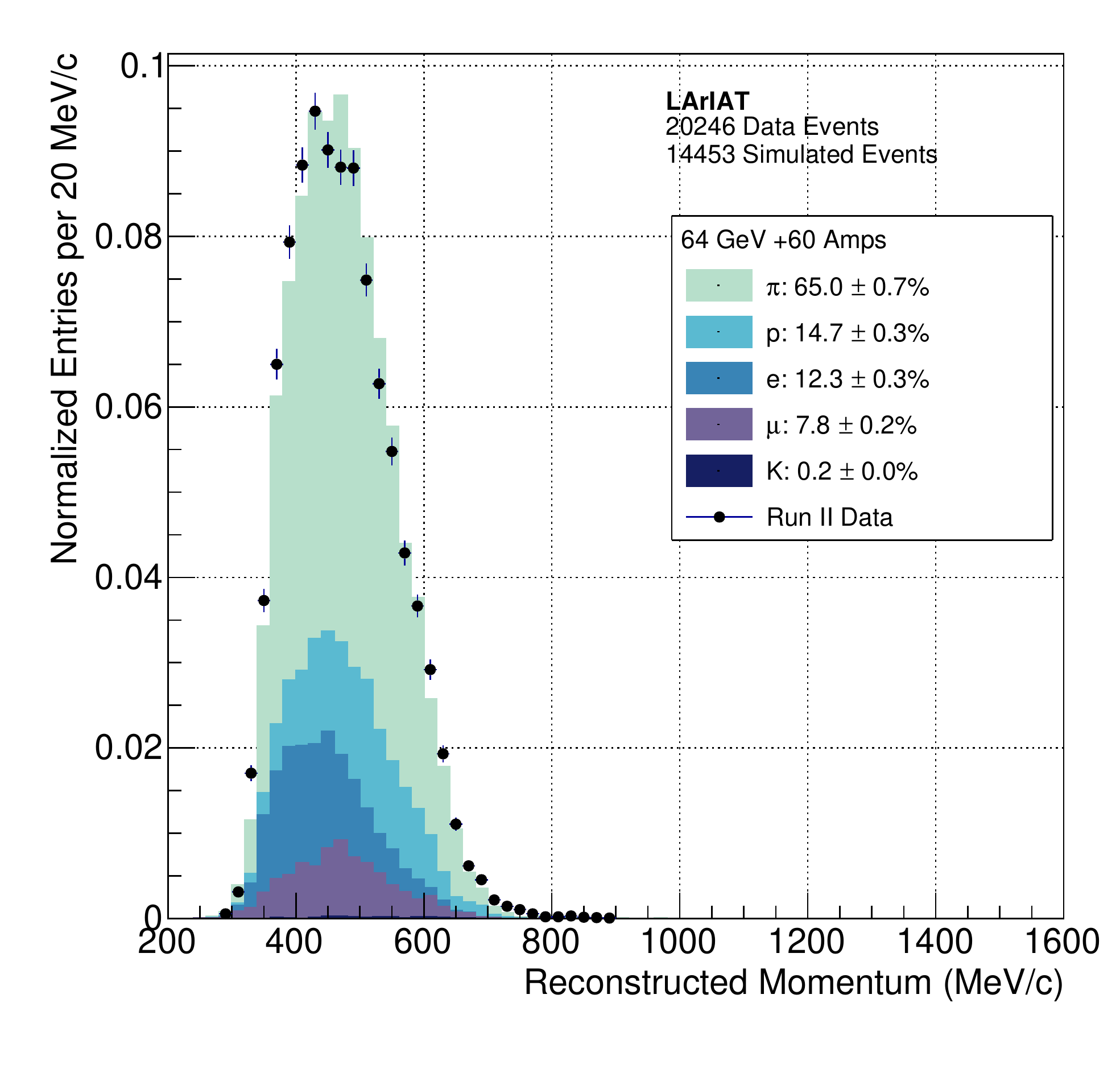}
\includegraphics[width=0.49\textwidth]{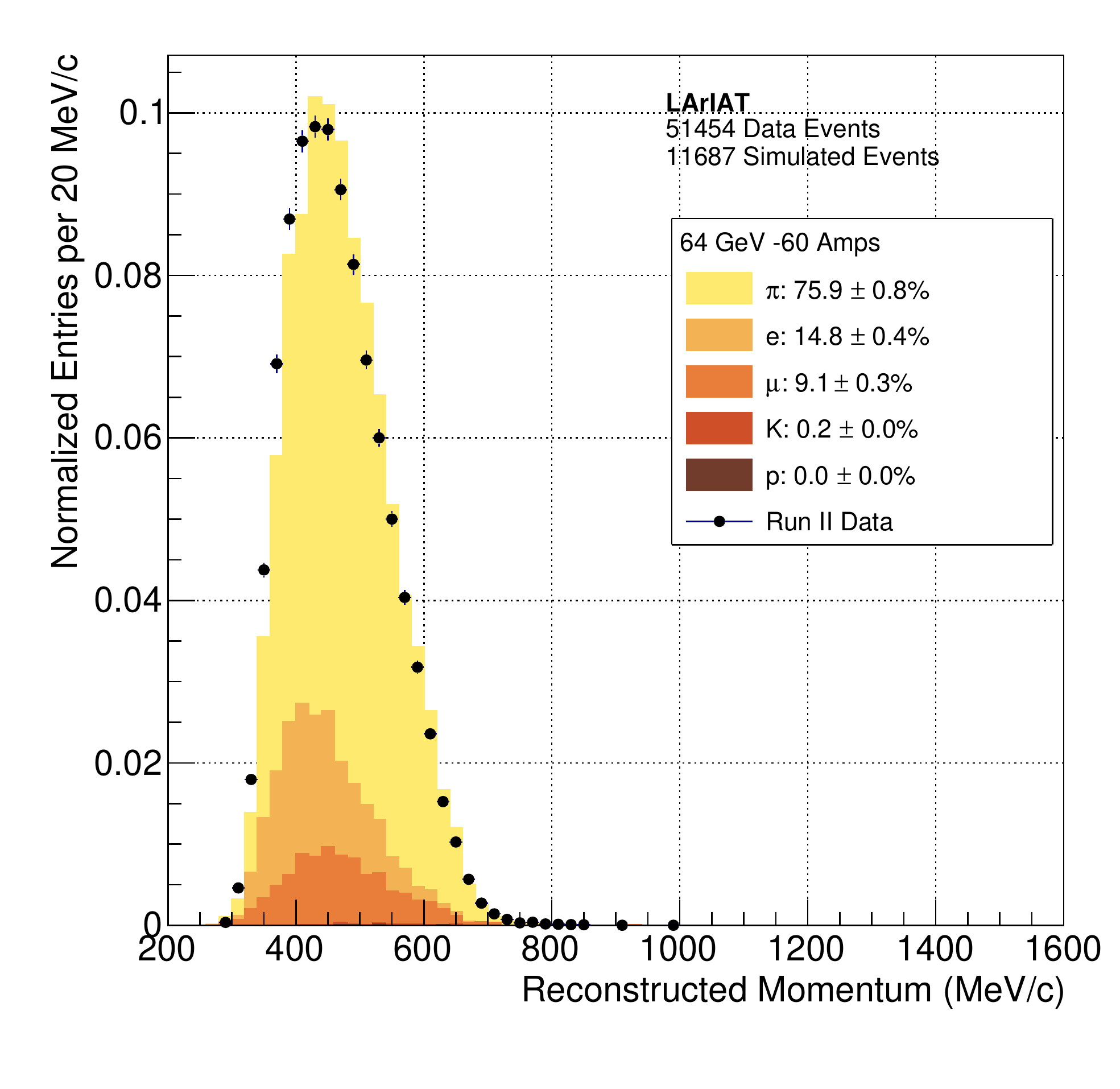}
\caption{Momentum spectrum and composition of the tertiary beam just downstream of the spectrometer. The top (bottom) two panels show the high-energy (low-energy) tune for positive-polarity, tertiary beam (\emph{left}) and negative- polarity, tertiary beam (\emph{right}). The secondary beam energy was 64 GeV for both tunes. In the high-energy tune, the tertiary bending magnet current was 100~amperes, while in the low-energy tune it was 60~amperes.}
\label{fig:momentum-tertiarybeam}
\end{centering}
\end{figure}

For most of the
LArIAT data-taking, the secondary beam
momentum was fixed at 64~GeV/c. For $p>200$~MeV/c, the positive tertiary beam is comprised mostly
of pions and protons with smaller fractions of electrons, muons and kaons. Figure~\ref{fig:momentum-tertiarybeam} shows stacked momentum spectra for the nominal 64~GeV/c secondary beam setting and with the tertiary line set for positively charged particles (left) and negatively charged particles (right), for two different tertiary beam settings: the higher-energy tune where the tertiary beam magnets are set to 100~A (upper) and the lower-energy tune where the magnets are set to 60~A (lower). The simulations to determine beam composition and momentum spectra were performed using G4beamline~\cite{g4beamline}, with 64~GeV/c pions impinging on a copper target at the upstream end of the tertiary beamline, and with the tertiary beam magnets set to the polarity and current values indicated in the figure legends.

\subsection{Tertiary beam spectrometer and particle ID system}\label{sec:Spectrometer}

Although the bending magnets are configured similarly to those of the MINERvA T-977 test beam calibration~\cite{MinervaTestbeam},
the geometry of the beamline has been optimized
for LArIAT. The 13 degree angle between the tertiary and secondary beamlines,
combined with a 10 degree bend through the pair of  dipole magnets,
provides a source of particles with momenta that can be tuned in the range
between 0.3 and 1.4 ~GeV/c.
Moreover, as noted in the introduction, a set of wire chambers and time-of-flight
scintillator paddles, also shown in figure~\ref{fig:tert-layout}, provide tracking, momentum determination and particle identification (PID).

\subsubsection{Time of flight system}\label{sec:TOF}

The LArIAT time-of-flight (TOF) detector system consists of two scintillator paddles which bracket the spectrometer, as seen in figure~\ref{fig:tert-layout}. The upstream
paddle is small (10~cm $\times$ 6~cm) and the downstream
paddle is larger (14~cm $\times$ 14~cm). Lightguides are mounted
on all four edges of each paddle. 
In Run~I and Run~II (which correspond to different configurations of the detector systems, discussed in detail in section~\ref{sec:DataCampaigns}), each of the paddles was read out
by two PMTs. The long axis of the
downstream paddle was directed horizontally and read out with PMTs placed at both ends.
The upstream paddle was rotated by 45 degrees with respect to the horizontal
and its two PMTs were mounted to the left side. In Run III, four more PMTs were
added to the system. The upstream paddle was read out with the four original
PMTs and the downstream paddle with four additional 
PMTs. The measured efficiency of each paddle was greater than 99\%.

During data-taking cycles, signals from the TOF PMTs were sampled at 1~GHz
with a CAEN V1751 digitizer~\footnote{CAEN Electronic Instrumentation, S.p.A., Via della Vetraia, 11, 55049 Viareggio LU, Italy. \url{http://www.caen.it}} and 10-bit samples were stored in a circular buffer.
In response to an experimental trigger, a 28.7~$\mu$s window of samples,
starting approximately 8.4~$\mu$s before the trigger, was written to the
output. 
The amplitude of the TOF signals was typically 200~mV in the upstream counter but only 50~mV in the downstream counter. The signals have a rise time (10-90\%) of 4~ns and
a full width, half-maximum of 9~ns. The rate in the upstream counter was
typically 15~kHz and much less, approximately 400~Hz, in the downstream counter.

Because the shape of signals from each TOF PMT is highly uniform, the
time of the pulses is determined using an oversampled template derived
from the data itself. The DC
offset, or pulse pedestal, is taken from samples which are tens of microseconds from the pulse.
The template is stretched vertically to match the pedestal-subtracted
pulse amplitude and moved horizontally to match the time. The pulse
time-pickoff resolution is better than 100~ps and the relative amplitude
resolution is better than 2\%. Because the average interval between  pulses
in the TOF system is large compared to its dead time, pulse pile up is
not a significant concern. Given the uniform width of the pulses produced
by any given PMT ($\sigma\sim400$~ps), the pulse width can be used to flag
events where two pulses overlap closely in time. 

The time-independent offsets in each channel, the cable delays and PMT transit times,
were calibrated with cosmic rays. 
The two scintillators were placed on top of one another
with the PMTs at their nominal voltages. In an attempt to account for the small
time of flight between the paddles, the calibration was made in two
configurations. In the first configuration, the paddles were placed in their nominal positions. In the second configuration, the positions of the two paddles were
switched. In principle, the two average time differences should be equal
and opposite and, based strictly on the separation of the paddles, the time
difference should have been no more than 0.1~ns. Using the calibrated
offsets, the average time difference in one configuration was 0.31~ns and
0.11~ns in the other.   The apparent offset of a few
tenths of a nanosecond is a good estimate of the reproducibility
of the measurement and is comparable to the intrinsic timing resolution
of 0.25~ns.

\begin{figure} [htb]
\begin{centering}
\includegraphics[width=0.8\textwidth]{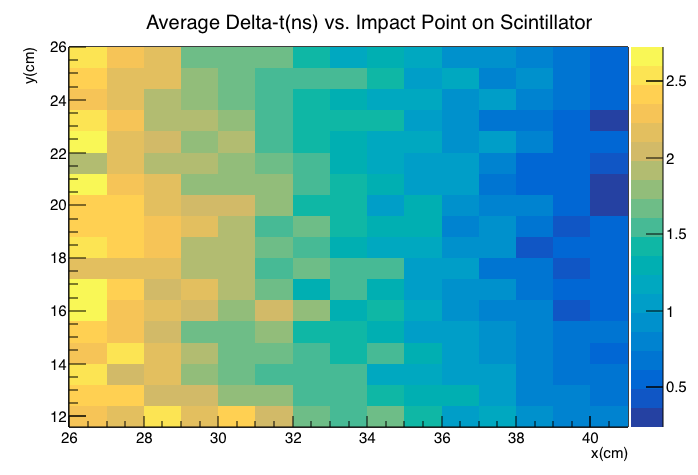}
\caption{
Difference in arrival time at the two downstream PMTs of the TOF system for impact points on the scintillator paddle.
}
\label{fig:tofdelay}
\end{centering}
\end{figure}

As indicated above, there are also variable delays associated with the
passage of light through the scintillator. If a rectangular scintillator
is read out at both ends, the average of the measured times is a
good estimate of the true time of the particle's passage. Alternatively,
the time {\it difference} between the signals should vary linearly along
the line connecting the two scintillators.
To determine a particle's time of arrival, pulse times from the multiple
PMTs mounted on each paddle had to be combined. In Run I and Run II,
with two PMTs
mounted on each paddle, a simple average was used. In the case of the upstream paddle, with PMTs mounted on two adjacent edges of the rectangular scintillator, the average does not correct for the fact that light from particles which strike the scintillator far from the two PMTs has a larger transit time than the light produced by particles which strike close to both PMTs. In the case of the
downstream paddle, with PMTs mounted at opposite ends, the average
corrected for the fact that transit time to the two PMTs depended on the
particles' point of impact on the scintillator. A plot of the time of arrival
{\it difference} between the two downstream PMTs vs the longitudinal
impact point,
made with the help of track segments from the downstream wire chambers, is
shown in figure~\ref{fig:tofdelay}. As expected, the time difference
changes systematically
across the face of the scintillator. The effective speed of light in the scintillator, approximately 12~cm/ns, is consistent with previous results~\cite{Charpak}.

Taking the average minimizes TOF uncertainties arising from optical path differences
in the scintillator. However, even for a set of particles which
pass through a single small area of the paddle, the times of pulses
registered in the two PMTs are still spread by approximately 300~ps, an
uncertainty which is probably caused by transit time jitter in the PMTs
themselves. This jitter is evident in both the upstream and downstream
detectors.

Although climate-control for the LArIAT enclosure in MC7 is fairly primitive, there
is little sign of systematic timing drift over longer periods. The average
time differences between pairs of PMTs reading out the same scintillator
drifted by no more than 150~ps over the 3-4 months of a data-taking period.

\subsubsection{Magnets}
The pair of spectrometer magnets in the LArIAT tertiary beam are Fermilab's ``NDB series'' dipole electromagnets\footnote{Device details for NDB magnets: \url{https://www-tdserver1.fnal.gov/proeng/Series/qrySeriesDetails.asp?qsEnterSeries=NDB}}, originally used in the antiproton ring. The
magnets have a gap height of 14.224~cm, a gap width of 31.75~cm and an
iron length of 46.67~cm. Since the horizontal aperture presented by the magnets is wider
than that of the wire chambers (which are approximately 12.5~cm on a side), only the
central part of the magnet is used. Over this limited region, there is
negligible variation in the field integral. The field intensity in one
of the pair of magnets was measured using two Hall probes, both calibrated
with NMR. 

The second magnet, having been constructed to the same standard and having the same
history, is assumed to have a very similar response.

The magnets were air- and water-cooled during operations and their temperatures
monitored. For all data-taking periods, LArIAT collected data with both positive and negative polarity and explored different momentum ranges by changing the magnets' current settings. Each magnet setting is labeled by the charge of the particles and the magnitude of the current in amperes; the latter sets the approximate range of particle momenta. For example, $+$60A indicates a data set containing positive particles with a momentum between approximately 200 and 800~MeV/c. Due to concerns with overheating, the current was limited to 100~A, corresponding to a maximum usable field at the magnets' center of
3.5~kG.

\subsubsection{Multiwire proportional chambers}

\begin{figure}[htb]
\begin{centering}
\includegraphics[width=0.75\textwidth]{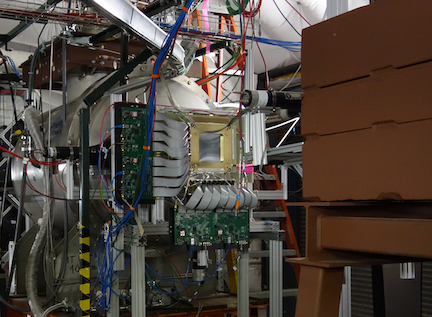}
\caption{One of the multiwire proportional chambers in the LArIAT tertiary beamline.
}
\label{fig:wirechamber}
\end{centering}
\end{figure}

The wire chambers are modeled on the Fenker chambers~\cite{Fenker:1983km}, long in use at Fermilab,
with additional grounding to improve the signal to noise ratio in the
electronic readout. The chambers, one of which is shown in figure~\ref{fig:wirechamber}, have an effective aperture of 128~mm in both the horizontal and vertical directions. The wire spacing
is 1~mm, with 128 wires in each view. The chambers run on a mixture of
85\% argon and 15\% isobutane. Each chamber's voltage was set at approximately 2450  volts.
The chamber efficiencies, measured using a cosmic telescope, were better than 99\%.
Since the signals have
limited timing resolution, the spatial resolution 
is determined largely by the spacing of the wires.

The raw signals are amplified, shaped and discriminated by ASDQ chips~\cite{ASDQchip} mounted directly on the frame of the wire chamber. The ASDQ features an effective differential input and its shaping circuit cancels both the ion and preamplifier tails. A baseline restoration circuit removes DC offsets and provides very uniform trigger thresholds across the inputs.
Short, flat cables connect the output of the discriminators to a new multi-hit TDC~\cite{Sten}.
The TDC provides a maskable, fast-OR output that can be used in a first-level trigger. The TDCs have a time resolution of 1.18~ns/bin (1/16$^{\textrm{th}}$ of the Main Injector's RF period), and can accept multiple hits per wire. These specifications are more than adequate for the LArIAT experiment, where the maximum rates and timing requirements are modest. There is sufficient memory on the TDC board to store a full spill of data (approximately 100 events) and sufficient time between spills to read it all out. A specially-designed
controller provides power to the TDC board and serves as a data interface, using the LVDS standard, between the board and the rest of the data acquisition system. The configurations of the TDC and ASDQ boards: the time window for accepting hits, timing offsets, the front-end thresholds and pulse-shaping parameters, are programmed through the controller.

\begin{figure} 
\centering
\includegraphics[width=0.7\textwidth]{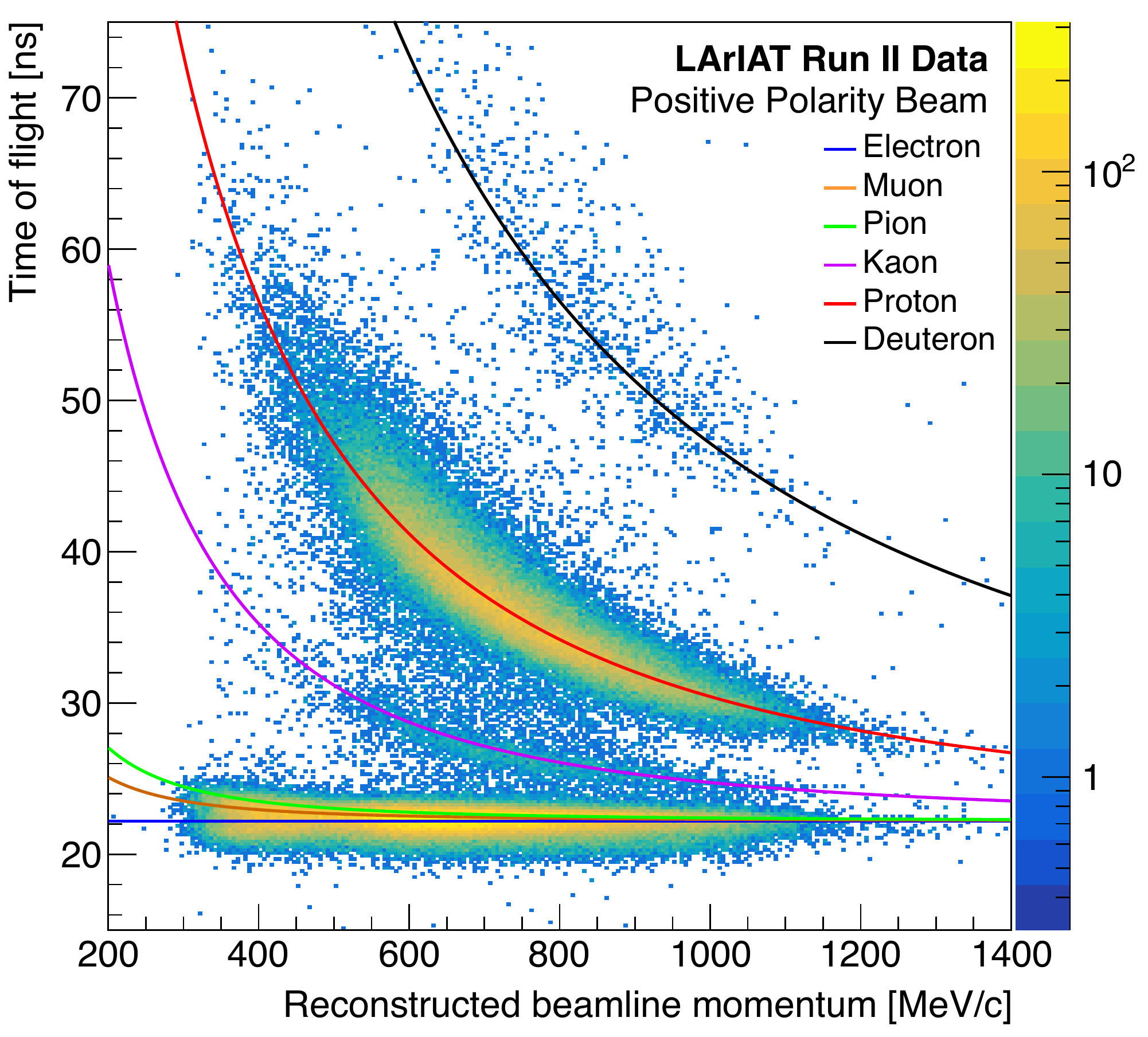}
\caption{\label{fig:tof_mom} 
Measured time of flight vs. momentum in the beamline spectrometer system for data collected with a positive-polarity tertiary beam. The colored lines show the expectations for different particle species.
}
\end{figure}

The combination of time-of-flight information and momentum determination via tracking through the wire chambers provides the ability to separate particle species in the beamline before they reach the TPC. This is shown in figure~\ref{fig:tof_mom} for LArIAT positive-polarity tertiary beam data, along with the expectations for the most common particle species in the beam.

\subsection{Auxiliary systems}\label{sec:Auxiliary}
The auxiliary beam line systems: the Aerogel Cherenkov detectors, the punchthrough detector, muon range stack (MuRS) and cosmic ray paddle detectors were used in particle identification. 
\subsubsection{Aerogel Cherenkov detectors}\label{sec:Aerogel}
A gel is a mixture in which liquid is dispersed within a solid matrix. In an {\it aerogel}, the liquid component of the gel is replaced by a gas. The resulting material is an ultra-light solid with low thermal conductivity and a low index of refraction which depends on the density. In the aerogel threshold Cherenkov detectors used by LArIAT, the density of each is chosen so that over a given momentum range, muons emit Cherenkov radiation while pions do not. The two threshold aerogel detectors used by LArIAT have different indices of refraction (1.103 and 1.057), allowing for muon-pion separation over two different momentum ranges. Both aerogel detectors are placed at the exit of the downstream collimator, between the two downstream wire chambers. 

The lower refractive-index detector consists of two stacks, each stack with four aerogel tiles, presenting a cross sectional area of $83~{\textrm{mm}}\times 94~{\textrm{mm}}$ to the beam. The two stacks are oriented side-by-side in the beam direction, giving a total thickness of 188~mm along the beam. The array of aerogel tiles is wrapped with a light diffusion sheet and viewed from the side by two photomultiplier tubes.

\subsubsection{Punchthrough detector and muon range stack}\label{sec:Punchthrough_MuRS}

The identification of particles which pass without stopping through the TPC 
is augmented by two detectors placed just downstream of its far end.
The {\it punchthrough} detector is composed of four identical scintillator
slabs, optically coupled through tapering light guides to 3-inch (76~mm)
PMTs. In the rhomboidal region where two slabs overlap, passing particles
encounter two layers of scintillator. The paddles were calibrated {\it in situ},
during beamline tuning.  Analog signals from the PMTs are
discriminated and coincidences of the overlapping pairs are formed. The
resulting logic signals are used in forming the LArIAT trigger.

\begin{figure} 
\centering
\includegraphics[angle=270,width=0.9\textwidth]{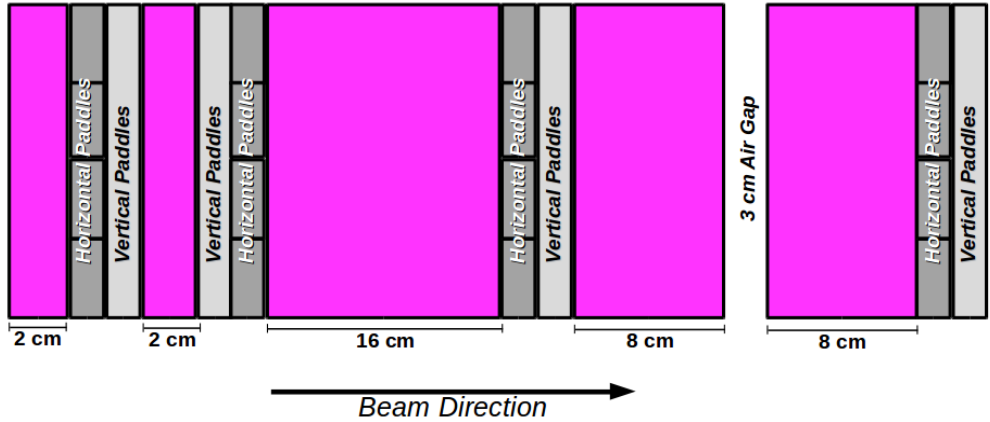}
\caption{\label{fig:MuRSSchematic} 
Side view of the upstream half of the muon range stack. Double layers of scintillator are sandwiched between steel plates. In this view, the segmentation of the horizontal paddles is visible.
}
\end{figure}

  The muon range stack (MuRS), the last detector in the LArIAT beamline, makes
a more detailed energy measurement of through-going particles and can be
used to distinguish muons from pions. The muon range stack, shown schematically in figure~\ref{fig:MuRSSchematic}, 
consists of eleven steel slabs of various thicknesses, 
with layers of scintillator in some of the gaps between them.
Muons, which experience no strong interactions, penetrate deeper into the
stack than pions of the same momentum. The depth of penetration is a
reasonable measure of the incident muon's kinetic energy.

Each of the four active layers in the MuRS consists of two planes of four
scintillator paddles, wrapped in reflective foil and black plastic.
The paddles are each 5~cm wide.
The paddles in the upstream 
plane of each pair are aligned horizontally
and those in the downstream plane are aligned vertically so that the position of
a particle which leaves a hit in both planes will be determined to roughly
${\frac {5} {\sqrt{12}}}$~cm in both X and Y.

Each of the horizontal paddles is read out with a single PMT through a
fishtail-shaped wave\-guide, similar to that used in the punchthrough veto
system. The vertical paddles, added at the end of Run~II, are read out
with pairs of 1-inch (25~mm) photomultiplier tubes, 
whose apertures sit flush against the surface of the scintillator.
Since this readout scheme reduces the fraction of light
that reaches the PMTs, relative to the wave-guides used for the
horizontal paddles, a signal from either PMT on a vertical bar is considered
sufficient to form a hit.

\subsubsection{Cosmic ray paddle detectors}\label{sec:CosmicRayPaddles}

\begin{figure} 
\centering
\includegraphics[width=0.6\textwidth]{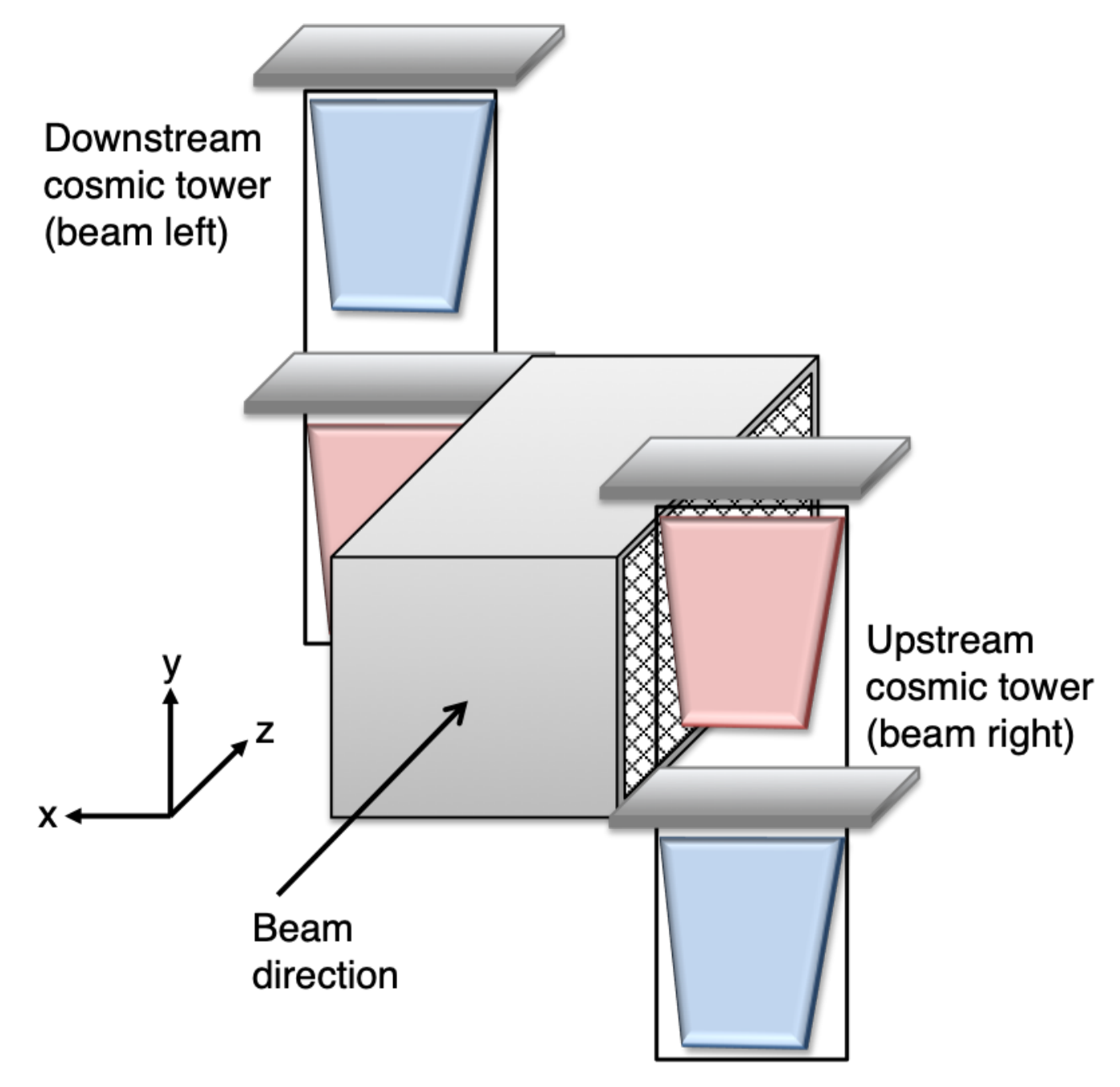}
\caption{\label{fig:cosmic_towers} 
Schematic view of the upstream and downstream cosmic towers that are used to select cosmic ray particles crossing the body diagonal of the TPC. Triggers are formed for coincidences of the upstream lower and downstream upper scintillator paddles (shown in blue), or the upstream upper and downstream lower scintillator paddles (shown in red). Not to scale.
}
\end{figure}

LArIAT's cosmic ray trigger consists of two {\it cosmic towers}, shown in figure~\ref{fig:cosmic_towers}, which frame the cryostat.
One stands beam right and upstream of the cryostat. The other stands beam left and downstream of the cryostat.  Each cosmic tower is composed of two
paddle assemblies, upper and lower. Each paddle assembly consists of
four paddles: a matched pair which stand upright and a second matched pair
which lie across the top of the assembly. The logical "AND" of discriminated
signals, from vertical paddle assemblies at either end, selects
cosmic ray muons which cross the TPC along one of its diagonals.
Signals from the horizontal paddles provide a veto for
downward-going cosmic ray air showers. A large fraction of the particles
selected by the cosmic ray muon trigger cross both the anode 
and the cathode. The tracks provided by these triggers have relatively
uniform ionization density, and because their trajectories cross a body diagonal of the active volume, they also span the full range of possible drift lengths. The attenuation of the ionization signal as a function of
drift length is used to monitor the level of electronegative contaminants
in the liquid argon. The cosmic tracks also provide an {\it in situ}
calibration sample for the calorimetry and electric field studies outlined
in section~\ref{sec:ChargeCalibration}. The total trigger rate for the coincidence of these cosmic towers is 0.032~Hz, or roughly two cosmic ray muons per minute.

The paddles have a trapezoidal shape\footnote{These paddles were recovered from the CDF detector during its decommissioning. Their shape is merely a consequence of the original design for that experiment.}  and are each enclosed in an aluminum
case. Light from the paddles is read out with
wavelength-shifting optical fibers, which run along one of the long sides,
and are optically coupled to a low voltage, Zener-diode Hamamatsu
H5783 PMT. Signals from the PMT are amplified and
discriminated by a custom PMT amplifier and discriminator (PAD) circuit,
mounted at one end of the paddle. The discriminated signal is sent
through a CAT5 cable  to a Control and Concentrator Unit (CCU), which
creates an ECL output for the trigger. The CCU also provides high voltage
to the PMT and the threshold level for the PAD circuit.

Candidate paddles were selected from a pool of more than 300 scintillating
counters, inherited from the CDF detector. Both the efficiency and accidental
rate as a function of voltage were measured for each counter. The paddle
under test was sandwiched between four sample counters: two above and two
below.
The efficiency was defined as the ratio between the 5-fold
coincidence rate and the 4-fold coincidence rate of the sample counters alone.
The accidental rate was defined as the number of 5-fold coincidences
registered per unit time, with the signal of the paddle under
test delayed by 5~$\mu s$. 
The candidate paddles with the highest efficiency and
lowest single count rate were selected for inclusion in the system. The efficiency of all the paddles selected was greater than 99\%.

\section{The {LArIAT} detector}\label{sec:LArIATDetector}

The heart of the experiment is the liquid argon time projection chamber, which was designed to serve as a development test bench for the next generation of detectors for the US neutrino program. Multiple readout wire planes make possible 3D imaging. Fine wire spacing and low noise cold electronics provide superb tracking resolution. The details of the LArTPC components, the cryostat, supporting cryogenic infrastructure used for cooling and purification of the liquid argon, and the light collection system are all described in this section.

\subsection{Cryogenic system}\label{sec:Cryo}
In the following sections, we describe the principle features of the cryogenic system: the cryostat, as well as the liquid argon purification and monitoring systems.
\subsubsection{Cryostat}

\begin{figure}[tb]
\centering
\includegraphics[width=0.49\textwidth]{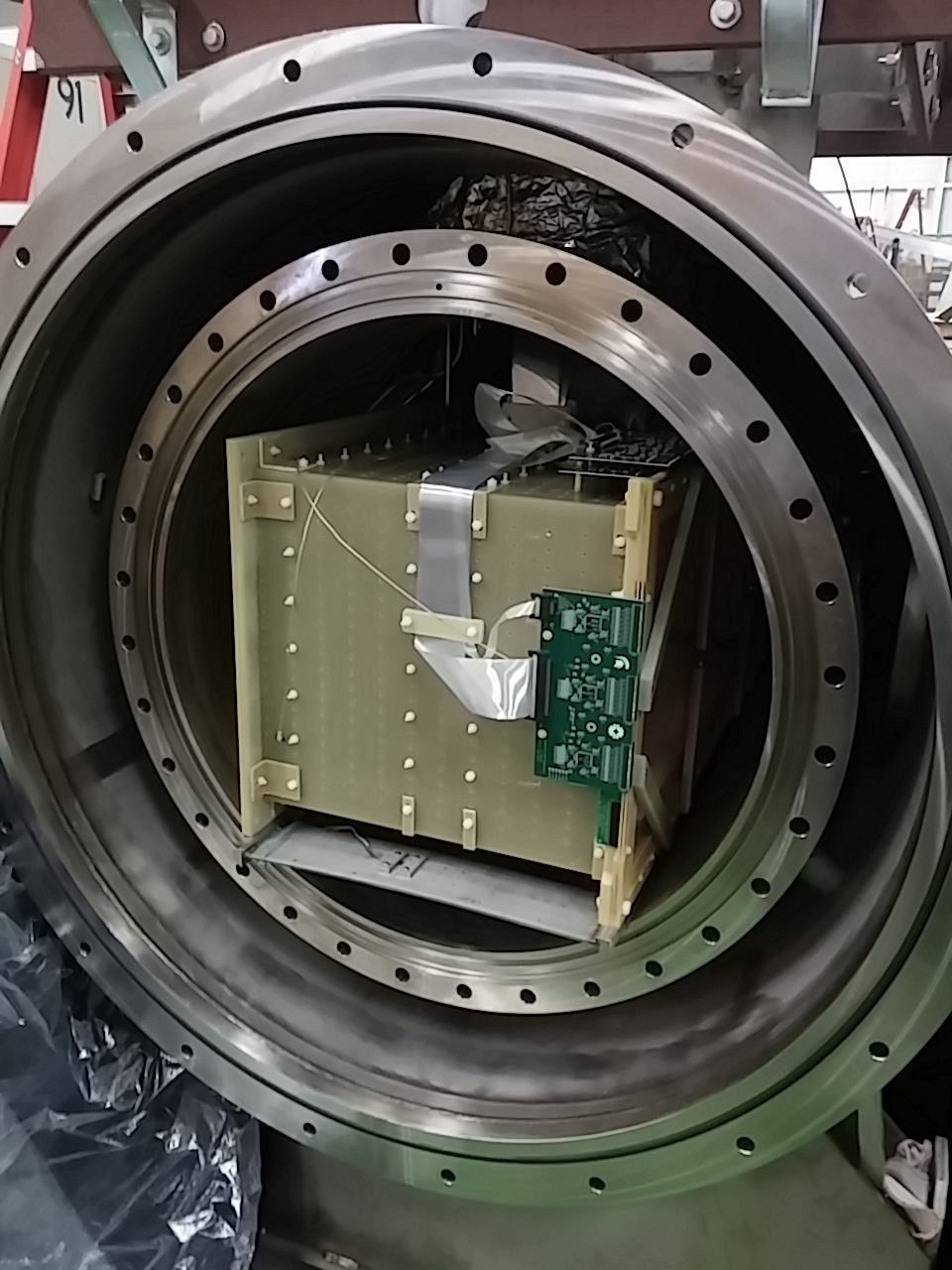}
\includegraphics[width=0.49\textwidth]{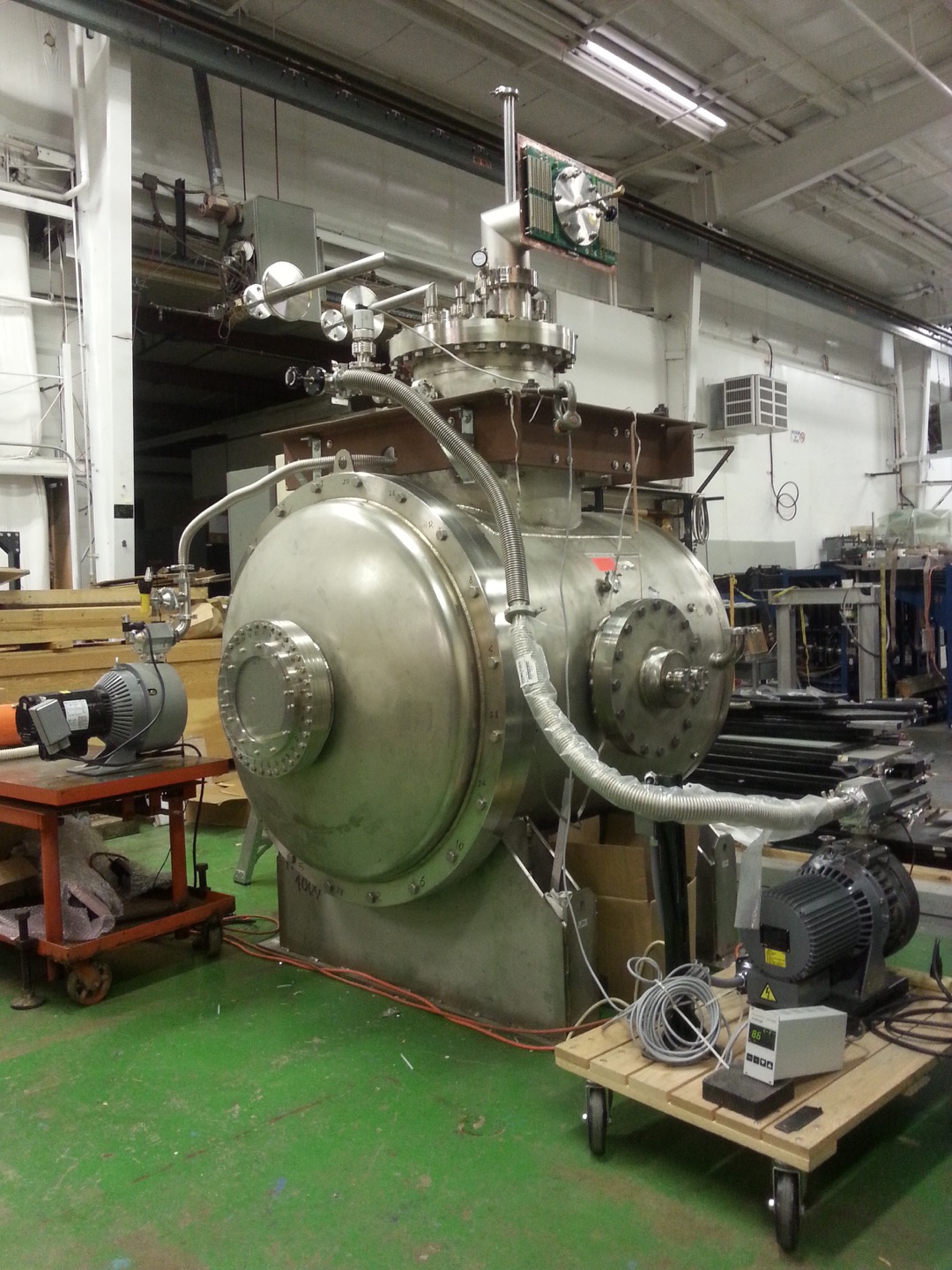}
\caption{\emph{Left}: The LArIAT cryostat, open to air, with the TPC placed in the inner volume. \emph{Right}: The LArIAT cryostat, sealed during initial commissioning, prior to installation at the Fermilab Testbeam Facility.}
\label{fig:LArIATCryostat}
\end{figure}

The LArIAT cryostat, shown in figure~\ref{fig:LArIATCryostat}, consists of an inner volume containing the purified liquid, and an outer volume serving as a vacuum jacket, with layers of aluminized
mylar superinsulation. Both the inner and outer cryostat vessels are
cylindrical with convex end caps. The main axis of the cryostat is
horizontal and oriented parallel to the beam. The inner vessel
is 30~inches (76.2~cm) in diameter and 51~inches (130~cm) in length, corresponding to a total liquid
argon volume of 550~L, or a mass of 0.76~t. The internal volume is accessed
through the end caps on
the inner and outer vessels. In addition, the cryostat has a wide neck, or {\it chimney},
protruding from its top at mid-length, which serves as
an access path for signal cables from the LArTPC and the internal instrumentation,
as well as for the high voltage feedthrough.

The cryostat, inherited from the ArgoNeuT experiment,  was modified to minimize the amount of material upstream of the TPC's
active volume.
 On the original ArgoNeuT cryostat, the stainless steel
end caps of both the inner and outer vessels were 3/16" (4.8~mm) thick.
For the LArIAT experiment, the outer vessel endcap was modified to include a flange hosting a 9~inch (22.9~cm) diameter, thin titanium beam window at its center.  In ArgoNeuT, upstream
of the TPC active volume, there was a region  of uninstrumented liquid argon,
approximately 15~cm long.  To
minimize the uninstrumented liquid argon for LArIAT, a hollow, concave volume
(the {\it excluder}) was installed on the inner vessel's front end cap, extending in toward the TPC. With
these modifications, the total thickness of the uninstrumented region
upstream of the TPC's active volume  was reduced from 1.8 radiation
lengths to less than 0.3.

To accommodate signal and high-voltage bias connections for the scintillation light detection system (described in section~\ref{sec:PhotonSystem}), CF-flanged apertures were added to the side ports of both the inner and outer cryostat vessels.

\subsubsection{Liquid argon system}

LArIAT's liquid argon is supplied by a commercial dewar, positioned
outside the experimental enclosure. Because drifting ionization
electrons attach themselves readily to oxygen and water, and the
slow component of the argon scintillation signal is quenched by nitrogen,
the levels of these contaminants must be minimized.
The liquid argon delivered by the vendor is relatively free of all
three -- none of the contamination levels exceed 1~ppm.
However reaching the desired maximum
contamination level of 100 parts per trillion (ppt) requires further purification.

The argon is delivered from the commercial dewar to the cryostat
through 1-inch (2.54~cm) diameter, Schedule~10 steel piping, 
insulated with 20~cm of polyurethane foam.
All oil and grease
was removed from the pipes before they were welded into the system. The argon passes through the pipes to
the purification system, modeled on the Liquid Argon Purity Demonstrator (LAPD) ~\cite{LAPD} system, which consists of a single 77-liter filter filled halfway with a 4\AA~molecular sieve\footnote{Sigma-Aldrich, P.O. Box 14508, St. Louis, MO 63178 USA.}. The  molecular sieve removes mainly water but can also remove small amounts of
nitrogen and oxygen. The remaining volume of the filter
contains BASF-CU-0226~S\footnote{BASF Corp., 100 Park Avenue, Florham Park, NJ 07932 USA}, a highly-dispersed copper oxide impregnated in a high surface-area alumina, which removes oxygen and, to a lesser extent, water.
The filter is insulated with a vacuum jacket and aluminum radiation
shields. When they become saturated, impurities can be removed from the filter media with heated gas, as outlined
in ref.~\cite{LAPD}.
After one volume of liquid
argon has made a single pass through the filter, electron drift times in the
TPC are typically several milliseconds long, much longer than the maximum
drift time of 300~$\mu$s.

\begin{figure}[htb]
\centering
\includegraphics[width=0.95\textwidth]
{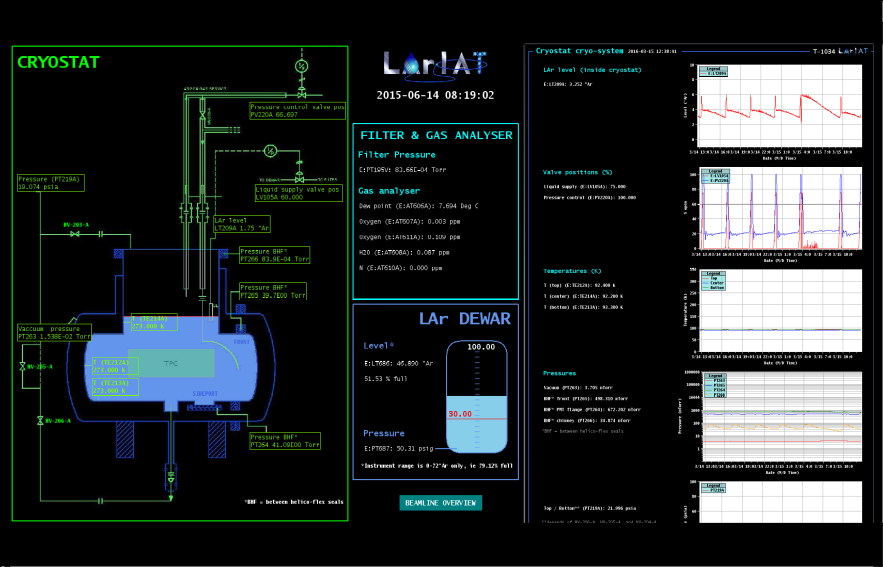}
\includegraphics[width=0.95\textwidth]
{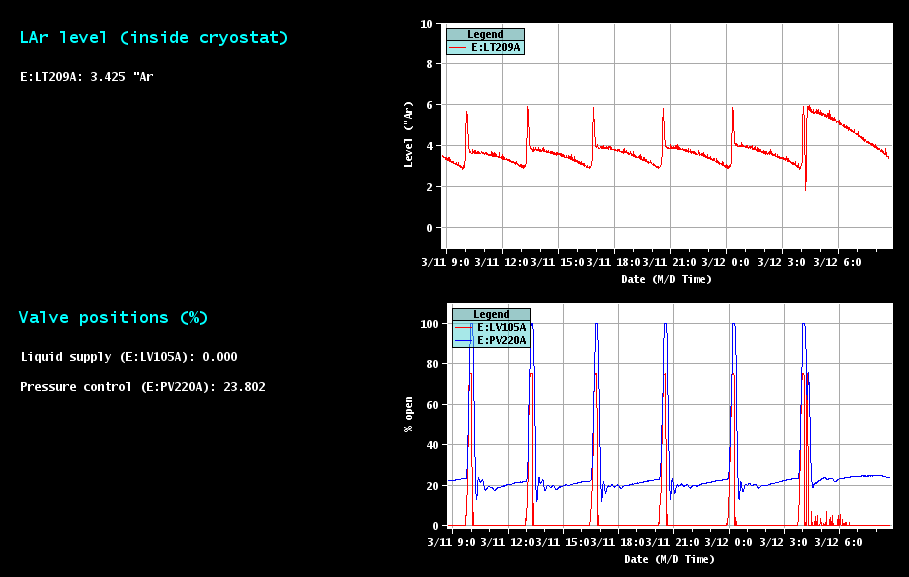}
\caption{\emph{Top}: Screenshot of the LArIAT cryostat monitoring page, showing the argon levels both inside the cryostat and in the supply dewar. Trend plots of the levels over approximately 24 hours are also shown. \emph{Bottom}:  The position of the liquid valve, which allows argon to flow into the cryostat, as well as the corresponding pressure changes inside the cryostat. The typical time of a fill/vent cycle was slightly more than 3 hours.}
\label{fig:LArIATCryoMonitor}
\end{figure}

The filtered argon is directed into the inner cryostat via a liquid
feedthrough mounted on the top of the outer cryostat. A vertical pipe,
connected to the feedthrough, deposits the argon at the bottom of the inner cryostat. The argon level, temperatures, and pressures, both in the commercial dewar and within the cryostat, are continuously monitored. Figure~\ref{fig:LArIATCryoMonitor} shows output from the sensors in both the commercial dewar and the cryostat.

During operation, the argon in the cryostat boils at the surface and is vented to the atmosphere. Whenever the argon level falls below a threshold, argon is directed from the commercial dewar into the cryostat, thereby
ensuring that the TPC high voltage feedthrough and cold electronics
are always submerged. During normal operations, the liquid level inside
the cryostat is replenished several times per day, as shown by the cycling liquid levels and valve positions shown at the bottom panel of figure~\ref{fig:LArIATCryoMonitor}.

Inside the cryostat, a rough temperature profile of the argon is determined with Resistance Temperature Detectors (RTDs)\footnote{Minco strip-sensing RTDs, type S651PDZ36A. http://minco.com} deployed at the bottom, middle, and top of the cryostat. In figure~\ref{fig:LArIATCryoMonitor}, these probes are
labeled as TE213A, TE314A, and TE212A, respectively. 
Establishing the temperature is essential for determining the electron mobility which, in turn, helps determine the electron drift speed. The temperature was stable to +/-0.4K, except during the times when the cryostat was being refilled. During refills, the environment inside the cryostat was somewhat turbulent, as indicated by pressure and temperature readings during the fills, and also apparent in noise seen on the induction and collection plane wires (thought to be due to physical vibrations of the wires).  Data collected during cryostat filling times were automatically flagged as bad quality starting when the automated fill valve began to open and stopping 10 minutes after the fill valve closed  (allowing time for the turbulent liquid to settle). These flagged data were not used in analysis.

\subsection{Time projection chamber}\label{sec:TPC}

The LArIAT TPC can be divided into three major subcomponents:  1) the high voltage
system for the drift field, 2) the cathode and the field cage which shapes
the uniform drift field, and 3) the wire planes which provide charge-sensitive
readout for the detector.

\subsubsection{High voltage}

\begin{figure}[tb]
\centering
\includegraphics[width=0.65\textwidth]{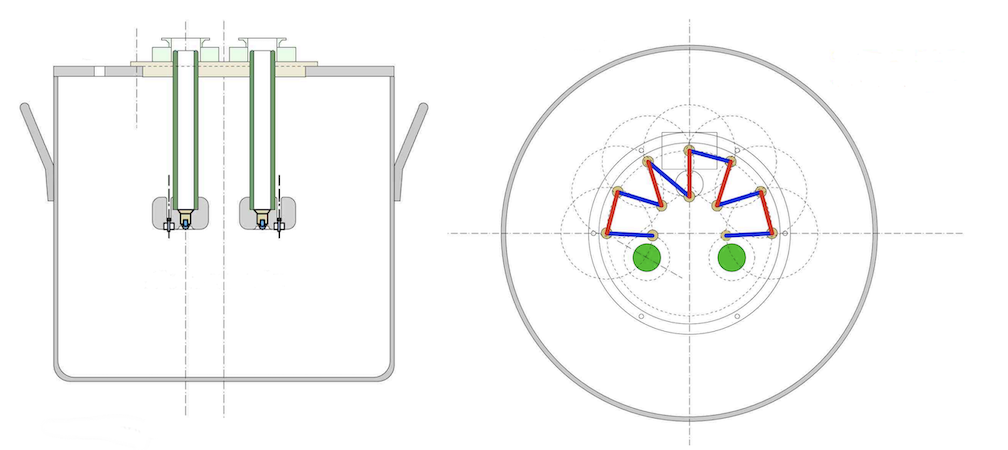}
\caption{Side and top views of the filter pots used in the LArIAT high voltage system. The blue and red segments are structures which hold the series resistors.}
\label{fig:FilterPot}
\end{figure}

Negative high voltage for the TPC cathode, typically 23.5~kV, is provided by a
Glassman LX125N16 power supply.
The output of the power supply, carried by special 1/2" high voltage cables, passes through two filter pots and is then
connected to the high voltage feedthrough on the top of the LArIAT cryostat. 
The two in-line filter pots limit the current draw on the power supply and
provide a low-pass filter that helps reduce voltage ripple on the cathode.
Also, in the event of a high voltage trip, the filter pots help limit the damage
that may be caused by energy stored in the system. Each filter pot,
as shown in figure~\ref{fig:FilterPot}, is a cylinder, 18.5~inches (47~cm) tall by 20~inches (51~cm) in diameter
and 3/16~inches (4.8~mm) thick. Their welded tops have an opening that allows
for a flange and O-ring to receive the high voltage cables. Within each pot, the
two leads on the flange are connected to four 10~M$\Omega$ resistors connected in series. The assembly is submerged in approximately 16~gallons (60~liters) of Diala transformer oil\footnote{Shell Diala S2 ZX-A Transformer Oil, https://www.shell.com/business-customers/lubricants-for-business/shell-diala-electrical-oils.html}, to aid in the suppression of corona discharge.

\begin{figure}[tb]
\centering
\includegraphics[width=\textwidth]{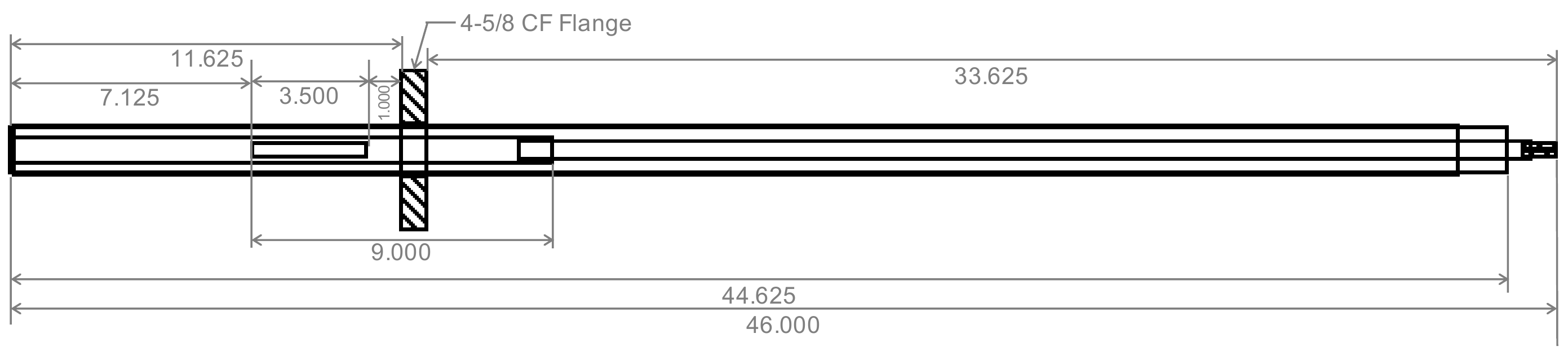}\\
\caption{Technical drawing of the LArIAT high voltage feedthrough with dimensions in inches. }
\label{fig:HVFT}
\end{figure}

The feedthrough brings the high voltage through the liquid argon volume to
the cathode of the TPC.  Modeled on a design used by ICARUS, the feedthrough
consists of a stainless steel inner conductor, surrounded by a tube of
ultra-high molecular-weight polyethylene, which was cryofit inside a stainless
steel, outer ground tube. The feedthrough enters the cryostat from a
dedicated 4-5/8~inch (11.7~cm) Conflat flange at the top of the cryostat. The end of the
inner conductor is attached to the cathode by a flexible conductor bolted
at both ends. A technical drawing of the feedthrough is shown in figure~\ref{fig:HVFT}.
The high voltage system was tested at voltages up to 32.8~kV (corresponding to
drift fields ranging from 0 to 700~V/cm) and the feedthrough itself was tested
up to 60~kV -- no electrical breakdown was observed.

\subsubsection{Cathode and field cage}

\begin{figure}[htb]
\centering
\includegraphics[width=0.6\textwidth]{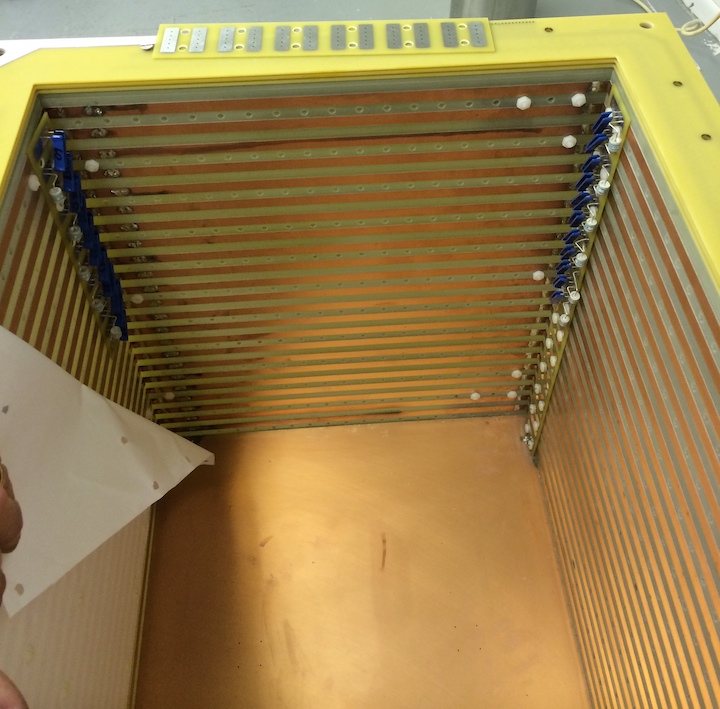}\\
\caption{Field cage during assembly, prior to installation of reflector foils. The resistor chains and gas discharge tubes are visible in the two corners.}
\label{fig:TPC_fieldcage}
\end{figure}

The TPC active volume (47~cm $\times$ 40~cm $\times$ 90~cm) is the volume enclosed by the cathode and
field cage structures. The rectangular field cage structure of the TPC, shown in figure~\ref{fig:TPC_fieldcage},
is composed of copper-clad G10 pieces, with 1~cm wide, horizontal copper
strips, and gaps of 1~cm;
the G10 pieces form the outer walls of the TPC. The four inner walls of the field
cage are connected electrically such that each copper strip forms a complete
loop around the drift volume.  Four 1~G$\Omega$  resistors are arranged in parallel
between the strips of the field cage, for an effective strip-to-strip
resistance of 250~M$\Omega$. The voltage changes uniformly from the cathode to the
anode, providing a uniform electric field throughout the TPC active
volume. EPCOS A71-H45X gas discharge tubes
are used for surge protection. The cathode for the first two running periods of LArIAT was composed of a single piece of copper-clad G10 with a copper area that
exactly matched the aperture of the field-cage structure. In subsequent
LArIAT runs, as a test of a design proposed for the SBND experiment, the original cathode was replaced with a cathode made of stainless steel mesh. 

\begin{figure}[htb]
\centering
\includegraphics[width=0.6\textwidth]{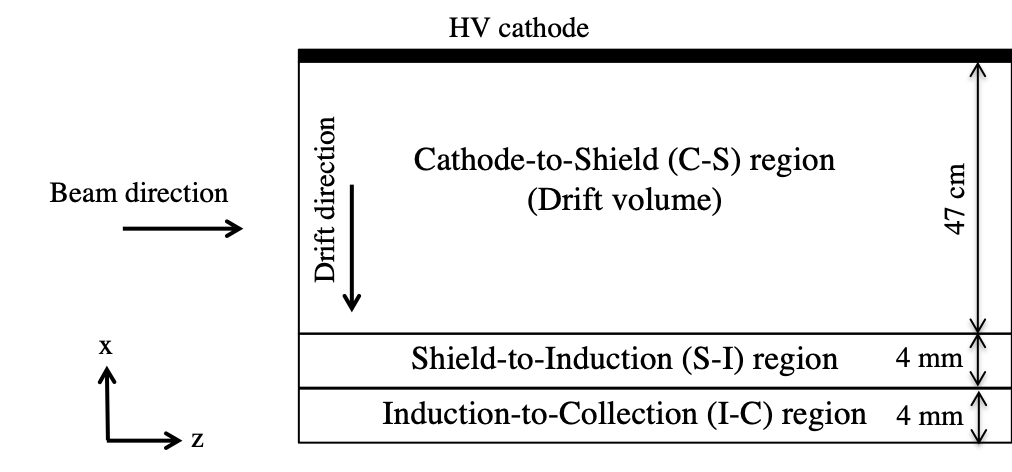}\\
\caption{Schematic diagram (not to scale) of the three drift regions inside the LArIAT TPC: the main drift volume between the HV cathode and the shield plane (C-S), the region between the shield plane and the induction plane (S-I), and the region between the induction plane and the collection plane (I-C).}
\label{fig:driftregions}
\end{figure}

As shown in figure~\ref{fig:driftregions}, the TPC contains three drift volumes
with different electric fields. The main drift volume is the region
between the HV cathode and shield plane (C-S). There is a second drift region
between the shield and induction planes (S-I) and a third between
the induction and collection planes (I-C). The shield plane wires are aligned vertically (along the $y$-axis in the figure), while the induction and collection plane wires aligned at $\pm 60$ degrees with respect to the horizontal $z$-axis in the figure.
The bias voltages for the wire planes are chosen such that the electric field
in these regions satisfies the {\it charge transparency} condition: 100\%
of the drifting electrons are transmitted through the shield plane
and the induction plane, and are then collected on the collection plane. A detailed description of the measurement of
the electric fields in each region is presented in section~\ref{sec:EFieldMeasurements}. The default
voltages by run for each of the planes are shown in table~\ref{tab:wireplane-voltages}, along
with the wire pitch. The voltages were tuned to maintain transparency when the (I-C) and 
(S-I) spacing changed.

\begin{table}[htb]
\centering
\caption{Nominal voltages for the LArIAT running periods.}
\label{tab:wireplane-voltages}
\begin{tabular}{l c c c c c}
\hline \hline
\textbf{Run Period} & \textbf{Pitch (mm)} & \textbf{Cathode (V)} & \textbf{Shield (V)} & \textbf{Induction (V)} & \textbf{Collection (V)} \\
Run I      & 4 & -23164 & -298 & -18.5 & 338 \\
Run IIA\&B & 4 & -23164 & -298 & -18.5 & 338 \\
Run IIIA   & 5 & -23164 & -325 &   0   & 423 \\
Run IIIB   & 3 & -23164 & -298 & -18.5 & 338 \\
\hline
\end{tabular}
\end{table}

\subsubsection{Wire plane assembly}
The wire planes are built on oversized, printed circuit boards with overall dimensions of $50~\text{cm}\times100~\text{cm}$, and a rectangular cut-out in the center, corresponding to the active region of sense wires. The cut-out dimensions for the wire planes used in Run~I, Run~II, and Run~IIIA were $40~\text{cm}\times90~\text{cm}$, while for Run~IIIB the cut-out was reduced in size ($30~\text{cm}\times70~\text{cm}$) because, with 3~mm wire spacing and the ability to digitize only 480 wire signals with our existing readout system, only a smaller spatial area could be recorded. The wire planes for all running periods of the experiment were assembled and tested at Fermilab's Lab~6. This included winding the wires, then hand-soldering both the wires and passive components of the circuit (resistors and capacitors) to the printed circuit boards.

Before being transferred to oversized printed circuit boards, on which they
are installed in the TPC, the copper-beryllium wires\footnote{Little Falls Alloys, Inc., Beryllium-Copper \#125 alloy wire, 0.006-inch (0.152~mm) diameter} of the LArIAT wire planes are wound to the
correct spacing and tension on an independent winding machine. Two
rectangular steel frames are bolted back-to-back on an apparatus that rotates
at a constant speed. As the frames rotate, wire
is drawn from a spool and wrapped around the two frames at constant tension,
monitored by a tensiometer in the device paying out the spool. Once the requisite
number of wires has been wound, the wires are attached to the steel
frames with five-minute epoxy and cut, yielding two separate wire
planes, one on each frame.

\begin{figure}[ht]
\centering
\includegraphics[width=0.49\textwidth]{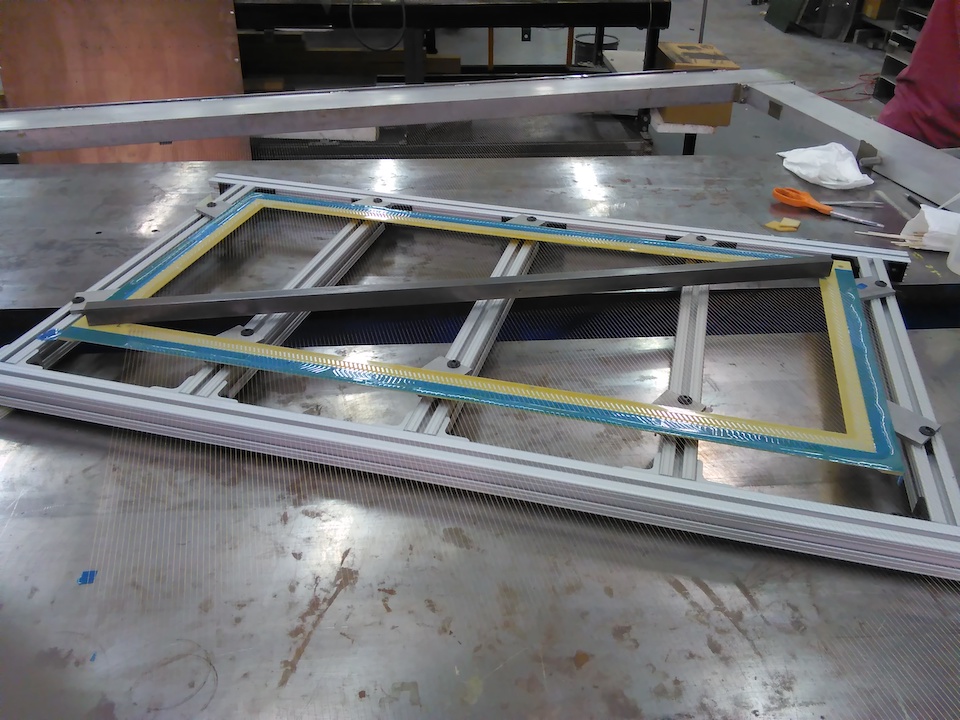}
\includegraphics[width=0.49\textwidth]{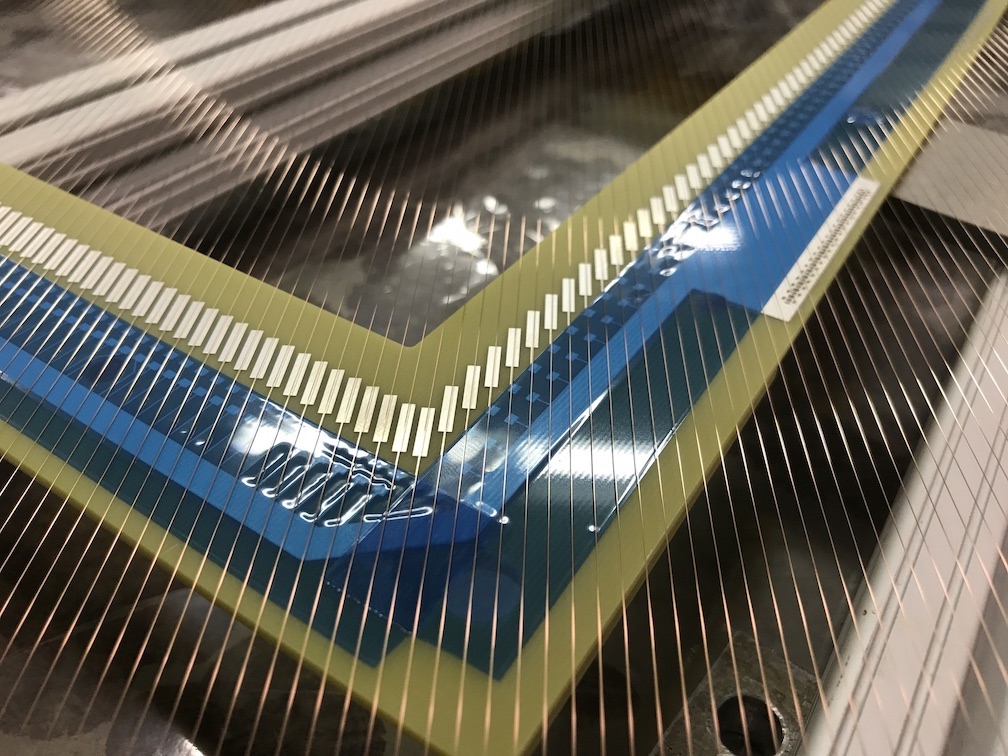}
\caption{\emph{Left}: One of the wire planes during transfer, with the wires not yet cut from the winding frame. A steel bar, anchored along the diagonal, ensured that all
wires made good contact with the board. \emph{Right}: Close up view of wires aligned with solder pads prior to attachment with epoxy and solder.}
\label{fig:wiretransfer}
\end{figure}

After the wires planes have been wound, they must be transferred from the
winding frames onto the G10 wire-carrier boards that are mounted into the
LArIAT TPC. To this end, the frames are unbolted from the winding
apparatus and laid over the board onto which the wires are transferred.
To prevent it from flexing under tension, the board is bolted to a rigid,
aluminum frame. Careful manual alignment is performed, using guide holes
drilled in the board for reference. The alignment frame in use during the wire transfer is shown in the left panel of figure~\ref{fig:wiretransfer}.

After the wires are aligned to the solder pads on the boards, they are attached to the board on the outside edge of the solder pads. A temporary epoxy strip, applied atop a protective strip of tape (seen in blue in the right panel of figure~\ref{fig:wiretransfer}), preserves the alignment.
The wires are then attached a second time, with a
permanent epoxy strip placed at the inside edge of the solder pads. This
epoxy\footnote{Andover Corporation, EPOLITE FH-5313 epoxy, \url{https://www.andovercorp.com}}, which is  slow-curing and safe for cryogenic conditions, is allowed
to cure for 12 to 24 hours. After curing is completed, the wires are cut behind the temporary strip and the winding frame is removed.

The wires are soldered to their pads using an alloy of 96\% tin and 4\%
silver but no flux core. 
The same soldering materials are used
to attach resistors and capacitors on the induction and collection planes. 
While the wires are being soldered, mylar sheeting is used to protect the active regions of the wire planes from evaporating solder flux. 
Once all wires have been soldered, the portion of the wires beyond the 
solder pads is cut away, and the temporary epoxy strip is removed. Finally, the entire board is cleaned with ethyl alcohol, to remove residual solder
flux.

Before being installed in the TPC, the assembled wire planes are tested. First, a simple, electrical continuity test ensures that every wire has a good electrical path to the board's connectors. Next, a 20~Hz square wave is injected onto each wire. The RC time constant is measured and compared with the nominal value of 48 ms, which is set by the 2200~pF capacitor and 22~M$\Omega$ resistor in the readout circuit.

\subsection{{TPC} electronics}\label{sec:MSUElectronics}

\begin{figure}[htbp]
\includegraphics[width=\textwidth]{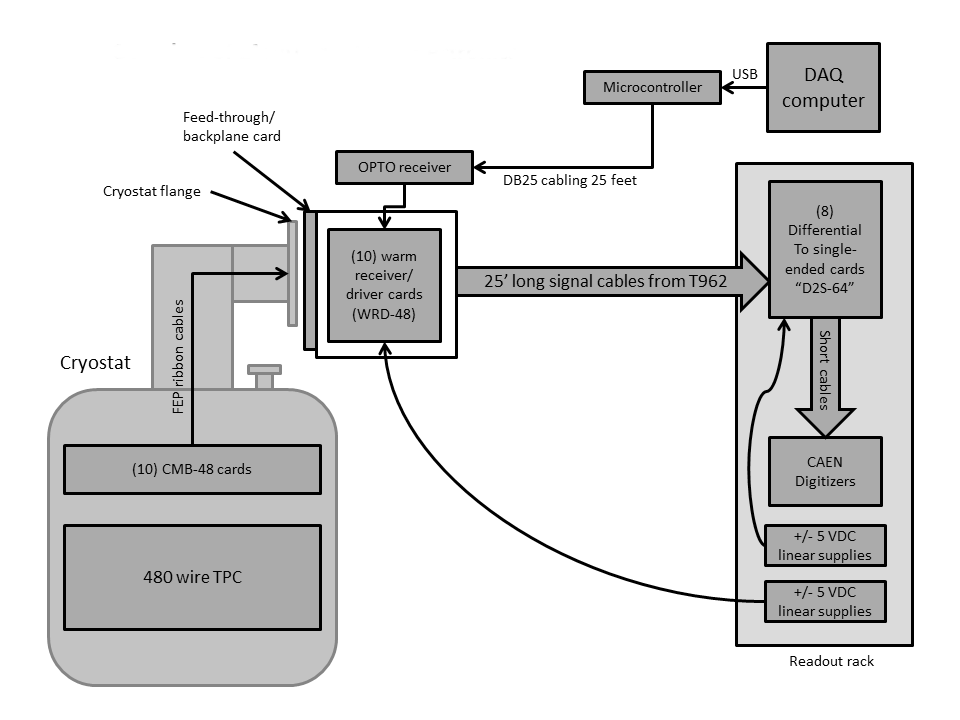}
\caption{Block diagram of the LArIAT front-end electronics. 
} 
\label{fig:FEelectronics}
\end{figure}

The TPC front-end electronics comprises 480 analog channels from the wire planes
to the signal digitizers. The front-end system also includes a digital control
system for the electronics mounted on the TPC, a power supply 
and a power distribution system. A block diagram of the system is shown in
figure~\ref{fig:FEelectronics}.

The electrical signals on the readout wires, produced without
amplification by drifting ionization charge, are usually quite small. To
optimize the signal-to-noise ratio (S/N), the signals are read out through
cold amplifier motherboards
hosting ASICs (LArASICs)~\cite{BNLASIC}, 
mounted directly on the frame of the TPC, inside the liquid argon
cryostat. 
The LArASICs are mixed signal devices where the analog amplifier characteristics
are controlled by digital parameters. At the typical gain setting (25~mV/fC),
the charge deposited on an individual wire from a passing 
MIP (3.5~fC) will generate a 90~mV output. The shaping time is 3~microseconds. Although the signals are sampled
every 16~ns by CAEN V1740 digitizers,
 special firmware provided by CAEN allowed LArIAT to store averages of 8 consecutive samples instead. The effective sampling period is 128~ns, but with a significant reduction in the electronic noise.

The maximum input range of the V1740 is 2~volts. Because
the collection signals are unipolar, an analog offset of 0.2~V was chosen, 
permitting the digitization of small negative excursions. The induction
plane signals are bipolar. To capture both the negative and positive
parts of these signals, a 1~V offset was chosen, placing the baseline at mid-range.
 The V1740's 12-bit samples and
maximum input range of 2 VDC yields a least significant bit of
0.5~mV. Since the analog path from the LArASIC to the digitizers is designed
to provide unity gain, the 90~mV MIP signal corresponds to an amplitude
of 180 ADC counts.

To minimize the number of charged particles passing through the digital electronics, the readout racks that host the CAEN digitizers are located 8~m from the
cryostat. Thus for practical reasons, the digitizers are referenced to a different ground from the TPC
electronics.   To prevent
problems caused by different grounds and noise pick-up along the
transmission line, the single-ended analog signals from the ASICs
are transmitted to the readout racks as differential analog pairs, with
each pair individually shielded.
The transmission path between the TPC and the digitizers includes
four interconnected cards. The first is a 48-channel, cold motherboard
(CMB-48) that mounts directly to the TPC and houses three,
16-channel LArASICs. The next is a single cryostat feedthrough (FT) card,
that carries the 480 signals, along with power and control lines, across
the cryostat boundary. The signals then pass through warm
receiver and driver (WRD-48) cards, which plug directly into the FT card.
The WRD-48 cards amplify the single-ended TPC signals, producing differential
analog outputs. These signals then pass through
8 high-quality, pleated foil cables to a set of D2S-64 cards. The D2S-64 cards
convert the differential signals into single-ended signals, canceling any
common mode noise, and providing sufficient current for direct input into
the array of CAEN V1740 digitizers.

The FT card is an 8-layer PCB, 254~mm  $\times$ 432~mm. 
To improve noise immunity, the outer layers of the card are mostly
uninterrupted ground planes. The wide traces enhance manufacturing
yield and physical robustness. To reduce capacitive cross-talk,
signal lines on adjacent layers are staggered. Ground fills are provided
whenever possible on all copper layers.

The FT card is sandwiched between an
ASA flange 
and sealed with O-rings. A stiff mechanical structure
and backplane-style connectors support the FT card and allow for the
direct insertion of WRD-48 cards. The design facilitates assembly and
possible repair work, while reducing the number of cables and connectors
needed.

The overall assembly of the FT card, the ten WRD-48 cards, and the WRD
card cage is light enough to be safely cantilevered  from a mechanical
structure rooted on the cryostat flange. The backplate of the structure
is a solid, copper sheet, which is electrically connected with the cryostat
flange through lock washers, and also connected to the FT card through 40
conductive standoffs. The copper plate and the cryostat flange are
defined to be the electrical ground of the TPC readout electronics.

\subsection{Photon detection system}\label{sec:PhotonSystem}

\begin{figure}
\centering
\includegraphics[height=5cm]{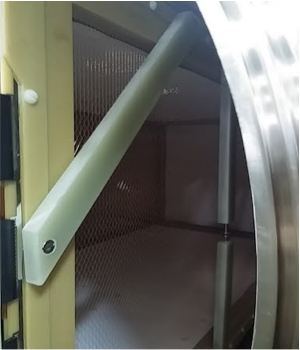}
\hspace{0.5cm}
\includegraphics[height=5cm]{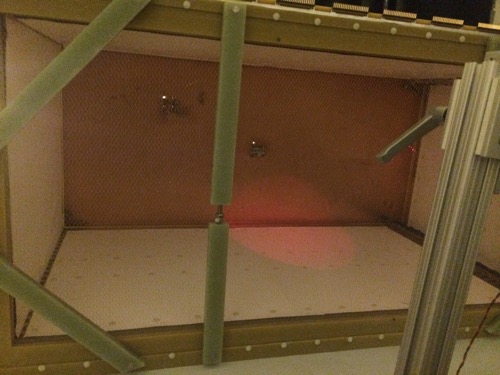}
\caption{\label{fig:lightsys_foils}The TPB-coated reflector foils mounted to the TPC's four field cage walls as viewed through the opening at the front of the inner cryostat (\emph{left}) and looking in from behind the wire planes (\emph{right}).
}
\end{figure}

In addition to the drift electrons from ionization, LArIAT collects
photons from liquid argon scintillation. Since liquid argon produces scintillation light in the 
VUV range, at 128 nm, that light must be shifted to
visible wavelengths before it can be detected by most photosensors. In LArIAT,
the scintillation light is wavelength-shifted within a thin layer of
tetraphenyl butadiene (TPB) which coats highly-reflective dielectric
substrate foils\footnote{Vikuiti\texttrademark, 3M Optical Systems Division, 3M Center, St. Paul, MN 55144-10003} lining the TPC's four field cage walls. A photo of the foils mounted inside the TPC volume is shown in
figure~\ref{fig:lightsys_foils}. The interaction of a VUV photon with the TPB induces the emission of one or more visible photons, which are then emitted back into the active volume where they can be detected by the photosystem.

\begin{figure}
\centering
\includegraphics[width=0.45\textwidth]{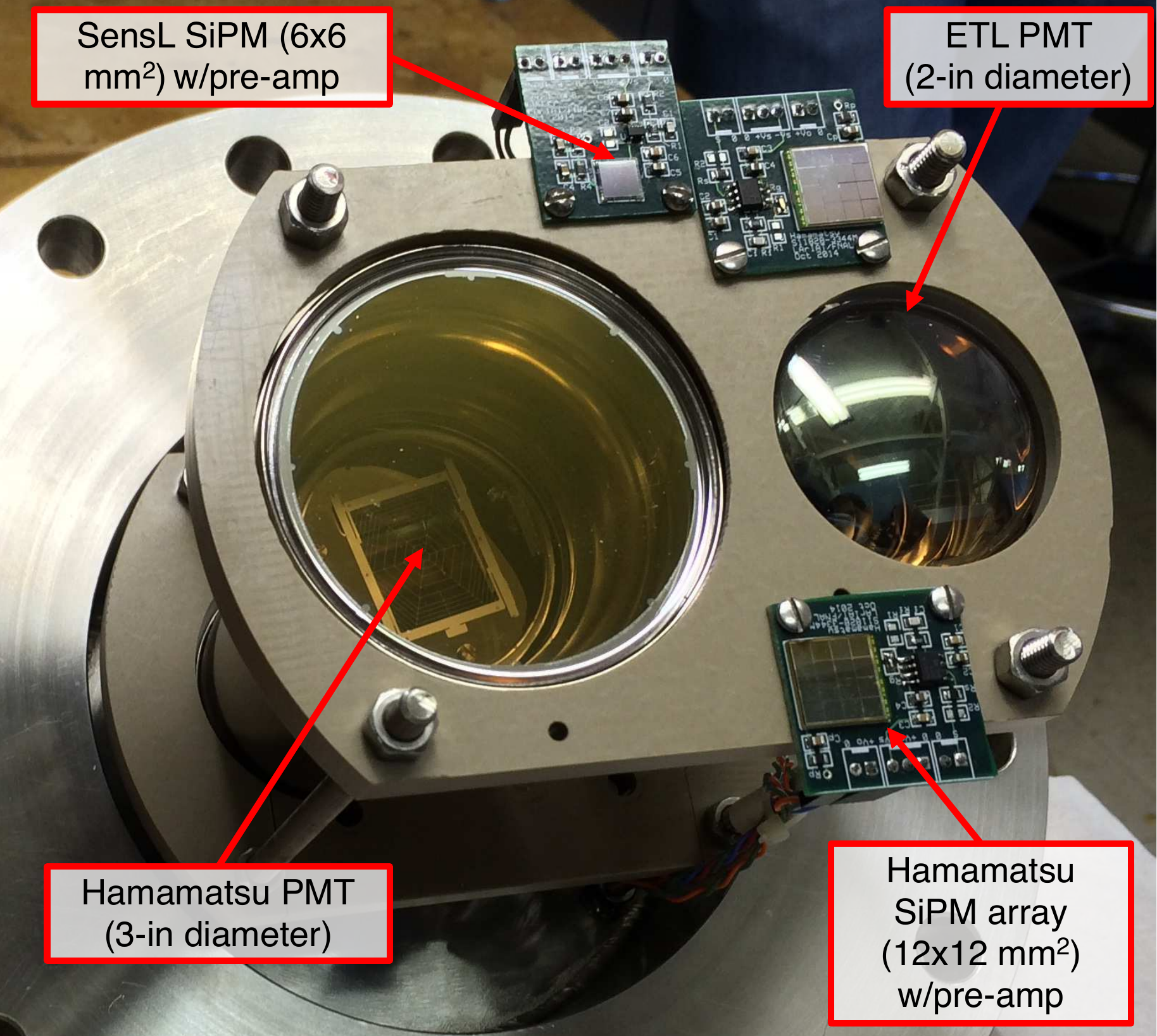}
\caption{LArIAT's photodetection system, which observes LAr scintillation light inside the TPC, is mounted to the side flange of the inner cryostat.}
\label{fig:lightsys_pmts}
\end{figure}

The LArIAT photodetectors are located just beyond the collection plane. 
A polyetherketone (PEEK) support structure, attached to a side access flange, holds in place the photodetectors, as shown in
figure~\ref{fig:lightsys_pmts}. The windows of all devices are held parallel to the wire planes with approximately 5~cm clearance.
For most of the LArIAT data-taking runs, the photodetection system
consisted of two PMTs: a 3-inch (76~mm) Hamamatsu R11065 and a 2-inch (51~mm) ETL D757KFL. The light yield of these PMTs was evaluated for the first two running periods based on simulations: 3.8 photoelectrons per MeV (PEs/MeV) for the ETL 2-inch PMT, and 14.8~PEs/MeV for the Hamamatsu 3-inch PMT. During parts of Run I and Run II, several silicon photomultiplier (SiPM) arrays were also deployed. The SiPMs consisted of two Hamamatsu S11828-3344M 4$\times$4 arrays (12~mm $\times$ 12~mm total active area) as well as one single-channel SensL MicroFB-60035 (with 6~mm~$\times$~6~mm active area) on custom cold-amplifier and readout boards, which were mounted along the edges of the PMT holder.

In Run~I, the divider circuits which provide voltages for the PMT dynodes were
configured for \emph{negative bias}, in that the cathode of each PMT was held at negative high voltage while
the DC-coupled anode floated near ground and connected directly to the output signal line. 
In Run~II, the dividers were altered to enable a \emph{positive bias} configuration in which the photocathode was grounded while the anode sat at positive high voltage. The new configuration minimized interference, observed in Run ~I, on wires near the charged photocathodes and outer PMT chassis. 
To provide AC-coupled signal readout in Run~II and beyond, 18~nF capacitors rated to 2~kV were soldered between the charged anodes and the output signal lines of each PMT. At liquid argon temperature and PMT operating voltages, the capacitances
are reduced to approximately one third of their nominal value~\cite{lightsys-capacitors}.

For Run~II, a TPB coating was added to the windows of two of the
photodetectors: the ETL PMT and the SensL SiPM. A solution of TPB dissolved in polystyrene was used for the coating. A second coat was added midway through the run. The TPB coating rendered the photosensors sensitive to VUV light emanating directly from the point of ionization, which enabled studies of scintillation light arrival times. 

In Run~III, as part of an effort to test new photosensors for liquid argon detectors, a prototype ARAPUCA~\cite{lightsys-arapuca} was installed in place of the SiPMs. The ARAPUCA captures photons in a box with  highly-reflective inner surfaces (reflectivity $>$98\%),  resulting in high photon-detection efficiency, even with limited photosensor area. The key to the
ARAPUCA's photon-trapping mechanism is the use of dichroic short-pass optical
filters, placed on the entrance window, which are largely transparent to
wavelengths below a particular cutoff and highly reflective for wavelengths above. The
outside of the filter is laminated with a wavelength-shifting (WLS) coating which converts the VUV scintillation light to a wavelength just below the filter's cutoff, allowing
the light to enter the box. A second WLS coating covers the inner side
of the filter or optionally, the inside of the box. The internal WLS
layer enables light to convert a second time, to a wavelength
greater than the cutoff, so that it cannot not escape the box. The interior of the ARAPUCA
is viewed by light sensors which capture the trapped photons after a few
reflections.

\begin{figure}
\centering
\includegraphics[height=5cm]{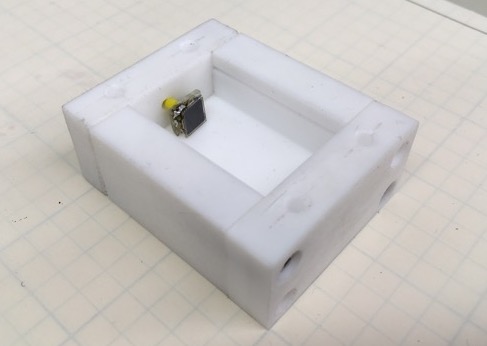}
\hspace{0.5cm}
\includegraphics[height=5cm]{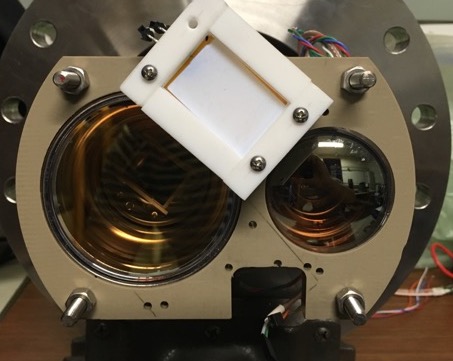}
\caption{The prototype ARAPUCA device: (\emph{left}) unmounted and without its dichroic
filter and (\emph{right}) with the filter installed, mounted on the PMT support structure.}
\label{fig:arapuca_photo}

\end{figure}

The prototype ARAPUCA installed in LArIAT, shown in figure~\ref{fig:arapuca_photo}, is
4.5 $\times$ 5.5 cm$^2$ with a total exposed filter area of
3.5 $\times$ 4.5 cm$^2$. The  outer surface of the
filter is coated with the wavelength shifter P-terphenyl
and the inner surface with TPB. The former emits photons with a wavelength of 350~nm and the latter emits photons at 
430~nm.The dichroic filter cutoff wavelength is 400~nm. The device is equipped with a single SensL MicroFC 60035 SiPM biased at +24V.
Its signals are sent out to the DAQ through a commercial preamplifier.

Signals from all photodetectors are routed from the side flange of
the cryostat to NIM fanouts, which provide the 50~$\Omega$ termination
needed to minimize reflections along the cables.
Discriminated copies of the PMT signals are used in the construction of
several light-based triggers and the raw waveforms are recorded at 1~GHz by a
CAEN V1751 digitizer.

The light detection system can also be used to study
nitrogen (N$_2$) contamination in liquid argon. The influence of
N$_2$ on liquid argon scintillation light emission was first observed
in the WArP experiment~\cite{WArP-nitrogen}: as the N$_2$ concentration rises, the
decay time of the liquid argon slow component decreases significantly,
from approximately one microsecond to hundreds of nanoseconds.
The slow component of argon scintillation arises primarily from the de-excitation of argon excimers in the triplet state. In the presence of nitrogen contamination, de-excitation also proceeds through the collision ${\text {Ar}}^*_2 + {\text N}_2 \rightarrow 2{\text {Ar} }+ 2{\text N}$, a non-radiative process which competes with the first. The lifetime
of the triplet-state excimers is effectively decreased:
\begin{equation}
{\frac{1}{\tau^\prime}} = {\frac{1}{\tau_0}} + k C_N,
\label{eq:lightfit}
\end{equation}
\noindent
where ${\tau_0}$ is the lifetime in the absence of nitrogen and $k$ describes the change in the decay rate; that change is directly proportional to the
concentration of the nitrogen contamination, $C_N$.

\begin{figure}
\centering
\includegraphics[width=0.6\textwidth]{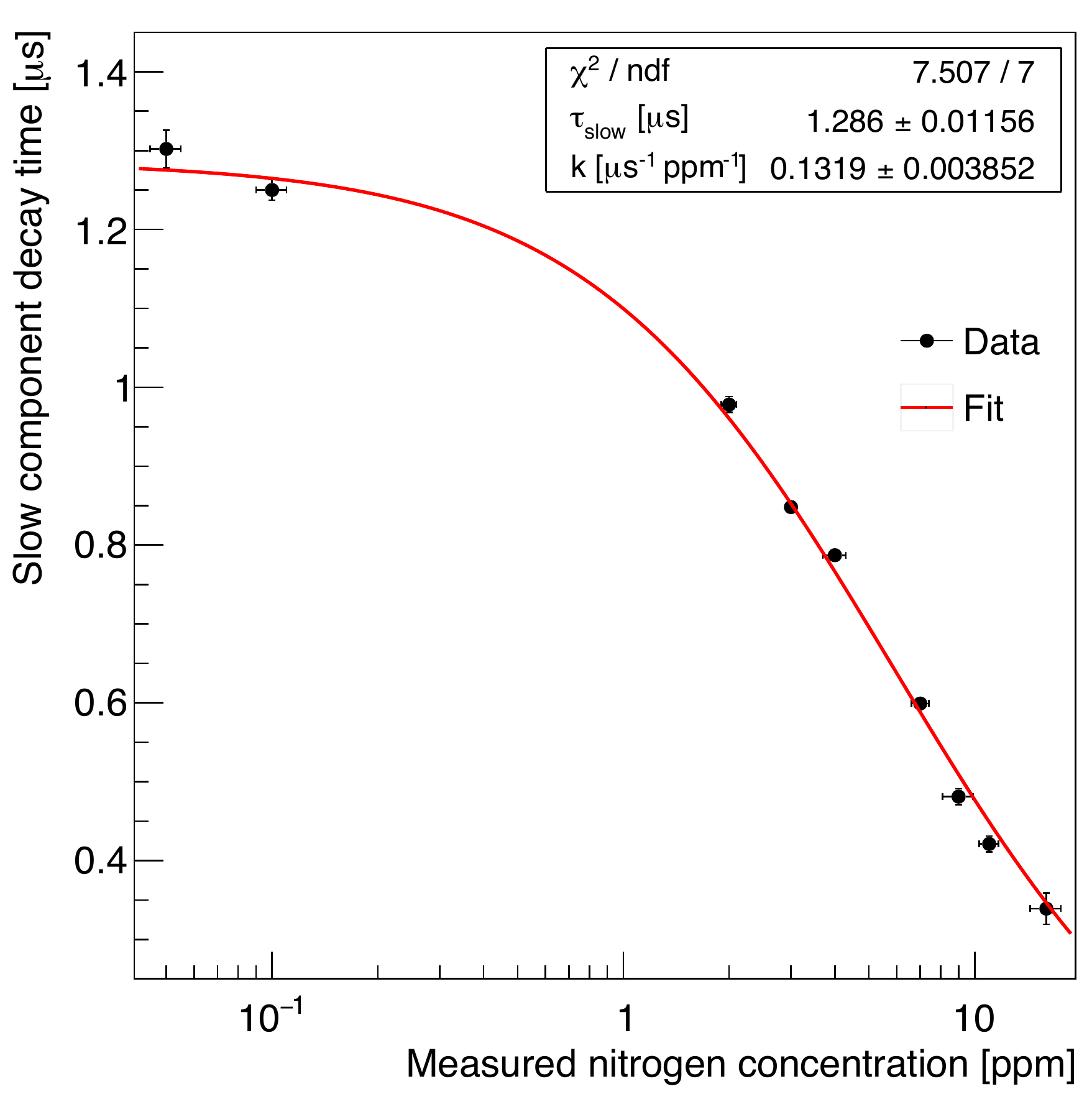}
\caption{ 
Slow component decay time measured with ETL PMT as a function of the concentration of nitrogen in the liquid argon as measured by the LArIAT gas analyzer, with best fit to eq.~\ref{eq:lightfit}.
}
\label{fig:light_nitro}
\end{figure}

To determine $k$, waveforms produced by the ETL PMT in Run~I, selected for periods of approximately uniform N$_2$ concentration, were fit to an exponential function in the range of \mbox{0.4-2.0}~$\mu$s following the start of each pulse. As shown in figure~\ref{fig:light_nitro}, the late-light lifetime increases as N$_2$ concentration decreases, exactly as expected.  The results of a fit
to eq.~\ref{eq:lightfit} are:
\begin{equation}
    \begin{aligned}[b]
    k &= 0.132 \pm 0.004 \, \text{ppm}^{-1} \mu\text{s}^{-1} \\
    \tau_0 &= 1286 \pm  12~{\text{ns}},
    \end{aligned}
\label{eq:lariatfit}
\end{equation}
which are approximately consistent, within statistical uncertainties, with results from WArP:
\begin{equation}
    \begin{aligned}[b]
    k \, \text{(WArP)} &= 0.11 \pm 0.05 \, \text{ppm}^{-1} \mu\text{s}^{-1} \\
    \tau_0 \, \text{(WArP)} &= 1260 \pm 10~{\text{ns}}.
    \end{aligned}
\label{eq:warpfit}
\end{equation}

\subsection{Digitization and data acquisition}\label{sec:Digitization}
\begin{figure}[htb]
\centering
\includegraphics[width=0.75\textwidth]{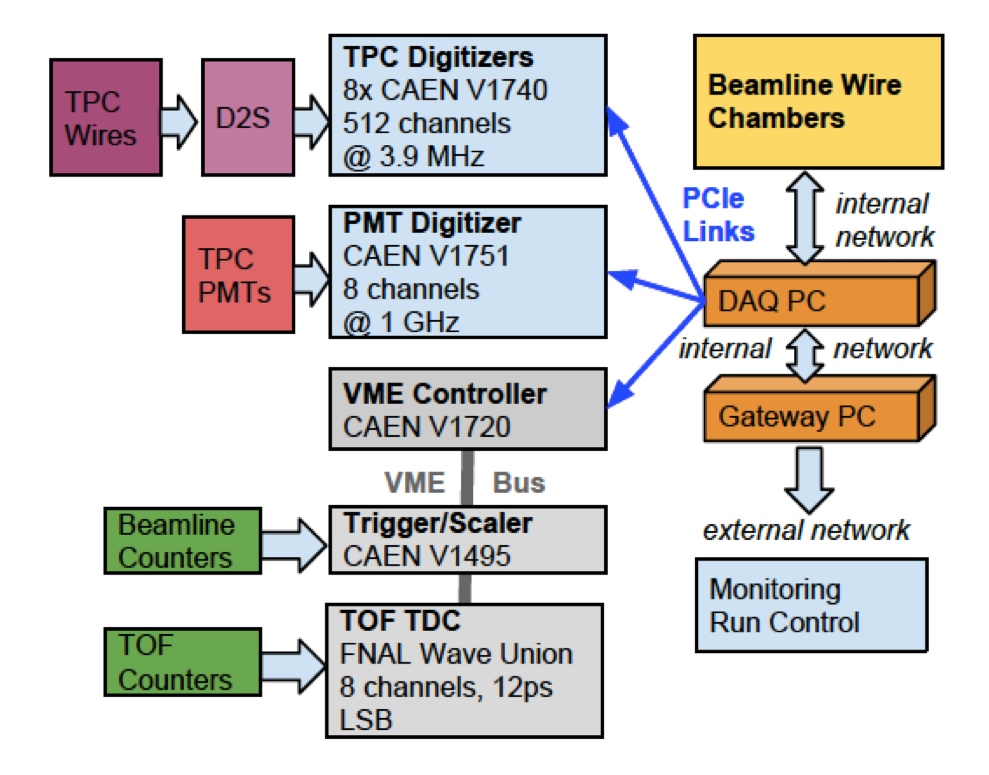}
\caption{
{Summary of LArIAT DAQ and data flow.}
}
\label{fig:daq_flow}
\end{figure}

In this section, details of the digitization of the detector signals are presented.
These include signals from the induction and collection planes of the TPC, scintillation
light detectors, the beamline time of flight detectors, and wire chambers. 
Details of the timing of the beam and data acquisition windows and the trigger
system that drives them are also presented. An overview of the system is shown in figure~\ref{fig:daq_flow}. The detectors, pictured at left, feed data to
the front-end electronics, which then pass digitized data to the
data acquisition server, pictured to the right. 

\begin{figure}
\centering
\includegraphics[width=\textwidth]{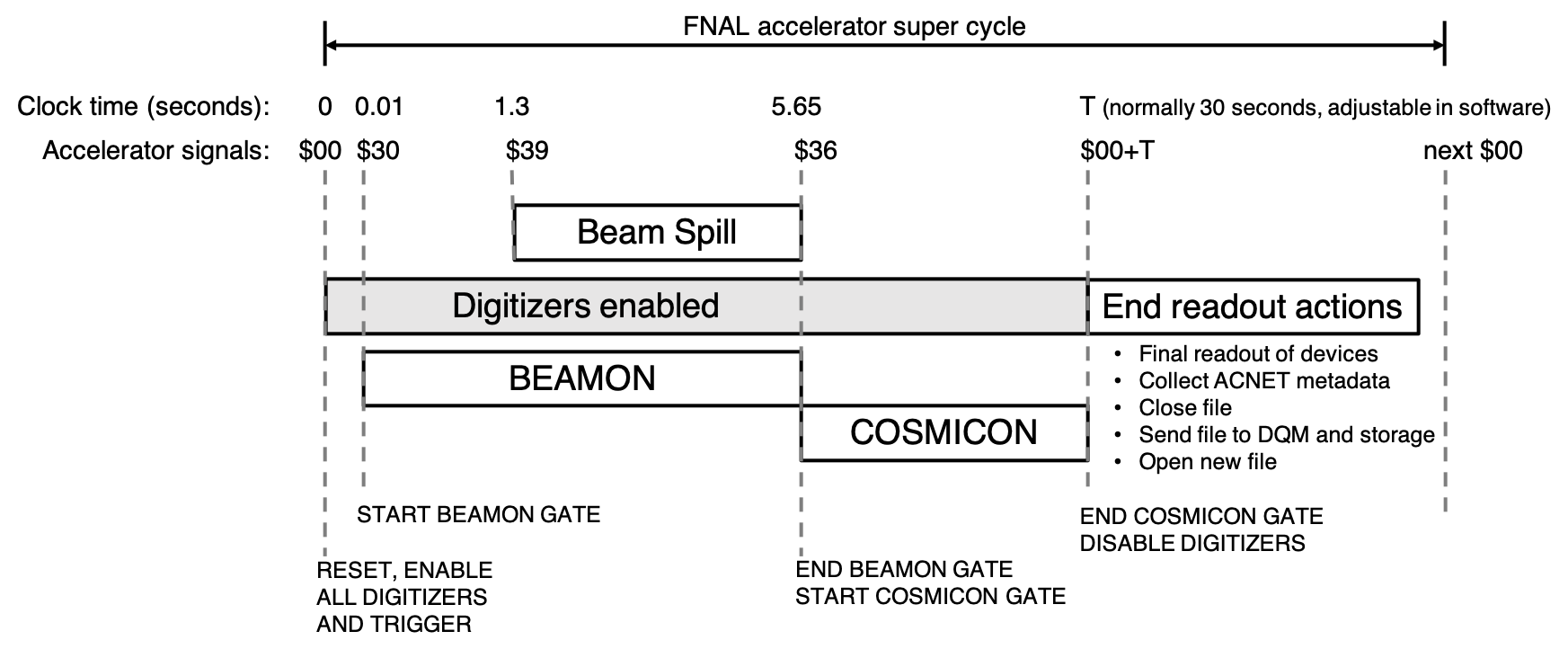}
\caption{
{The LArIAT DAQ timeline during a beam super cycle at FTBF (not to scale). All systems and data fragments are synchronized to the \$00 accelerator signal at the start of each super cycle. BEAMON and COSMICON are inputs to the trigger. }
}
\label{fig:lariat_timing}
\end{figure}

Every 60.5 seconds, as part of the Fermilab accelerator complex super cycle,
beam is slowly extracted from the Main Injector and sent to LArIAT.
LArIAT collects beam data during the 4.2 seconds between the \$39 and
\$36 accelerator signals shown in figure~\ref{fig:lariat_timing}, and collects cosmic ray events for an additional 24 seconds following. During the spill setup time between markers \$30 and \$39, LArIAT collects random pedestal
triggers at 10 Hz. Each digitization element uses the \$30 accelerator signal as a common start, which resets local timers and provides a means to synchronize data buffers across disparate hardware. 

The data acquisition system was designed to collect events at rates as high as
100~Hz during the beam spill, with little or no dead time. The rates
of cosmic ray and pedestal calibration triggers are relatively insignificant -- 
digitized beam signals from the TPC dominate the experimental data volume.
With a maximum electron drift time of approximately 300~$\mu$s in the TPC, and padding
before and after the drift window, 393~$\mu$s of waveform data are collected for
each wire. The CAEN V1740 sampling period of 128~ns implies 3072 12-bit
samples per wire per event. With a trigger rate of 100~Hz, the total data
rate for all 480 wires would be approximately 2~Gbytes/sec. In practice, the typical rate was only 50-100 triggers per 4.2~second spill, followed by a low rate of cosmic triggers between spills. Nonetheless, the 56 seconds of beam-off time per cycle was needed to read out the roughly 8~Gbytes of data collected during the spill.

CAEN V1740 digitizers were chosen to read out the wire planes, primarily
because of their high channel-density, 12-bit dynamic range,
variable effective sampling frequency and time window, and large event-memory
buffers. Various sampling periods were tested but most of the data were taken
with a period of 128~ns. The LArIAT sampling rate is significantly higher
than that of other, larger LArTPCs, which improves the resolution of the wire
chamber signals. 
As noted above, the fast scintillation signals from the time-of-flight and light detection
systems were read out with 1~GHz CAEN V1751 digitizers, which produce 10-bit data samples. The overall event size is dominated by the V1740 data from the wire planes.

\begin{figure}
\centering
\includegraphics[width=\textwidth]{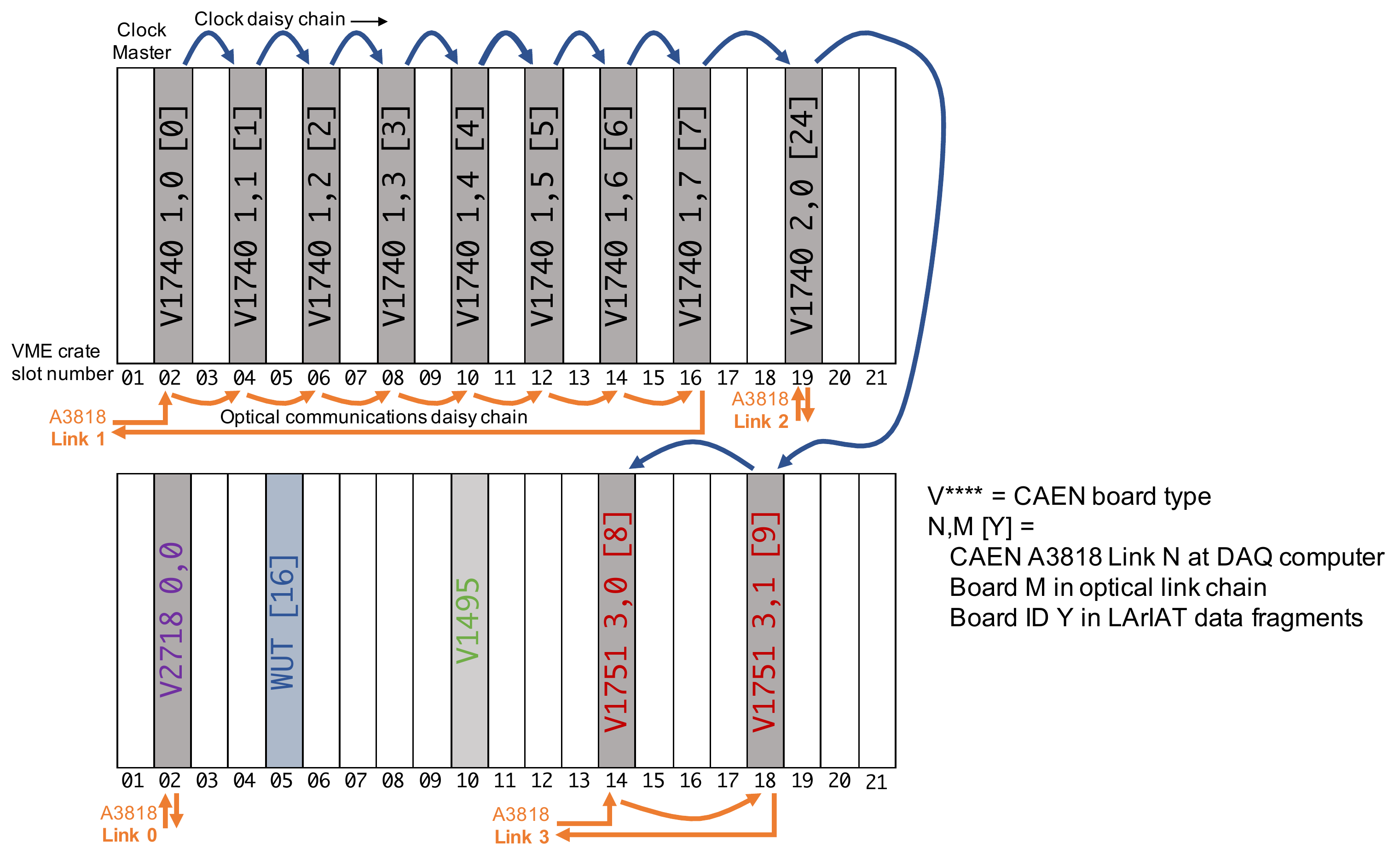}
\caption{
{The LArIAT digitizer timing and clock configuration.}
}
\label{fig:lariat_digitizers}
\end{figure}

The main LArIAT readout system, represented schematically in figure~\ref{fig:lariat_digitizers}, consists of two VME crates. A common sampling clock was produced by the first V1740
in the upper-left corner of the figure and then daisy-chained to all
the other CAEN digitizers. The synchronization procedure recommended by CAEN 
was applied to correct for board-to-board clock skew to the level of
approximately 150~ps.

Optical transceivers on each digitizer, implementing the CAEN CONET-2 protocol,
provide the control, configuration and data readout path. A four-port
CAEN A3818 CONET-2 controller card, attached to the system PCIe bus, serves
as an interface between the digitizers and the primary data acquisition
server. The configuration of the A3818 ports and associated digitizer modules,
connected in a daisy-chain, is also shown schematically in figure~\ref{fig:lariat_digitizers}. Because of
their large memory buffers and an extended post-spill readout time, the
boards were read out without data loss or dead time.

The CAEN V2718 VME crate controller shown in the bottom left of the same figure was used to configure the CAEN V1495 general-purpose logic board, which generates
the experimental triggers. 
In general, the VME crates were used only to supply power to
the modules. The configuration of the
V1495 was the only DAQ operation that made use of the VME data and protocol
lines.

As noted earlier, signals from the four beamline wire chambers were digitized
by a set of sixteen time-to-digital converters (TDCs), custom designed at
Fermilab and supported by FTBF staff.  The sixteen TDCs feed data to a central, custom TDC controller.  Data and configuration of the TDCs were directed through the ethernet port on the TDC controller.  Like the CAEN digitizers, the wire chamber readout and control software ran on the primary data acquisition server.

LArIAT was among the first experiments at Fermilab to use the {\texttt{artdaq}} package~\cite{artdaq} as the foundation of its data acquisition system. The package provides executables, in which {\texttt{art}}~\cite{artframework} analysis and output modules can be embedded,  which process live events passing through the DAQ.  In LArIAT, {\texttt{art}}'s output
module was embedded in an {\texttt{artdaq}} executable and was used to write data
in {\texttt{art}}-readable ROOT~\cite{ROOT} files. These ROOT files were then passed to nearline and
offline workflows, consisting of additional {\texttt{art}} modules, for further analysis.

The LArIAT data acquisition system ran on one, dedicated, 8-core server
using the Scientific Linux 6 operating system\footnote{\url{http://scientificlinux.org/about/}}. Because LArIAT
was a small, short-term experiment, the operator interface was designed to
be simple. There are no complex graphical interfaces. The {\texttt{artdaq}} executables are controlled by a single BASH script: data-taking is started and stopped with simple commands to the BASH shell.

The configuration and readout code were written in C++, taking advantage
of its object-oriented programming structure. Each type of digitizer
board was represented by its own driver class. An instance of that class
is created at run time for every physical board in the system.
In the case of the CAEN digitizers, a parent
driver class implemented features shared by all the models. Various
child classes, inherited from the parent, implemented features unique
to particular models. A top-level readout class performed class instantiation and implemented the overall run-flow logic. 

\begin{figure}[htb]
\centering
\includegraphics[width=0.8\textwidth]{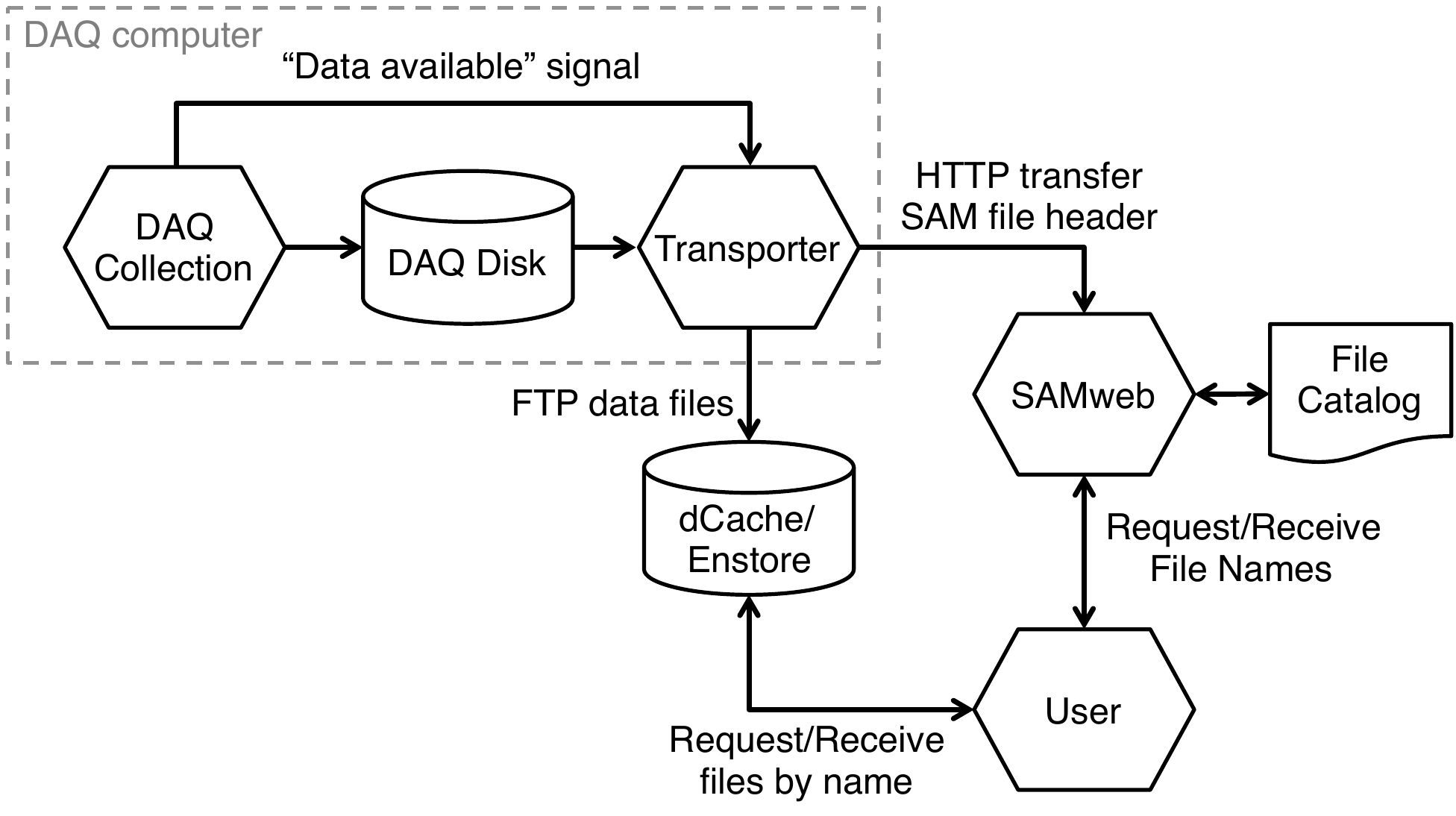}
\caption{
Overview of data archiving processes.
}
\label{fig:data_archiving}
\end{figure}

Figure~\ref{fig:data_archiving} shows an overview of the data storage and archiving process. The
basic unit of data storage is one beam {\it spill} in one file, with
approximately one file written per minute. For each run, the DAQ also writes
a run configuration file. At the beginning of each subrun, the DAQ writes a
small file containing the start time of the subrun and a flag describing the
run type. Once the data from a subrun are written to a local disk, a
python script generates the appropriate metadata for that file. The metadata includes information about beam conditions,
drawn from the accelerator's database at the time of the subrun.
Two database tables, one containing the run's configuration and the other
containing accelerator data for the subruns, are also updated by the script.
The data file and associated metadata file are transferred to local storage. Finally, the files are copied to long term storage by Fermilab's File Transfer Service.

\subsection{Trigger and readout system}\label{sec:Trigger}
Readout of the front-end data buffers is triggered by a CAEN V1495 module.
The V1495 is a powerful, easily-configurable coincidence module, featuring a
user-programmable FPGA. Sixteen logical inputs (32 in Run III) are sampled
at 100~MHz and are checked for matches to
any of the sixteen user-defined patterns in the trigger menu. If a trigger
pattern persists for two consecutive clock ticks, the V1495 generates
a readout request. 
In addition to signals from the accelerator, NIM-standard logic inputs
are derived from a variety of LArIAT detector
systems: from the beamline instruments, the cosmic ray paddles, and the
cryostat's scintillation detectors. LArIAT has taken
full advantage of the V1495's great flexibility: as experimental conditions
and detector configurations change, inputs to the trigger card
and its internal logic are modified accordingly. At the start of each
run, the configuration of trigger inputs is automatically logged in a database.

Three critical gates are derived from the accelerator control
signals: a brief, 1~second window before the beam arrives, a 4.2~second beam-on window,
and a 24 second beam-off window which is often used to collect cosmic
ray events. There is also a pulser input, running independently of the
experimental cycle, which is useful for collecting
events with zero bias.

Inputs from the time-of-flight (TOF) and wire chamber (WC) systems help generate the
basic, charged-particle trigger.
Coincident activity in the two systems suggests
that a charged particle has traversed the tertiary beamline
and entered the TPC and moreover, that measurements of the time of flight and
momentum can be used for particle identification. On each of the TOF
paddles, a coincidence (within 20~ns) of pulses from all its PMTs is formed.
The coincidence of the upstream and downstream signals is then formed by the V1495.
Each of the TDC modules used to read out the wire chamber system
provides a fast, logical-OR
of its inputs, indicating that one or more of the signal wires exceeded the
variable threshold. Using NIM logic units, the four OR signals from each
chamber are ORed together -- a logical-true level corresponds to significant
activity in any of the wire chamber's 256 wires. The summary OR signals are routed
to the V1495's FPGA, which generates a trigger whenever pulses from at least
three of its inputs fall within a 100~ns coincidence window.

\begin{figure}
\includegraphics[width=\textwidth]{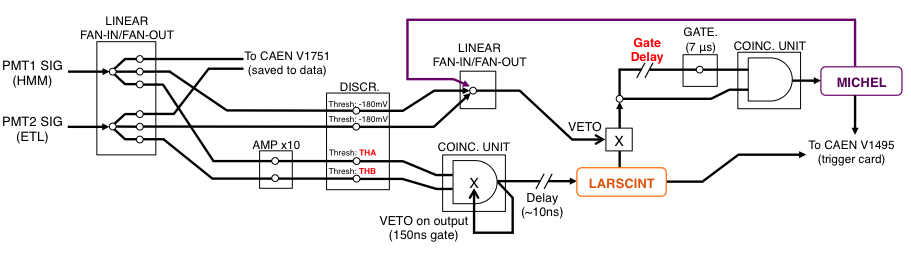}
\caption{\label{fig:michel_logic} A schematic diagram of the trigger logic used to select Michel electron events during the cosmic readout window of the LArIAT supercycle.  The two PMT signals refer to the Hamamatsu (``HMM'') and ETL PMTs described in section~\ref{sec:PhotonSystem}.  For some data-taking periods in Run II, unamplified PMT signals, discriminated at 180~mV, were used to veto events that might have saturated the 200-mV dynamic range of the V1751 digitizer.  The discriminator thresholds used on the amplified (x10) PMT signal copies (\emph{ThA}, \emph{ThB}) as well as the length of the gate delay, were adjusted between run periods.}
\end{figure}

Pulses from the cryogenic PMTs are used to construct several trigger
inputs from the TPC. The coincidence of discriminated pulses from the two PMTS (within 100~ns) indicates that ionizing radiation was produced in the TPC. {A \it Michel} electron trigger, indicating the decay of a stopping muon,  is formed by a delayed coincidence of two consecutive scintillation logic pulses, separated by a time interval ranging from 300~ns to 7~$\mu$s. Figure~\ref{fig:michel_logic} shows a schematic diagram of the logic comprising the Michel electron trigger, which is active outside of the beam spill window.

LArIAT also collects triggers from cosmic ray muons that do not stop inside the TPC, where a minimally-ionizing
particle crosses the TPC along the body diagonals between one upper (lower)
assembly in the upstream cosmic tower and one lower (upper) assembly
in the downstream tower, as illustrated in figure~\ref{fig:cosmic_towers}. The trigger for each body diagonal is formed
from coincidences of the two signals in each of the two corresponding
assemblies. The logical OR of the two possible triggers is routed to the V1495.

Other trigger inputs are formed with discriminated signals from the PMTs of the punchthrough scintillator paddles, located directly downstream of the TPC, and with signals from the PMTs installed in the muon range stack (MuRS). A single logic level, indicative of coincident activity in at least two overlapping paddles, is formed from the punch-through scintillator paddles. Discriminated signals within each of the four layers of the MuRS are combined in a similar fashion -- a logic level true indicates that there were pulses within a 20~ns coincidence window in 
at least one pair of overlapping paddles.

\subsection{Trigger decision}

The V1495 may be configured with up to sixteen trigger patterns and
sixteen veto patterns derived from the trigger input signals. A trigger pattern is defined as the AND of one or more defined inputs, and may include the NOT of the AND of further inputs. 
Veto patterns are defined in the same way, but they have a very different effect. When any of the trigger patterns fire, a "fast trigger" signal is issued and an adjustable countdown is initiated. If the countdown is complete before a veto pattern fires, a "slow trigger" signal is also issued on a separate hardware channel. If a veto pattern fires during the countdown, it vetoes the slow trigger signal. The fast trigger signal prompts readout of all the shorter data buffers, which includes the V1751 modules, the V1495 itself and the wire chambers' controller. TPC wire signal buffers, which stretch longer in time and are more numerous, are read out only when the slow trigger is issued.

\section{Monitoring and operations}\label{sec:Monitoring}

During the data-taking periods described in section~\ref{sec:DataCampaigns}, the LArIAT experiment's status was controlled and monitored by LArIAT shifters, using the tools described in this section.
\subsection{Slow controls monitoring}\label{sec:Synoptic}

The LArIAT slow controls system provides a live display of the current
experimental conditions, including information from the beamline
instrumentation, the cryogenic system, and the TPC. It uses the Fermilab
accelerator controls network, ACNET~\cite{ACNETURL}, and is based on
Synoptic\footnote{Java Synoptic Toolkit for ACNET, \url{http://synoptic.fnal.gov/about}}, a Java-based, graphical user interface, designed
to display ACNET data in a user-friendly way.  The ACNET protocol is used
to communicate with hardware components, providing an interface for
controlling and monitoring device voltages, currents, and temperatures.
The system includes an integrated data logger, Lumberjack, which allows direct
access to trend plots through the Synoptic displays. For offline
use, the slow-controls data are also stored in a PostgreSQL database managed by Fermilab's database group\footnote{IF Beam database, \url{https://cdcvs.fnal.gov/redmine/projects/ifbeamdata}}.
These data are later used in selecting good runs for physics analyses.

\begin{figure}[htb]
\centering
\includegraphics[width=\textwidth]{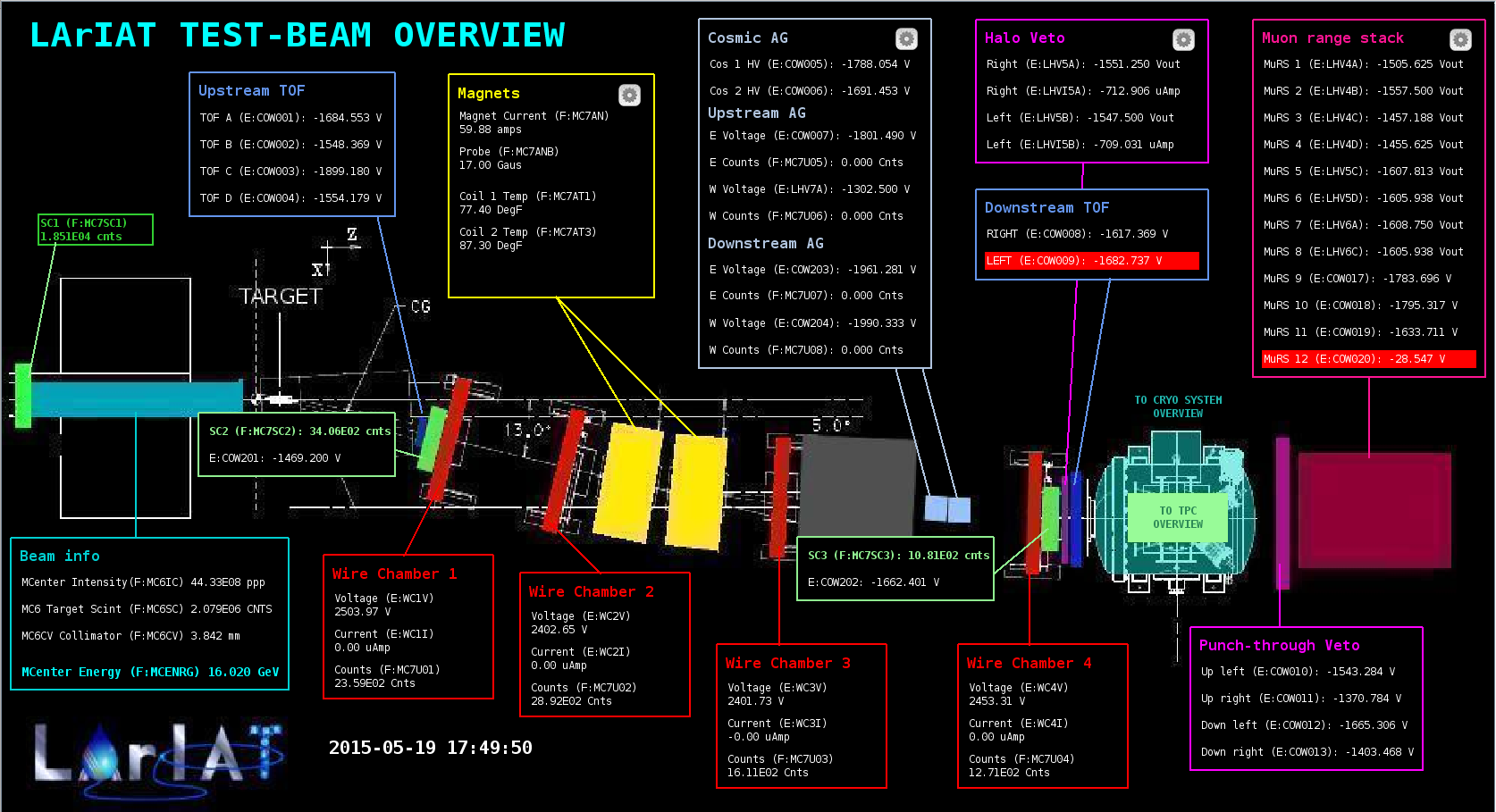}
\caption{Screenshot of the LArIAT main slow-controls monitoring page
showing secondary beam information, status and controls of tertiary
beam line instrumentation, cryogenics system, and TPC. Alarms are indicated by a red highlight box behind the white
text, as seen in the downstream TOF and muon range stack sections at right.}
\label{fig:LArIATBeamOverview}
\end{figure}

Synoptic reads the information from the ACNET devices and displays it
in two different ways: restricted access displays which allow control of
the devices via a Java application, and view-only, online displays
accessible through any web-browser. The flexibility of Synoptic allows
the creation of interconnected displays, which show the overall status of
the experiment (as in figure~\ref{fig:LArIATBeamOverview}) as well as the
detailed information and/or controls available for each sub-system
(e.g., figure~\ref{fig:LArIATCryoMonitor}). Plots are interactive, allowing
the user to choose the time range of the data displayed, from the last
few seconds up to the full data set since the device was connected, which
can be months or even years. Several alarms are implemented, alerting
the shifters and experts whenever a value is out-of-range, when a system
is malfunctioning, or when an interlock to the experimental hall or the detector systems has been activated.

\subsection{{DAQ} monitoring}\label{sec:DAQmonitoring}

\begin{figure}[htb]
\begin{centering}
\includegraphics[width=\textwidth]{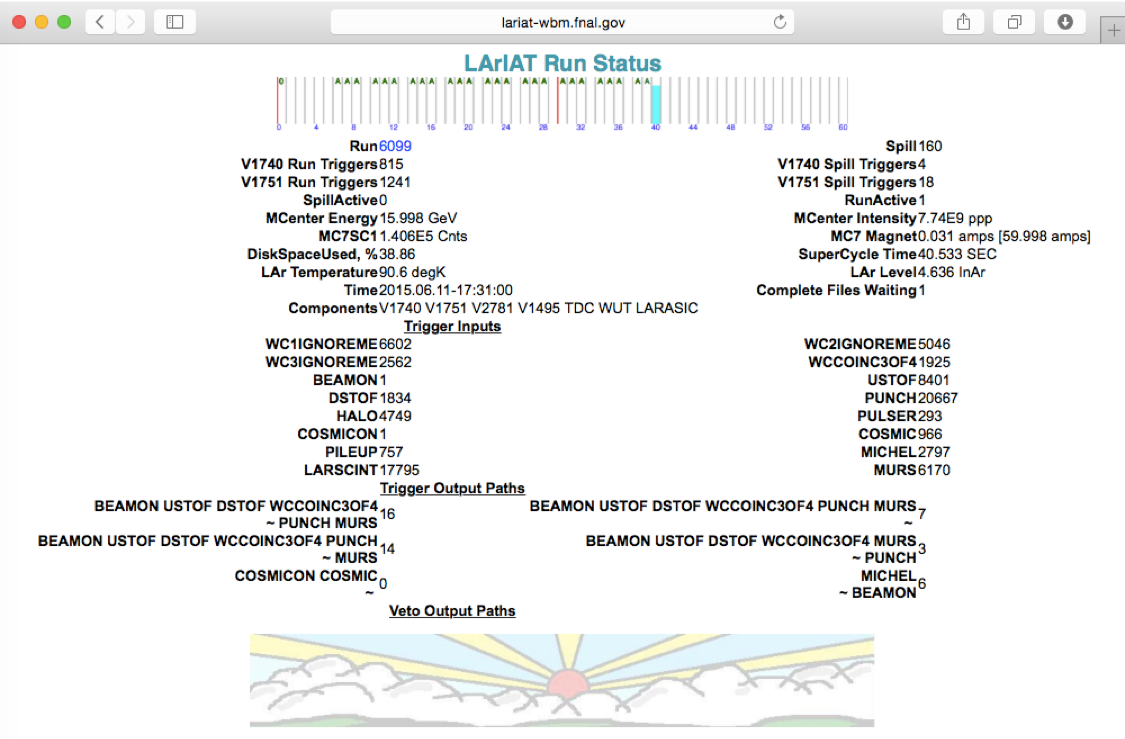}
\caption{ 
Succinct run status web page.
}
\label{fig:run_page}
\end{centering}
\end{figure}

The DAQ keeps track of a number of low-level quantities as it is writing data to disk. All of these quantities are displayed on a simple web page (shown in figure~\ref{fig:run_page}) that is updated in real-time,  which enables easy monitoring of the run status during data acquisition.

Some of the low-level quantities displayed include: relative time in the Fermilab accelerator complex supercycle, total number of triggers in the event per CAEN digitizer, total number of detectors triggered during a beam spill, trigger patterns and the number of times a particular trigger pattern was satisfied during a beam spill, as well as beam conditions and environmental parameters of the TPC, as seen in the figure. 
The web page issues an audible alarm and changes some of the page's text color to red if data-taking is not detected for a period of longer than 2 minutes.

\subsection{Data quality monitoring}\label{sec:DQM}

\begin{figure}[htb]
\centering
\includegraphics[width=\textwidth]{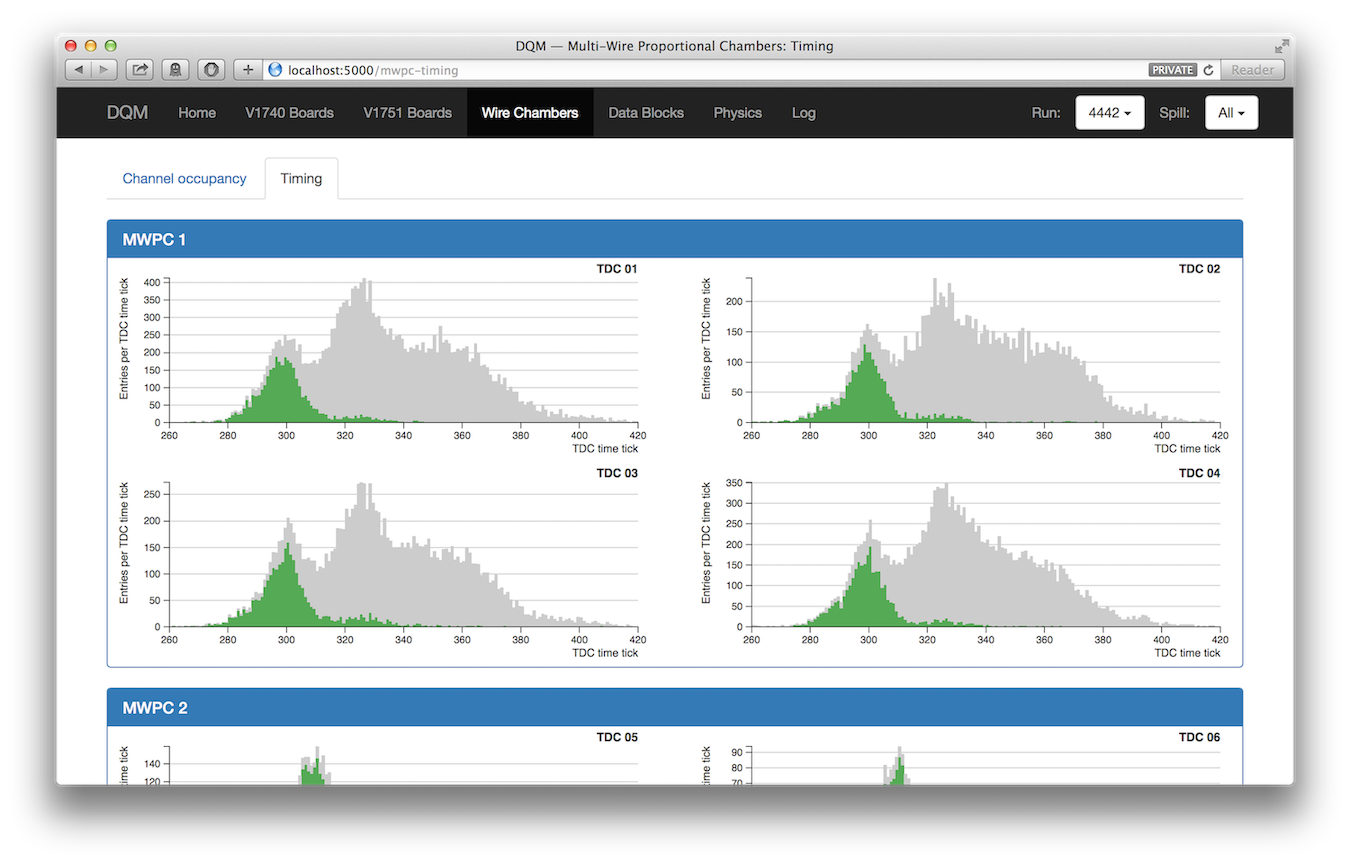}
\caption{
{MWPC panel from the DQM monitoring page}
}
\label{fig:dqm_page}
\end{figure}

The LArIAT data-quality monitoring (DQM) system is primarily used to perform near real-time
checks on low-level quantities in data.  The DQM system
processes the data on a spill-by-spill basis and a quick analysis is done.  The
results of the analysis are recorded in a PostgreSQL database and displayed on an interactive web page, as shown in figure~\ref{fig:dqm_page}. The front-end of the DQM consists of a website running on Flask, a microweb framework written in Python\footnote{Flask: \url{https://palletsprojects.com/p/flask}}.  
The DQM page allows shifters and experts to view the results from the most current collection of beam spills, typically with a 2-minute delay, as well as results from past collections of beam spills.

The low-level quantities monitored by the system include:
\begin{itemize}
  \item the number of data fragments recorded by the CAEN digitizer boards and the wire chamber controller ({\it data fragments} are pieced together to form an event),
  \item the number of recorded data fragments that are used to build a TPC event,
  \item the pedestal mean and pedestal RMS on the CAEN digitizer boards (this includes the readout from the TPC wires, light-collection PMTs, and PMTs of various beam line detectors),
  \item the hit occupancy and timing plots on the multi-wire chambers.
\end{itemize}
\noindent
There is also a high-level time-of-flight plot, which provides a crude 
monitor of the tertiary beam composition. 

In addition, the DQM system has a series of alarms that are activated whenever a
low-level quantity is found outside its tolerance  during the most current beam spill.  The 2-minute feedback for beam and detector conditions allows a quick response by the shifter or detector experts, minimizing beam and detector down time.

 The Online Event Viewer displays in near real time a two-dimensional representation (wire number vs. time tick) of LArIAT TPC events on both the induction and the collection planes. The raw pulses collected by the DAQ on each wire are plotted as a function of drift time, resulting in an easily-read image of the TPC event. The Event Viewer's summary of TPC operation provides immediate feedback for troubleshooting a number of common issues. For example, it is easy for the LArIAT shifter to spot high occupancy events, a decrease in argon purity, or upticks in the level of electronic noise.

\section{Data-taking campaigns}\label{sec:DataCampaigns}

LArIAT commissioned its beamline in the summer of 2014. The cryostat, TPC 
and light collection system were installed for Run~I, which lasted from April
of 2015 until the summer shutdown in July, 2015. The wire pitch in the TPC was 4~mm
and the light collection system consisted of a pair of cryogenic PMTs. The
configuration for Run~II, which started in February 2016, was similar to
that of Run~I. However, important improvements were made to both the
wire readout and light collection systems (discussed more in section~\ref{sec:PhotonSystem}) which resulted in significant
improvement in data quality. 
Silicon photomultiplier devices (SiPMs) were also tested as alternatives to more conventional PMTs during Run~II. Run~II ended with the summer shutdown in July 2016.

It became clear that testing additional experimental configurations would be helpful for refining the designs of the SBND and DUNE LArTPCs, and so a number of experimental modifications were tested in Run~III, which extended
from March to July of 2017. As a prototype test of the SBND cathode design, the original solid cathode (section~\ref{sec:TPC}) was
replaced with a semi-transparent wire-grid cathode and, to enhance the light
collection system, wavelength-shifting reflector
foils were installed on the cathode, sandwiched between the two cathode grids.
In addition, a new light detection device under consideration for DUNE, the ARAPUCA, was installed alongside the two cryogenic PMTs. In light of the DUNE design process, the impact of wire pitch on TPC performace was also considered interesting and to this end, the LArIAT TPC was first outfitted in Run~III with a set of 5~mm-pitch wire planes and later with a set of 3~mm-pitch wire planes.

The LArIAT data-taking campaigns and hardware configurations are summarized in Table~\ref{tab:data-campaigns}. 

\begin{table}[htb]
\centering
\caption{A summary of the hardware configurations in each of the LArIAT data-taking periods. The signal-to-noise (S/N) ratio that is reported in the last column is calculated using the most probable pulse amplitude for minimally-ionizing, through-going beam tracks over all collection plane wires. The reported S/N is corrected for average track pitch relative to the orientation of the collection wires.}
\label{tab:data-campaigns}
\begin{tabular}{l l l c c}
\hline
 \textbf{Run Period} &  \makecell[c]{\textbf{Dates}} & \textbf{Detector configuration} & \textbf{Num. triggers} & \textbf{TPC S/N} \\ \hline 
 
 \hline
 Engineering    
 & \makecell[tr]{Aug. 15, 2014\\-- Sep. 9, 2014}  
 & No TPC     
 & --- 
 & ---   \\

\hline 
Run I    
& \makecell[tr]{Apr. 30, 2015\\-- Jul. 7, 2015} 
& \makecell[tl]{4-mm wire pitch\\ETL 2-inch PMT}  
& \makecell[tr]{Beam: 202,000\\Cosmic: 564,000}   
& 30 \\

\hline
 Run IIA   
 & \makecell[tr]{Feb. 18, 2016\\-- May 20, 2016} 
 & \makecell[tl]{4-mm wire pitch\\ETL 2-inch PMT (w/TPB)\\SiPMs}  
 & \makecell[tr]{Beam: 1,887,000\\Cosmic: 3,207,000} 
 & 45\\

\hline
 Run IIB   
 & \makecell[tr]{May 24, 2016\\-- Jul. 7, 2016} 
 & \makecell[tl]{4-mm wire pitch\\Hamamatsu 3-inch PMT\\ETL 2-inch PMT (w/TPB)\\SiPMs}  
 & \makecell[tr]{Beam: 770,000\\Cosmic: 1,309,000} 
 & 45 \\

\hline
Run IIIA  
& \makecell[tr]{Mar. 14, 2017\\-- May 17, 2017}   
& \makecell[tl]{5-mm wire pitch\\Hamamatsu 3-inch PMT\\ETL 2-inch PMT\\ARAPUCA\\Mesh cathode} 
& \makecell[tr]{Beam: 1,098,000\\Cosmic: 1,772,000} 
& 53 \\

\hline
Run IIIB 
& \makecell[tr]{Jun. 24, 2017\\-- Jul. 7, 2017}  
& \makecell[tl]{3-mm wire pitch\\Hamamatsu 3-inch PMT\\ETL 2-inch PMT\\ARAPUCA\\Mesh cathode w/reflector foil} 
& \makecell[tr]{Beam: 188,000\\Cosmic: 319,000}
& 36 \\
\hline
\end{tabular}
\end{table}


\section{Beamline and detector performance}\label{sec:DetectorPerformance}

LArIAT results depend critically on a detailed understanding of both the beamline and the TPC. Critical issues for the former include the calibration of the time of flight and wire chamber systems, which determine the momentum and particle ID. Other important issues include separation of pions and kaons, pile up, and beam halo. The TPC requires its own set of studies: determining the electric field and the drift velocity of the electrons, balancing the response of the individual wires in the induction and collection planes, determining calibration constants for the translation of wire signals into charge deposition, as well corrections for losses to impurities. We provide an overview of the analysis of these various systems in the following sections.

\subsection{Beamline momentum reconstruction} \label{sec:momentumscale}

The bending magnets, together with the wire chamber system, act as a momentum spectrometer. The direction of the magnetic field in the center of the magnets is along the vertical, $y$-axis, pointing up or down depending on the direction of the current in the magnets. The 3D positions
of the hits in the upstream wire chambers  provide a straight trajectory before the bending magnets, while the positions 
of the hits in the downstream wire chambers provide a straight trajectory afterwards. A charged particle traversing the beamline bends in the $xz$-plane, allowing the measurement of the transverse component of the particle's  momentum, $p_{xz}$.  The bend plane angles of the upstream and downstream trajectories, $\theta_{\textrm{US}}$ and $\theta_{\textrm{DS}}$, are measured relative to the $z$-axis.
The total bending angle of the particle, $\theta_\textrm{bend} = \theta_{\textrm{DS}} - \theta_{\textrm{US}}$, is then used to calculate the transverse momentum:
\begin{equation}
p_{xz}=\frac{ 0.2998 q \int{B\textrm{d}\ell}}{\theta_\textrm{bend}}
\end{equation}
where, with $q$ measured in elementary charges, a factor of 0.2998 converts the momentum units from $\mathrm{kg\cdot m/s}$ to MeV/c , and $q\int{B\textrm{d}\ell}$ is the $p_t$-kick, the change in the direction of the transverse component of the momentum. With modest errors from multiple scattering and energy loss, $p_{xz}$ and the particle's 3D trajectory in the downstream wire chamber pair can be used to determine the three components of the particle's momentum when it enters the TPC.

The momentum resolution is determined in part by geometric effects: uncertainties in the location of each wire chamber, the position of the wires within the chambers and most important, the spacing of the wires. For example, the distance between the first two wire chambers is approximately 1~m and the second pair about 2~m. With a wire pitch of 1~mm, the angular
resolution on a straight line between the hits should be a milliradian or less. Since the total bend, $\theta_\textrm{bend}$, is about 10 degrees (0.17 radians), the resolution on the bend angle is less than 1\%. The contribution to the momentum resolution, estimated with the Monte Carlo simulation program, was 0.8 \%. 


Other less significant factors affected the position resolution. 
While the rate of beam particles was fairly low, beam halo sometimes produced false hits, especially in wire chamber 1. In most cases, clusters with multiple wire hits were caused by cross talk. To minimize the effect of cross talk on the position resolution, the position of the cluster was taken to be that of the wire with the earliest hit. The residual effect on the momentum resolution was negligible.

The momentum resolution is also affected by energy loss and multiple scattering. At 500~MeV/c, energy loss is less than 1~MeV in the spectrometer and straggling is completely negligible.  For a 500~MeV/c pion, the typical scattering angle over 2~m of air is 2~mrad, a fractional resolution of about 1\%. The contribution of multiple scattering to the momentum resolution of the spectrometer was modeled with the Monte Carlo simulation program. For 500~MeV/c muons, typical of LArIAT measurements, the effect of multiple scattering is 1.3\%.  Since detailed tracking through the magnets isn't done, the uniformity of the field integral is also very important. Although remarkably constant for a broad range of trajectories, lower momentum particles, which curve more, have a larger integrated path length than higher momentum particles and hence bend more. The integrated magnetic field felt by particles which pass closer to the edges of the magnet's transverse apertures may be slightly different from those which pass through the center. These effects were important for the MINER$\nu$A test beam experiment~\cite{MinervaTestbeam}, which used a very similar tertiary beamline before LArIAT. Fortunately, LArIAT's effective aperture is much smaller than was MINER$\nu$A's, and the effect of magnetic field variation is negligible\footnote{Details on the MINER$\nu$A magnetic field map can be found in reference~\cite{MINERvA-fieldmap}}.
The overall momentum resolution, dominated by multiple scattering is $1.51\pm0.01$\%. 

While the absolute rate in the LArIAT beamline is relatively
low, the wire chambers sometimes register multiple hits, leading to an ambiguity in associating hits from one wire chamber with hits in the others. Therefore, two classes of reconstructed tracks are defined. In a {\it golden} track, where all four wire chambers have exactly one $xy$-hit, there is no ambiguity. Golden tracks provide a low-statistics, high-purity calibration sample. In a {\it high yield} track, multiple hits are allowed within a wire chamber. 
High yield tracks thus provide a high-statistics, lower-purity sample. Even for the high yield sample, only one track is reconstructed from each readout of the wire chambers. A temporal and spatial clustering algorithm is applied to the individual wire hits within a chamber, and the hit cluster that would form the straightest line with the other wire chambers in the non-bend ($yz$) plane is chosen for that track.

\subsection{Beamline spectrometer calibration}

By taking advantage of the known masses of the protons and kaons~\cite{PDG2018} in the beam, a data-driven calibration of both the TOF system and the beamline momentum scale was obtained. This calibration also provides the momentum scale systematic uncertainty, which is needed for all physics analyses that use the beamline momentum measurement. 

The TOF system is first used to select samples of protons and kaons with high purity. These samples are binned in momentum, where the momentum bin size is optimized to be as small as possible, while still providing sufficient statistics for obtaining the most probable value (MPV) of the TOF measurement, $\tau_i$, and momentum mean, $p_i$, for each momentum bin.  The simultaneous calibration of the TOF measurement and the momentum scale is obtained from a binned likelihood fit comparing the reconstructed particle mass $m_i$ with the corresponding PDG value, which is known to much higher precision:
\begin{equation}
 \chi^2=\sum_{i} \frac{(m^2_{PDG} - m_i^2)^2}{\sigma_i^2}
\end{equation}
where the sum is over all proton and kaon momentum bins,
\begin{equation}
 m^2_{i}=\alpha^2 {\frac{p^2_{i}}{c^2}}\left(\frac{(\tau_{i}+\Delta\tau)^2}{\tau_{l}^2}-1\right),
\end{equation}
and $\sigma_i$ is the uncertainty of $m_i^2$. The momentum scale $\alpha$ and the TOF offset, $\Delta\tau$, are the free fit parameters, and $\tau_l=22.30 \pm 0.04$~ns is the time it takes light to travel the distance between the TOF counters along the beamline.  

\begin{figure}[htb]
\begin{centering}
\includegraphics[width=0.7\textwidth]{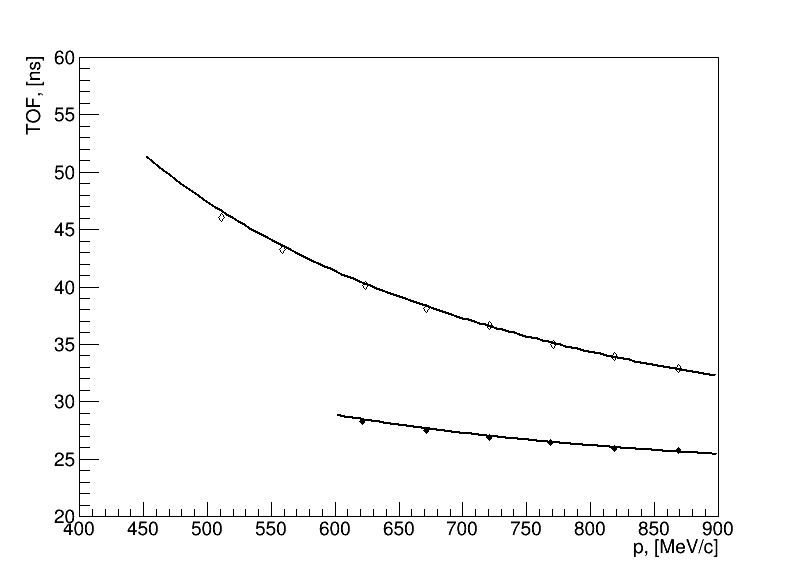}
\caption{TOF MPV in momentum bins after calibration for protons (\emph{top}) and kaons (\emph{bottom}) for the +100A sample. Errors on data points are too small to be visible. The lines represent the theoretical expectation for the time it takes protons and kaons to traverse the beamline as a function of momentum. 
}
\label{fig:mom_cal_100A}
\end{centering}
\end{figure}

Figure~\ref{fig:mom_cal_100A} shows the TOF MPV in momentum bins after calibration for 
the +100A beam sample.  The lines represent the time it takes a proton (top) or a kaon (bottom) to traverse the beamline, as a function of momentum. The momentum scale uncertainty obtained from the fit is 0.5\%. The calibration of the momentum scale is, in effect, a measurement of the $\int B\cdot dl$ of the bending magnets. The measured value agrees, within the quoted uncertainty, with  the magnetic field map produced by MINERvA for these magnets during MINERvA's test beam run. 

\subsection{TPC Reconstruction}\label{sec:TPCReco}
For event reconstruction, LArIAT uses LArSoft~\cite{larsoft}, a software package that provides a framework for the simulation, reconstruction and analysis of data from LArTPCs.  Here we summarize LArIAT's TPC data processing and reconstruction chain, used on both experimental and Monte Carlo data.

Event reconstruction takes ionization-induced signals
on the sense wires as input and produces as output three-dimensional
physics objects, with associated calorimetry. In the first step,
TPC {\it signal deconvolution} removes the response of the electronics
and transforms the raw, bipolar induction-plane pulses and unipolar
collection-plane pulses into unipolar, approximately-gaussian
pulses. Following deconvolution, the resulting pulses are grouped into
{\it hits} which indicate an energy deposition in the detector.
In this hit-finding process, an algorithm scans the deconvolved TPC waveforms for each
wire over the entire readout, looking for peaks above the waveform's
baseline. These peaks are fit with a gaussian function and the
parameters are stored. The area of the gaussian is proportional to the charge
collected on the wire while the location of the fitted peak corresponds to the arrival time
of the charge, which (together with the assumed electron drift velocity) is used to determine where along the drift direction the ionization occurred.

Based on their topology, collections of hits on each wire plane are then grouped into {\it clusters} using the TrajCluster~\cite{SignalReco} algorithm. TrajCluster forms clusters by identifying groups of hits in 2D wire-time space that form line-like trajectories. First, two hits are chosen to form a ``seed trajectory,'' and the algorithm then subsequently walks through other hits within proximity and identifies which ones belong to the  assumed trajectory. Several factors determine whether a given hit is added to a trajectory, including but not limited to: (1) the goodness of fit of the single hit; (2) the charge of the hit compared to the average charge and RMS of the hits already forming the trajectory; (3) the angle between the two lines formed by the collection of hits before and after the considered hit in the trajectory.

Collections of 2D clusters are matched between wire planes to form 3D \emph{track} objects using a set of ``projection-matching'' algorithms that were first developed for the ICARUS experiment~\cite{pmtrack}. This set of algorithms first identifies clusters in the induction plane and collection plane with similar distributions of hit arrival times, and constructs a tentative 3D trajectory using the endpoints of the matched clusters. Then, the tentative trajectory is projected onto the 2D planes and the parameters of the 3D track are adjusted to simultaneously minimize the distances between the projections and the 2D hits making up the individual clusters. Tri-dimensional tracking can use multiple clusters in one plane, but it can never break them into smaller groups of hits.


Similar procedures are used to group clusters together to form more spatially dispersed depositions of charge referred to as {\it showers}. The exact details behind the various shower-finding algorithms used in LArIAT are analysis-specific, and are not relevant to any of the calibrations described in this section.


A multi-pass reconstruction is done to account for the electric field non-uniformities near the upstream and downstream edges of the TPC.  In the first pass, reconstructed 3D hits that are within a few centimeters from the front and back edges of the TPC are identified and tagged; in the second pass, the reconstruction is run again with the tagged hits removed.  This iterative fiducialization procedure is done to minimize the effects of the field non-uniformities on the reconstruction.

Although LArIAT has used LArSoft as the basis of its offline reconstruction, analyzers tuned its clustering and tracking algorithms for sensitivity to tracks coming from a known beam direction. In neutrino experiments, a much broader range of initial angles must be considered. LArIAT's sensitivity to charged hadron interactions, which are flagged as changes in track direction, is not limited by the cosmic ray background. Unlike other LArTPC experiments which use LArSoft, the overlap of cosmic rays within the readout window  of the LArIAT TPC is negligible. To match LArIAT's single-pass sensitivity to beam-related tracks with small interaction angles, other LArTPC experiments must perform several reconstruction passes. Finally, the excellent performance of the LArIAT TPC cold electronics made noise mitigation and removal largely unnecessary. As a result, LArIAT physics objects were not distorted by the electronic filtering required by some other LArTPC experiments.

\subsection{Electric field and drift velocity measurements} \label{sec:EFieldMeasurements}
%

\begin{figure} [htb]
\centering
\includegraphics[width=0.6\textwidth]{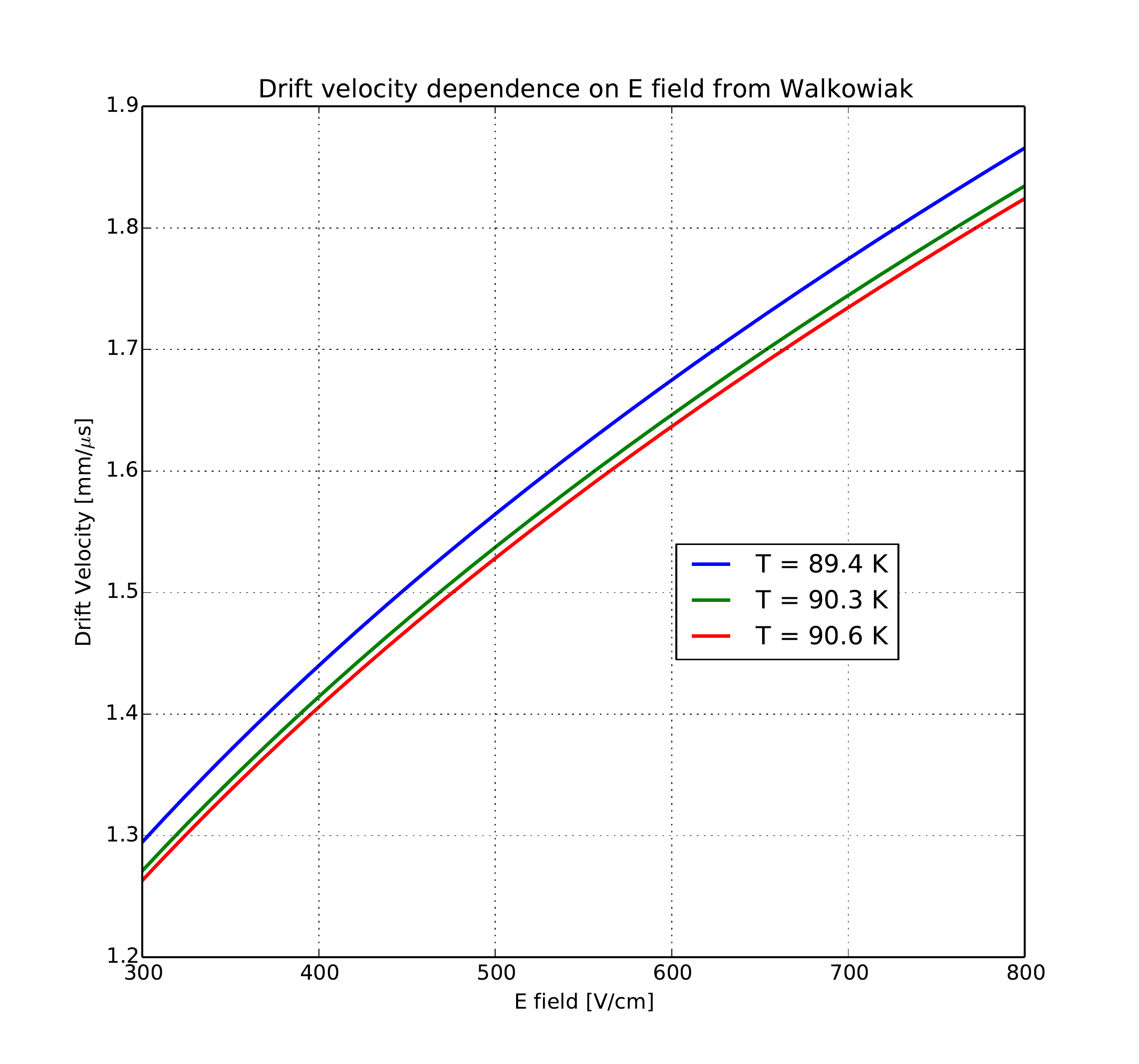}
\caption{Drift velocity dependence on electric field for several temperatures~\cite{walkowiak-driftvel}. The slope of the line at any point represents the electron mobility for that temperature and electric field.}
\label{fig:walkowiak}
\end{figure}

The electric field in the drift volume of the TPC must be known in order to correctly account for recombination effects, and to match TPC tracks with beamline tracks. Given its importance to the experiment, the electric field of the LArIAT TPC was determined using two approaches: a direct measurement of the potential difference across the relevant electrodes, and an indirect measurement, which makes use of tracks crossing the anode and cathode to measure the time difference, $t_{{\textrm{drift}}}$, between the earliest and latest hits of the track. In the
latter approach, $v_{{\textrm{drift}}}$ is simply $d_{{\textrm{drift}}}/t_{{\textrm{drift}}}$ where
$d_{{\textrm{drift}}}$ is the separation of the electrodes. The electric field is determined by the familiar relation
$v_{{\textrm{drift}}} = \mu(\mathcal{E},T)\cdot \mathcal{E}$, where
electron mobility, $\mu$, is a well-known function of the electric field and temperature. An empirical
formula for the dependence of $\mu$ on $\mathcal{E}$ and $T$ is described in~\cite{walkowiak-driftvel} and shown in figure~\ref{fig:walkowiak} for several argon temperatures. In LArIAT, this technique is subject to larger uncertainties than the direct measurement because of a hardware limitation, however the technique is presented here because it will be a useful tool in future experiments.

\begin{figure} [htb]
\centering
\includegraphics[width=0.6\textwidth]{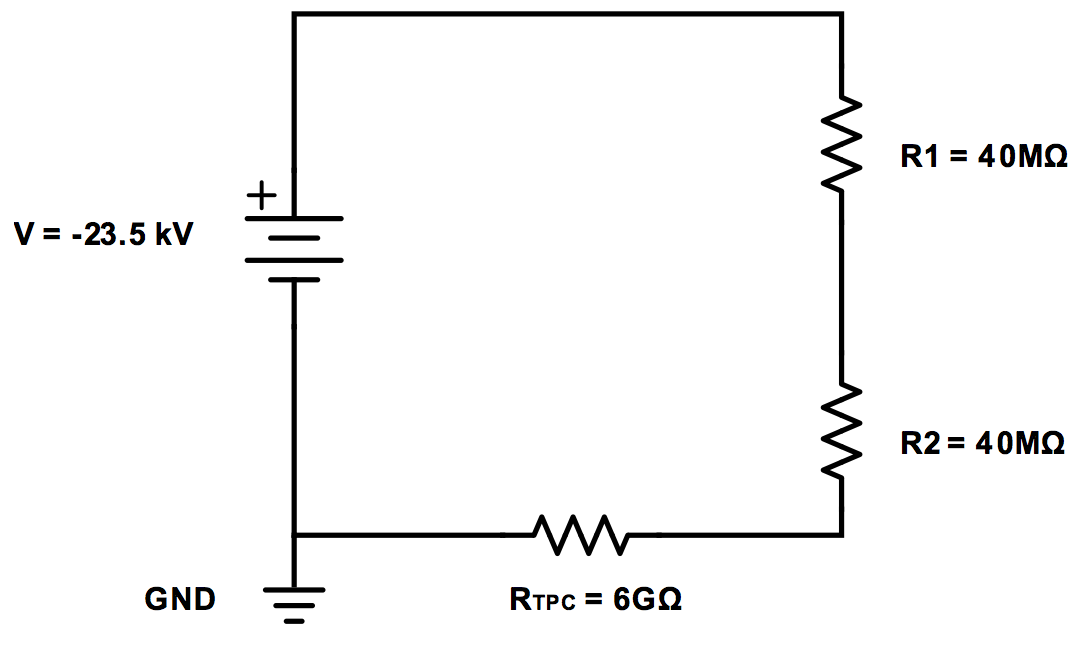}
\caption{Schematic diagram of the LArIAT TPC circuit, including the two in-line resistances (R$_1$ and R$_2$) from the filter pots, and the equivalent resistance of the TPC (R$_{\mathrm{TPC}}$).}
\label{fig:LArIATCircuit}
\end{figure}

The direct measurements are relatively simple. For example, the main drift
field depends on the potential difference between the cathode and the
shield plane. Applying Ohm's law to the circuit diagram in figure~\ref{fig:LArIATCircuit} determines the voltage applied to the cathode, $V_{BC}$:

\begin{equation} \label{eq:VBC}
V_{BC}=V_{PS} - (I \times R_{eq}) = -23.5\text{ kV} + ( 0.00417\text{ mA} \times 80\text{ M}\Omega ) = -23.17\text{ kV}, 
\end{equation}
\noindent
where $V_{PS}$ is the terminal voltage of the power supply, $I$ is the current and $R_{eq}$ is the equivalent resistance of the
two filter pots. The uncertainty in this measurement is dominated by fluctuations in the power supply current draw ($\pm0.0015\text{ mA}$). Then the electric field for Run~I, as an example, is
\begin{equation}
\mathcal{E} = \frac{|V_{BC} - V_{\text{shield}}|}{d_{{\textrm{drift}}}}
= 483.5 \pm 2.5\text{ V/cm}
\end{equation}


The indirect determination of $\mathcal{E}$
requires tracks which cross both cathode and anode.
Since tracks created by beam particles travel almost
perpendicular to the drift direction, the {\it anode to cathode piercing}
(ACP) tracks are selected exclusively from a cosmic ray sample. To 
minimize background noise, including Michel electrons from muon decay,
the events must contain no more than a single reconstructed
track. Note that this method does not rely on knowledge of the trigger time.
Because the passage of the cosmic ray is instantaneous on the
time scale of the drifting charge,
the maximum drift time for ACP tracks is simply the time difference between the earliest and latest hits.

\begin{figure} 
\centering
\includegraphics[width=0.6\textwidth]{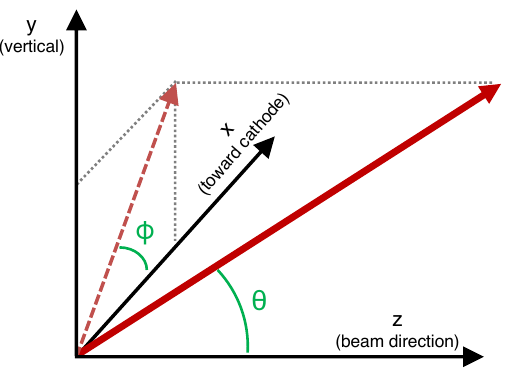}
\caption{Definition of angles in the LArIAT coordinate system. 
}
\label{fig:AngleDef}
\end{figure}

Further requirements enrich the fraction of ACP tracks in the cosmic ray
sample:
\begin{itemize}
\item To eliminate tracks which pass from top to bottom, the vertical
position (Y) of the first and last hits should be within 18~cm of the
vertical midplane of the TPC.
\item To eliminate through-going tracks, the horizontal position (Z)
of the first and last hits should be more than 2~cm, and less than 86~cm,
respectively, from the front face of the TPC.
\item To increase the probability of selecting a crossing track, the
track length should be greater than 48~cm.
\item For more reliable tracking, the angle of the track relative to the drift
direction ($\phi$ in figure~\ref{fig:AngleDef}) should be smaller than 50$^{\circ}$.
\item Similarly, the angle relative to the beam direction ($\theta$ in figure~\ref{fig:AngleDef}) should be greater than 50$^{\circ}$.
\end{itemize}

Some care must be taken in extracting the maximum drift times in each drift region. Since the wires of the shield plane are not read out, the time difference of the earliest and latest hits on a given plane (induction or collection) includes the time taken for the charge to drift 473~mm from the cathode to the shield plane plus the amount of time taken to traverse the 4~mm gap between the shield plane and the induction plane, where the electric field is necessarily higher than in the main drift region in order for the shield plane to satisfy the wire grid transparency condition~\cite{blum2008particle}. 

The voltages on the LArIAT shield, induction, and collection planes were extremely stable at their set values listed in table~\ref{tab:wireplane-voltages}, with fluctuations well below 0.1~V. The electric field in the shield-to-induction gap is thus calculated as $|V_{\textrm{shield}}-V_{\textrm{induction}}|/L_{\textrm{gap}}$, the voltage difference between those two planes, divided by the length of the gap (4~mm). The electric field in this gap is converted to a drift velocity via the relationship shown in figure~\ref{fig:walkowiak}. As seen in the figure, there is a strong dependence of the drift velocity on the temperature. Although the RTDs installed inside the LArIAT cryostat indicate that the temperature was stable to within 0.4~K during the operating periods, the absolute calibration of these RTDs is not known. However, the precision of the RTD measurements is trustworthy. To determine the absolute temperature, a pressure sensor in the cryostat was used to determine the equilibrium temperature at the gas liquid boundary. 

\begin{table}[htb]
\centering
\caption{Liquid argon temperatures and electric field measurements via the ACP track method}
\label{tab:cryo_temps}
\begin{tabular}{l c c}

\hline
{\bf Run  Period} & {\bf Temperature} & {\bf Electric field}\\
   & (K) & (V/cm)\\
\hline
Run-I  & $89.4\pm0.4$ &  $463.7\pm10.6$\\
Run-II & $90.3\pm0.3$ & $465.9\pm7.8$\\
Run-III & $90.3\pm0.3$ & $485.7\pm12.0$\\
\hline
\end{tabular}
\end{table}

\begin{figure}[htb]
\centering
\includegraphics[width=0.49\textwidth]{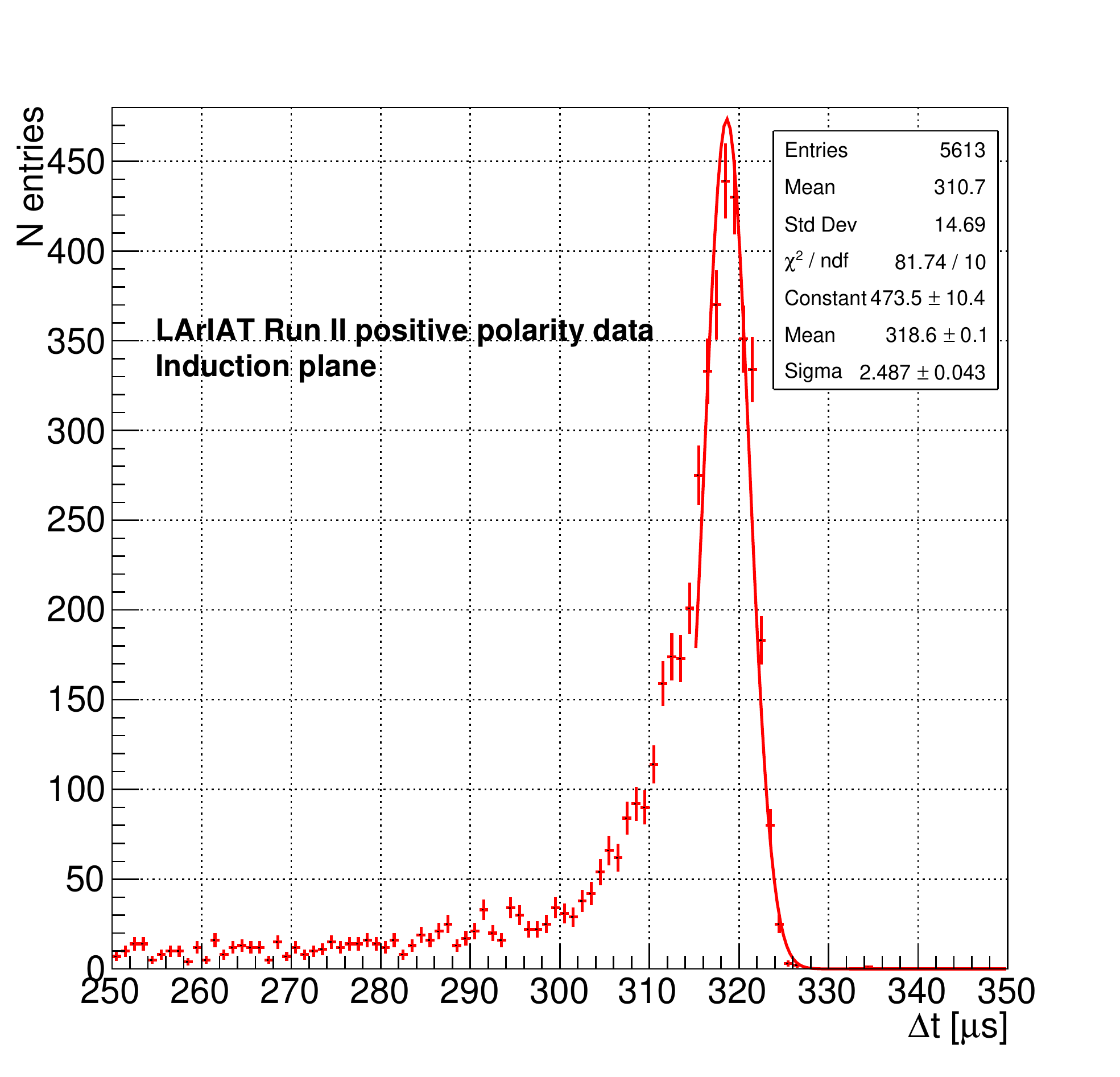}
\includegraphics[width=0.49\textwidth]{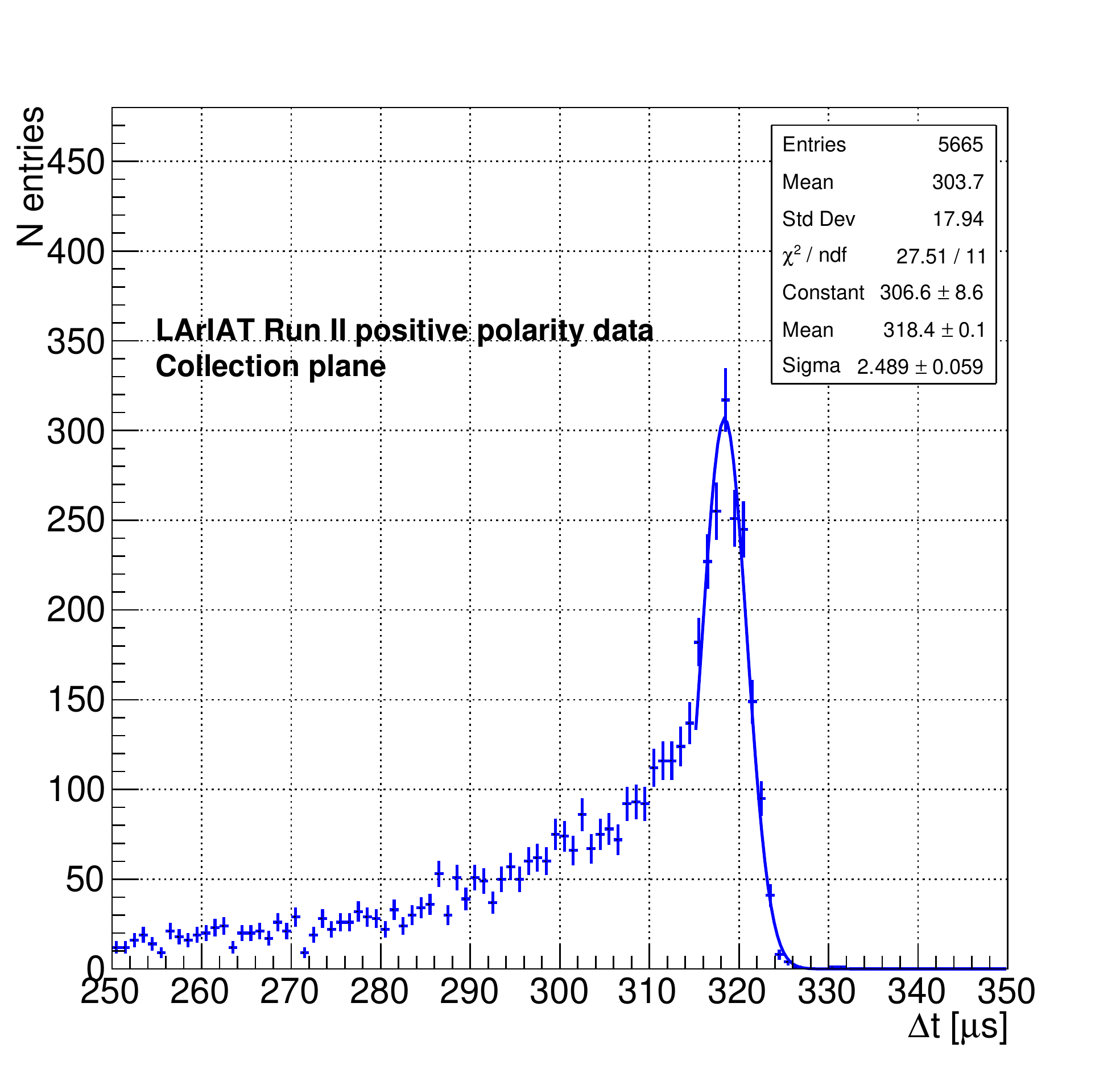}
\caption{Timing difference for Run II positive-polarity ACP tracks at the indcution plane ({\emph{left}}) and at the collection plane ({\emph{right}}).}
\label{fig:driftvel}
\end{figure}

The time difference of the earliest and latest hits seen on the induction plane is the sum of the drift times from cathode to shield plane and from shield plane to induction plane. Subtraction of $\sim2.3\mu$s removes the shield-to-induction gap, leaving only the time difference for the main drift volume. Figure~\ref{fig:driftvel} shows two spectra of maximal time differences for ACP tracks, before corrections of the shield-to-induction gap drift time. Both spectra are drawn from the Run~II positive polarity sample, with the induction plane data shown at left and the collection plane data shown at right. The maximum drift time, including the shield-to-induction gap, is taken as the mean of the fitted gaussian in the peak region of the histogram. The long tail at small $\Delta t$ indicates contamination from non-ACP tracks in the sample, and is not included in the fit. After correcting the induction and collection plane mean times by 2.3~$\mu$s, the drift speed is determined for the main volume by dividing by 473~mm, the distance between the cathode and the shield plane. As mentioned above, converting drift velocities to electric fields requires a known temperature. Table~\ref{tab:cryo_temps} reports the mean temperature values for each of the three major running periods, with uncertainties based on the RMS of the RTD measurements. The temperature uncertainties translate to a range of possible drift times which, in turn, introduce a large uncertainty in the electric field determination, as seen in the right-most column.
These ACP results are roughly consistent with the electric field from the direct method.

\subsection{Electron lifetime measurement}\label{sec:ElectronLifetime}
Electronegative contaminants in liquid argon, such as oxygen and water vapor, are harmful to physics measurements because they readily capture drifting ionization electrons. In the presence of such contaminants, the amount of charge collected at the wire planes depends on the distance the charge travels to reach the wire planes, or equivalently, the drift time.
For a given charge {\it deposited} in argon, the charge {\it collected} decays exponentially with drift time. The characteristic time constant, $\tau$, is called the \textit{electron lifetime}. In the LArIAT experiment, the electron lifetime is measured for each 24-hour period of collected data, and a database of lifetime corrections is filled for each day of data collection; these correction factors are applied to data used in physics analyses.

\subsubsection{Measurement principle}
The effect of contamination is measured with a data set of cosmic ray muon tracks
which cross a body diagonal of the TPC, tracks that are triggered by the cosmic towers described 
in figure~\ref{fig:cosmic_towers}.
A portion of each of these tracks will be far from the wire planes, and a portion of each track will be close.
Cosmic ray muons are minimum
ionizing particles (MIPs) and fluctuations in the amount of charge they deposit per unit
length follow a Landau distribution. In addition, over the momentum
range of cosmic rays in LArIAT, the most probable value (MPV) of that distribution
remains roughly constant. While energy deposited by these cosmic rays is uniform over
any plane perpendicular to the electric field, because of contaminants,
the amount of energy collected by the wire planes depends on the drift
distance. Ionization charge produced close to the wire planes will be
minimally attenuated, while ionization charge produced close to the
cathode will suffer maximal attenuation.

If the contaminants are uniformly distributed in the TPC, the fractional
loss of ionization is constant at every step, and the charge
deposited at the collection plane decreases exponentially with the
total drift length. And, since the drift length is directly proportional
to the electrons' (constant) drift speed, the charge collected
also decreases exponentially with the drift time. The electron lifetime, $\tau$, is the time over which the number of ionization electrons decreases by a factor $e$.

To determine $\tau$, the charge collected per unit length  ($\frac{dQ}{dx}$) for cosmic ray muon tracks
vs. drift distance is fit to the exponential function:
\begin{equation}\label{eq:electronlifetime_fit}
\frac{dQ}{dx}= \left( \frac{dQ}{dx}\right)_0 e^{-t_i/\tau}
\end{equation}
where $(dQ/dx)_0$ is the initial amount of deposited charge per unit length, and $t_i$ is the time of hit $i$ at the wire plane. The extracted value of $\tau$ is then used to correct the observed charge for tracks collected during the time period over which that $\tau$ was measured.

Two other factors also affect the signal on the wire planes:
electronic noise and electron diffusion. Both contribute a Gaussian
smearing to the measured charge, although the former is constant with time while the
latter grows as the square root of the drift time. Finally, although
unseen delta rays can produce small modifications, the distribution of $dQ/dx$
collected by the wire planes is well-described by the convolution of a
Landau function with a Gaussian.

\subsubsection{Determining the electron lifetime}

\begin{figure} 
\centering
    \includegraphics[width=0.95\textwidth]{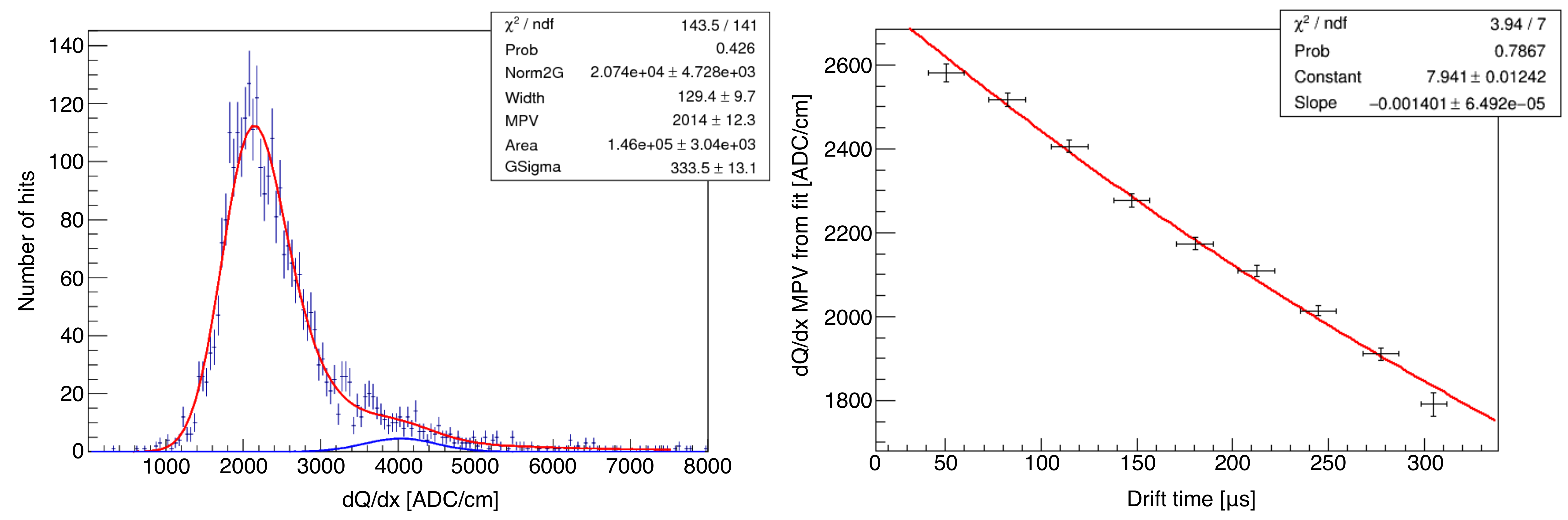}
\caption{\label{fig:elifetime_multitrack}
\emph{Left}: An example of a $dQ/dx$ distribution 
using hits in the drift time range of 229~$\mu$s to 262~$\mu$s. The distribution is fit to a Landau function convoluted with two Gaussian functions ($L \otimes G \oplus G$), overlaid in red. The blue Gaussian curve shows the contribution of delta rays that were not removed by the cut. \emph{Right}: MPVs of fitted $dQ/dx$ distributions for different drift time bins. The vertical error bars represent the statistical uncertainty of the fitted sample in each time bin. The horizontal error bars indicate the width of the drift time bin. Overlaid in red is the exponential fit from which the lifetime is determined. 
} 
\end{figure}

To characterize the attenuation of the accumulated charge, the TPC readout window of 393~$\mu$s is divided into 12 bins, each 33~$\mu$s wide. The size of the time bins is chosen to be small enough such that variations in quenching within the bin are negligible, but large enough to ensure sufficient statistics in each bin. For an electron lifetime $\tau = 800$~$\mu$s, the difference in $dQ/dx$ of two hits at the extremes of their drift time bin is approximately 4\%, which decreases to 2\% if $\tau = 1500$~$\mu$s.
For each 33~$\mu$s drift time bin, a histogram of $dQ/dx$ is filled with information from the corresponding segment of each cosmic muon track. Each histogram is then fit with a sum of two functions: a Landau convoluted with a Gaussian ($L \otimes G$) function, which includes the resolution effects described in the previous section, plus an extra Gaussian to account for any delta ray backgrounds that were not successfully removed. The left panel of figure~\ref{fig:elifetime_multitrack} shows a typical $dQ/dx$ distribution for one 33~$\mu$s-wide drift bin (229-262~$\mu$s), with a $L \otimes G \oplus G$ fit overlaid. The most probable value (MPV) of the Landau function is determined from the fit. The fitting process and determination of the MPV is repeated for each 33~$\mu$s drift time bin that has at least 500 entries, and those MPVs are entered into a second histogram as a function of drift time bin, as shown in the right panel of figure~\ref{fig:elifetime_multitrack}. An exponential fit to the MPV vs. drift time yields $\tau$, the electron lifetime.

To optimize the fits, a series of selection criteria is applied to both hits and tracks.
The effect of delta rays is minimized by excluding clusters of hits with large
energy deposition. A hit is considered part of a
delta ray candidate if its energy is 20\% higher than the MPV. 
Tracks with three or more consecutive delta ray candidate hits are flagged as delta rays
and their energy is removed. Because it is often difficult to associate their energies with the proper track, the first and last hits of each track are rejected.
In addition, any wire with multiple hits is removed from consideration. 

The event must contain only one reconstructed track with more than five hits.
A selection on both the vertical and horizontal track angles ensures that the
track does not lie along the beam line. It is important that tracks be
straight because the inclination of the track,
used to weight the charge of each hit, is computed only once for the track as
a whole. Therefore, the scalar product of the
direction vectors at the beginning and end of the track is required to be greater than
0.99. Also, a linear fit of the hit wire number vs. hit drift time, for both the collection and induction views, should have a normalized $\chi^2$ which is less than 10, that is, the track should be straight in both views.
Although in principle hits from either plane could be used, only hits from the collection plane are used in the electron lifetime analysis because deconvolution of the unipolar collection plane signal is simpler.


\begin{figure}
\centering
  \includegraphics[width=\textwidth]{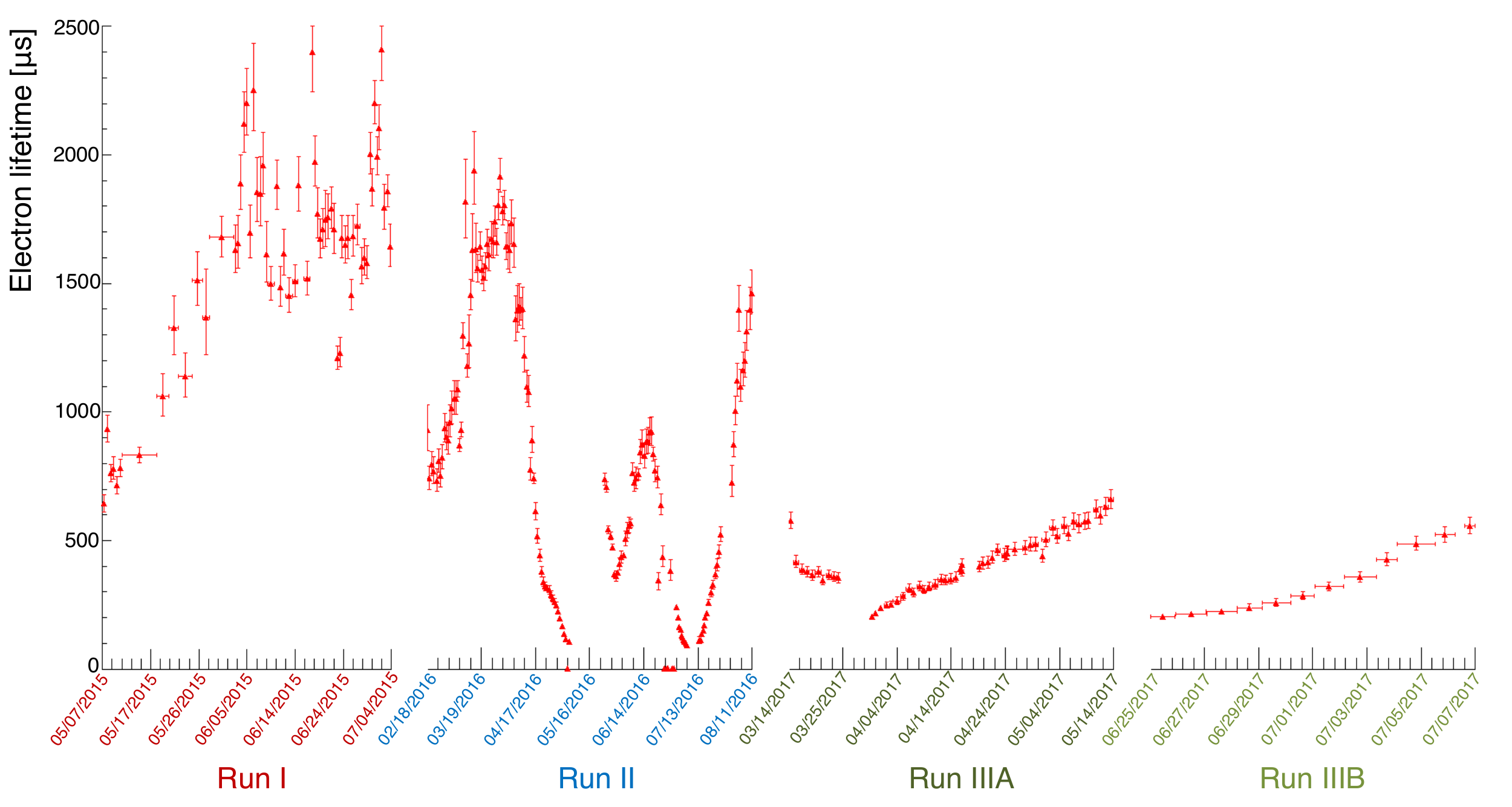}
\caption{The electron attenuation lifetime measured using cosmic muon tracks over LArIAT's three data-taking periods. Note that the scale of the horizontal time axis differs between the run periods. Gaps are generally due to periods when the filters needed to be regenerated, necessitating the emptying of the cryostat.} 
\label{fig:elifetime_allruns}
\end{figure}

The lifetime is determined by accumulating information from crossing
cosmic muon tracks in a given time period, usually about 24 hours.
The electron lifetime measurements over the three data-taking
periods of LArIAT, as seen in figure~\ref{fig:elifetime_allruns},
show characteristic variations. The variation in electron lifetime can be traced to several factors: the quality of the argon that was delivered once per week to refill the external argon supply dewar, the saturation level of the filtration system, and the sensitivity of the measurement technique.  For long lifetimes, $\tau > 1.2$~ms, where there is relatively little signal attenuation over the full range of drift times, the errors on the lifetime fit can be as large as 100~$\mu$s. While low signal attenuation leads to larger uncertainty on $\tau$, the corrections themselves are small. The uncertainties on corrected $dQ/dx$ are 3 to 5\%.

For smaller electron lifetimes, $\tau<1000$~$\mu$s, the large attenuation of $dQ/dx$ with drift length results in a robust exponential fit. In these areas of the plot, the observed lifetime variations reflect the status
of the filter system. Dramatic decreases of electron lifetime over a short time period, as in March 2016, are indicative of filter saturation. The filter material can be regenerated, however, and after regeneration, the lifetime slowly increases as the less pure argon inside the cryostat is diluted by top-offs of cleaner argon from the filter (as in April 2017, for example).  Fluctuations of the corrected dQ/dx value related to the electron lifetime are characterized by the statistical error of the exponential fit, typically between 2 and  4\%.
 
It is important to note that these fluctuations in the correction describe the worst-case scenario: the maximal $dQ/dx$ correction applied to charge leaving from the cathode.  The correction for charge produced near the anode wires will be minimal. Also, the conversion from $dQ/dx$ to $dE/dx$ depends on the assumed theoretical model, usually the Birks or Box model.  Finally, the effect of fluctuations in $dQ/dx$ for MIPs is different from that for strongly-ionizing particles. For both groups of particles, in an average over many tracks, such as the plots shown in figure~\ref{fig:dedx_calibration_fits}, random fluctuations will sum to zero.

\subsection{Charge response calibration}\label{sec:ChargeCalibration}

Understanding the charge response of wires in a LArTPC is essential for 
calorimetry. To calculate the number of drift electrons that were collected 
to produce a pulse (or "hit") on a wire, an integrated-ADC-to-electron conversion 
factor is applied. This conversion factor is a property of the readout electronics and signal filtering process.  The charge response can be determined via charge injection studies in a
dedicated test stand, or by looking at the charge deposition per unit length
from known particles in data. 

The charge-response
calibration for the LArIAT TPC consisted of two independent steps. First, the relative gains of the wires
(channels) in each plane are measured using MIPs which cross the TPC. A set of channel-specific calibration constants,
derived from these data, are used to bring all channels to a uniform gain.  
The second stage of calibration uses the well-known distribution of energy loss along trajectories of charged particles (of known type) traveling through liquid argon to determine the conversion factor needed to translate integrated wire signals into quantities of collected charge. No attempts were made to correct for E-field non-uniformity along a single wire.

\subsubsection{Channel non-uniformity calibration}\label{sec:wirebywire}

The response to charge varies from wire to wire on both the collection and induction planes. To calibrate and correct for variations
in response, the relative gain of each channel is measured using beam muons which pass close to the wire planes. Since muons from LArIAT's beamline are minimally-ionizing, the amount of charge they deposit per unit length follows a well-known distribution and does not vary significantly over the length of the track.
To avoid complications from the finite electron lifetime, the correction was
constructed with muon tracks which traverse the length of the TPC, parallel to, and no more than 5~cm distant from, the wire planes. For an electron lifetime of
1.5~ms, typical of Run~I, the charge lost over 5~cm is approximately
2\%, which is negligible in comparison to the roughly 20\% width of the typical $dQ/dx$
distribution.

A number of selection criteria are applied to the muon tracks used in this calibration sample.
The track is required to start within 2~cm of the upstream face and end no more than
2~cm from the downstream face. In the case of the 4~mm and 5~mm wire spacing data, this means that the track is always at least 86~cm long. For the 3~mm spacing data,
with a smaller instrumented volume, the track is required to be at least 60~cm
long. The selection criteria on angles $\phi$ and $\theta$ ensure that the tracks are  directed approximately parallel to the beam line.

\begin{figure}[tb]
\centering
\includegraphics[width=0.49\textwidth]{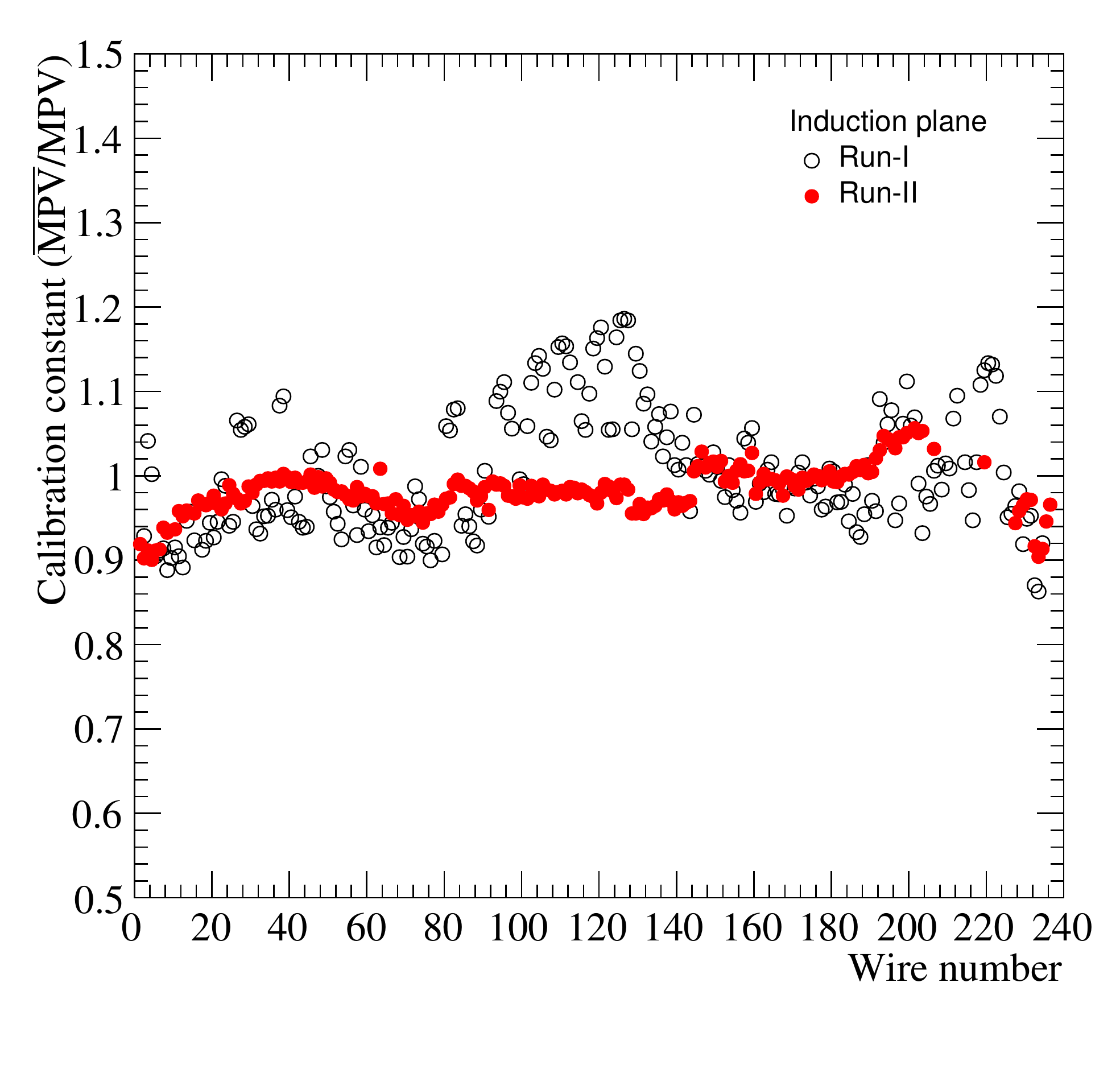}
\includegraphics[width=0.49\textwidth]{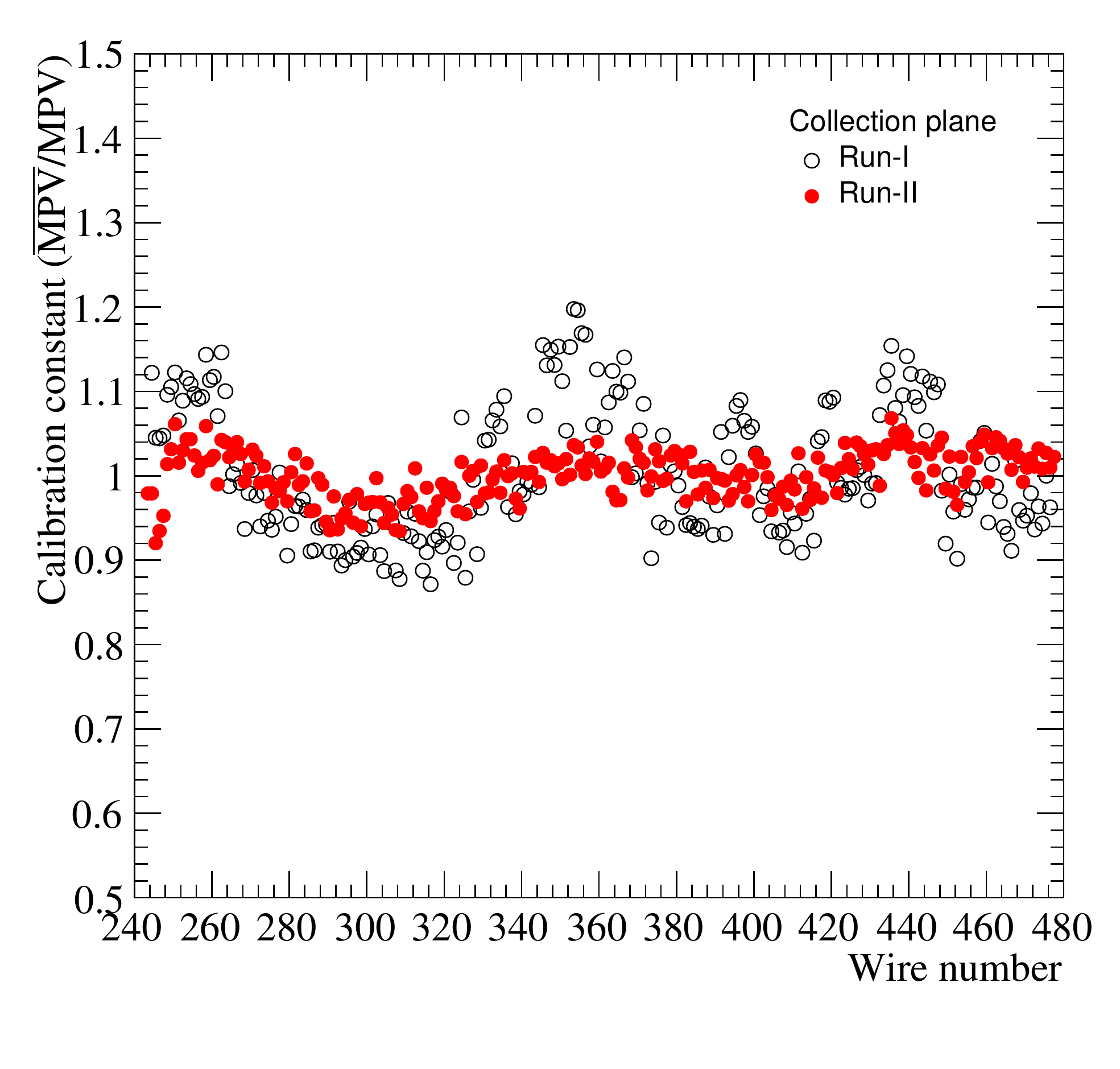}
\caption{
Calibration constants for each wire in the induction plane (\emph{left}) and collection plane (\emph{right}). Open black circles show Run I data; red filled circles show Run II data.} 
\label{fig:wirebywire_calibration}
\end{figure}

Energy depositions from selected muon tracks  are used to fill
the calibration histograms, one histogram for each wire. The charge is
corrected for any non-zero track pitch, which increases the effective $dx$. 
As with the evaluation of the electron lifetime, hits that
appear to come from delta rays are removed from the tracks.
The distribution is fit to a Landau function convoluted
with a Gaussian function, plus an extra Gaussian function to account for missed delta rays.
The MPV of the fit is used as a calibration constant, which corrects
for the observed wire-by-wire variations in response. To be precise, the
calibration constant, $C_i$, for each wire is defined as
\begin{equation}\label{eq:calconst_wirebywire}
C_{i} = \frac{\langle MPV \rangle}{MPV_{i}}
\end{equation}
where $MPV_i$ corresponds to wire $i$ while $\langle MPV\rangle$ is an average of
all the wires in that plane. During event reconstruction,
the raw charge is scaled by the corresponding calibration constant. Calibration constants for the induction plane wires in Run~I and Run~II are shown in the left panel of figure~\ref{fig:wirebywire_calibration}. Calibration constants for the collection plane wires are shown in the right panel of the same figure. Better wire response uniformity was seen in Run~II due to improvements made to the TPC readout and, as discussed in section~\ref{sec:PhotonSystem}, improvements to the light collection system. These improvements minimized electrical interference from the PMTs on nearby wires.

\subsubsection{Calibration of calorimetric response}\label{sec:caloconsts}

Once the electron lifetime corrections (section~\ref{sec:ElectronLifetime}) and the relative channel non-uniformity gain corrections (section~\ref{sec:wirebywire})  are applied, the calorimetric response of the TPC to the passage of charged particles must be calibrated. To understand the calibration procedure, we first review the formula for calculating energy deposited in liquid argon.  The linear density of deposited energy along reconstructed tracks in the TPC is calculated as
\begin{equation} \label{eq:dEdx_from_dQdx}
\frac{dE}{dx} \, \text{[MeV/cm]} =
    \frac{W_{\text{ion}}}{R(dE/dx,\mathcal{E})} \times 
    \frac{dQ}{dx} \, \text{[e}^-\text{/cm]},
\end{equation}
where $W_{\text{ion}} = 23.6$~eV is the average energy needed to ionize an argon atom, and $R$ is the \emph{recombination factor}, or the fraction of electrons that recombine with ions at the deposition site. The recombination factor is itself a function of the energy deposition density as well as the applied electric field, $\mathcal{E}$.

An integrated-ADC-to-electron conversion constant is used to calculate the number of drift electrons that have produced a pulse on a wire. This wire plane-specific calibration constant, $C_{\text{cal}}$, which is a property of the readout electronics and signal filtering process, represents the average number of integrated ADC units for each detected electron:
\begin{equation}\label{eq:Q_from_ADC}
Q \text{ [e}^-\text{] } = \frac{\text{Pulse area [ADC}\cdot\text{ticks]}}{C_\text{cal} \text{ [ADC}\cdot\text{ticks/e}^-\text{]}}.
\end{equation}

\begin{figure}
\centering
\includegraphics[width=0.55\textwidth]{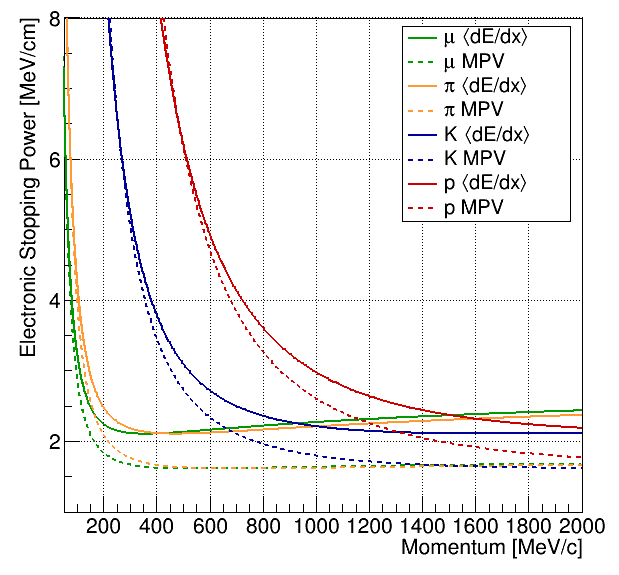}
\caption{Mean energy loss and MPV of energy loss ($\Delta_p / x$) predicted for muons ($\mu$), pions ($\pi$), kaons ($K$), and protons ($p$) in liquid argon. A thickness of $x = 0.4$ cm was used in the calculation of the MPV, corresponding to the wire spacing in Runs I and II.}
\label{fig:dEdx}
\end{figure}

The value of $C_\text{cal}$ can be determined from data by looking at $dQ/dx$ along tracks for which the particle type and momentum are well-known.  For this procedure, the $dE/dx$ along the track is reconstructed and compared to predictions from the Bethe-Bloch equation with the Sternheimer parameterization of density variations. The most probable value (MPV) of deposited energy is given by 
\begin{equation}
\Delta_p = \eta \Big[ \ln\left(\frac{2m_ec^2\beta^2\gamma^2}{I}\right) + \ln\left(\frac{\eta}{I}\right) + j - \beta^2 - \delta(\beta\gamma)  \Big],
\label{eq:mpv}
\end{equation}
where $\eta = \frac{1}{2}K(Z/A)(x/\beta^2)$, $x$ is the thickness of material traversed, and $I$ is the mean excitation energy~\cite{PDG2018}.  Both $\langle dE/dx\rangle$ and $\Delta_p/x$ are plotted as functions of momentum for several particle species in figure~\ref{fig:dEdx}.

In LArIAT, cosmic muons that stop within the TPC are used for the calibration.
Such muons are ideal candidates for this technique because, as they come to a rest, their $dE/dx$ is well-known and their energy can therefore be extrapolated at all points along the track. Stopping muons are identified by tagging an accompanying decay electron, as described in greater detail in ref.~\cite{LArIATMichel}.

The stopping muon track candidates are first formed with the standard LArIAT track reconstruction program, including relative channel-by-channel gain non-uniformity corrections, wire signal deconvolution filtering, wire hit reconstruction, trajectory-based clustering of hits, 3D projection-matching track reconstruction, as well as track calorimetry. Candidate events contain a single, stopping 3D track. A clustering algorithm, seeded by the charge deposition at the vertex, clusters the hits (on both collection and induction planes) associated with the muon and its decay electron.  Then calorimetric and spatial-tracking information is used to identify the boundary between the muon and electron hits within the cluster. As a muon slows down, the amount of energy it deposits per unit length increases dramatically, resulting in what is known as a Bragg peak. The outgoing decay electron is emitted in a random direction, often resulting in a visible kink within the cluster, which divides the electron hits from those of the relatively straight, muon track.  The charge deposition profile and local linearity along the cluster are used, respectively, to identify the Bragg peak and the kink within the cluster of muon and electron hits that marks the beginning of the electron track -- a technique adapted from an analysis of Michel electrons in MicroBooNE~\cite{microboone_michel}. A delayed double pulse (see section~\ref{sec:PhotonSystem}) in the light collection system, created by the stopping muon and subsequent decay electron, is also required. If time-coincident muon-electron boundary hits are identified on both collection and induction planes, they are used to form a 3D spacepoint, $P_{\textrm{boundary}}$, which marks the end of the muon track. 

The point $P_{\textrm{boundary}}$, found from the cluster profiling procedure described above, is not always consistent with the endpoint of the reconstructed 3D track, $P_{\textrm{trk}}$, since the 3D tracking algorithm is not specifically tailored to deal with the muon decay topology. To increase the fraction of well-reconstructed events, the separation of $P_{\textrm{boundary}}$ and $P_{\textrm{trk}}$,  projected along the direction of the reconstructed track, is required to be less than 2~mm.

Finally, to minimize the correction associated with recombination, requirements are made on the pitch of the track, the effective path length associated with each hit, and the angle between the track and the direction of the electric field.  The track's pitch (or the average distance between hits) must be less than 1.2~cm and its angle relative to the electric field must be less than 20~degrees. Neither cut removes many candidates.

Note that the $dQ/dx$ for the hits associated  with the muon and electron tracks must be corrected for the electron attenuation lifetime, $\tau$. For the muon, this means scaling up the raw $dQ/dx$ by a factor $e^{t_i/\tau}$, where $t_i$ is the drift time of hit $i$. The same is true of hits associated with the decay electron, except the drift time is corrected to account for the decay time delay.

\begin{figure}
    \centering
    \includegraphics[width=0.9\textwidth]{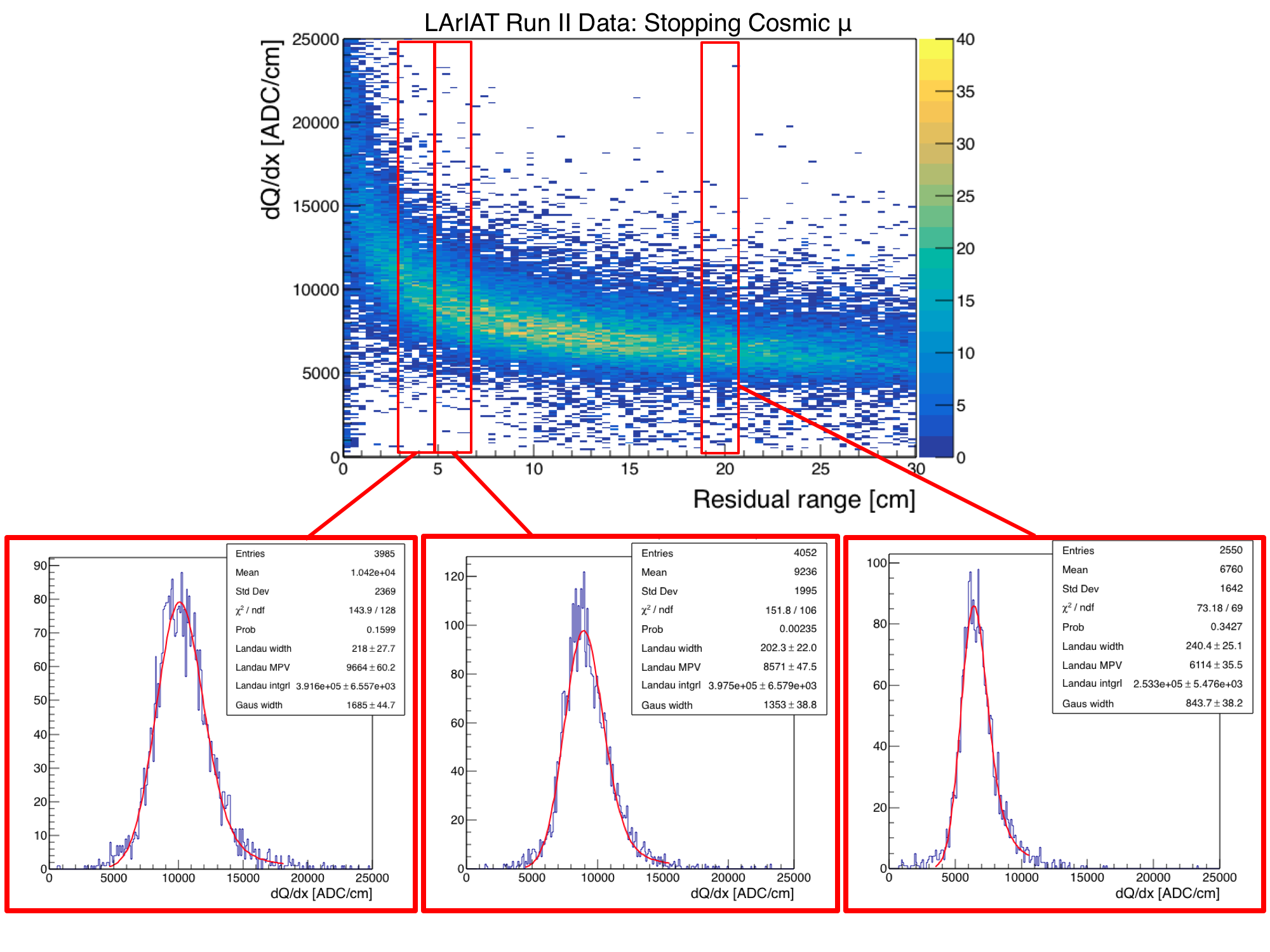}
    \caption{Reconstructed $dQ/dx$, in units of $\textrm{ADC}\cdot \textrm{ticks/cm}$ plotted against residual range along the selected stopping muon tracks. Example fits to ``slices'' of $dQ/dx$ are shown in the lower panels.}
    \label{fig:michel_calibdQdx}
\end{figure}

Using the final set of stopping tracks, $dQ/dx$ is plotted against residual range $r$, the distance to the end of the track, as a 2D histogram. Over narrow ranges in $r$, the  values of $dQ/dx$ are fit to a Landau function convoluted with a Gaussian, as shown in figure~\ref{fig:michel_calibdQdx}.

\begin{figure}
\centering
\includegraphics[width=0.5\textwidth]{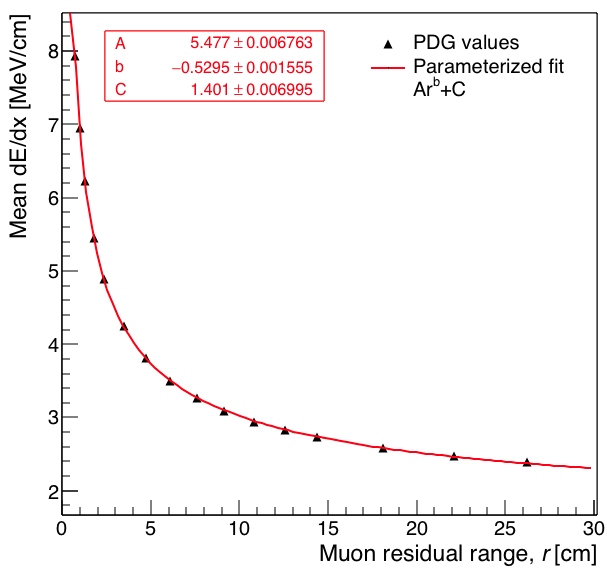}
\caption{Mean $dE/dx$ vs. residual range for muons in liquid argon. Data points are from ref.~\cite{PDG2018}. The red curve is a parameterized fit of the form $dE/dx = A r^b + C$.}
\label{fig:dEdx_rr}
\end{figure}

The mean $dE/dx$ as a function of residual range along a track is well described by a simple power law function:
\begin{equation}\label{eq:dEdx_rr}
\frac{dE}{dx} = Ar^b + C
\end{equation}
where the residual range, $r$, is measured in cm, and $A$, $b$, and $C$ are fit parameters. A fit to data taken with muons passing through liquid argon~\cite{PDG2018} yields
\begin{equation}
\begin{split}
A & = 5.477 \pm 0.007~{\textrm{MeV/cm}}\\
b & = -0.5295 \pm 0.0016 \\
C & = 1.401 \pm 0.007~{\textrm{MeV/cm}},\\
\end{split}
\end{equation}
as shown in figure~\ref{fig:dEdx_rr}.  By integrating eq.~\ref{eq:dEdx_rr} over the muon's trajectory, starting from its end point (where $r = 0$), we derive a relationship for its energy as a function of its range,
\begin{equation}
T_r = \left(\frac{A}{b+1}\right)r^{b+1} + Cr.
\end{equation}

Using the average 3D separation between hits ($dx$) from the data sample, eq.~\ref{eq:mpv} can be used to calculate the expected $dE/dx$ at each point along the muon track.  Then, eqs.~\ref{eq:dEdx_from_dQdx} and \ref{eq:Q_from_ADC} are used to derive the predicted $dQ/dx$ in units of ADC$\cdot$ticks/cm:
\begin{equation}\label{eq:dQdx}
\frac{dQ}{dx} \text{ [ADC}\cdot\text{ticks/cm]} = C_{{\textrm{cal}}} \times \frac{dE/dx}{W_{\text{ion}}} \times R(dE/dx,\mathcal{E}).
\end{equation}

\begin{figure}
    \centering
    \includegraphics[width=0.9\textwidth]{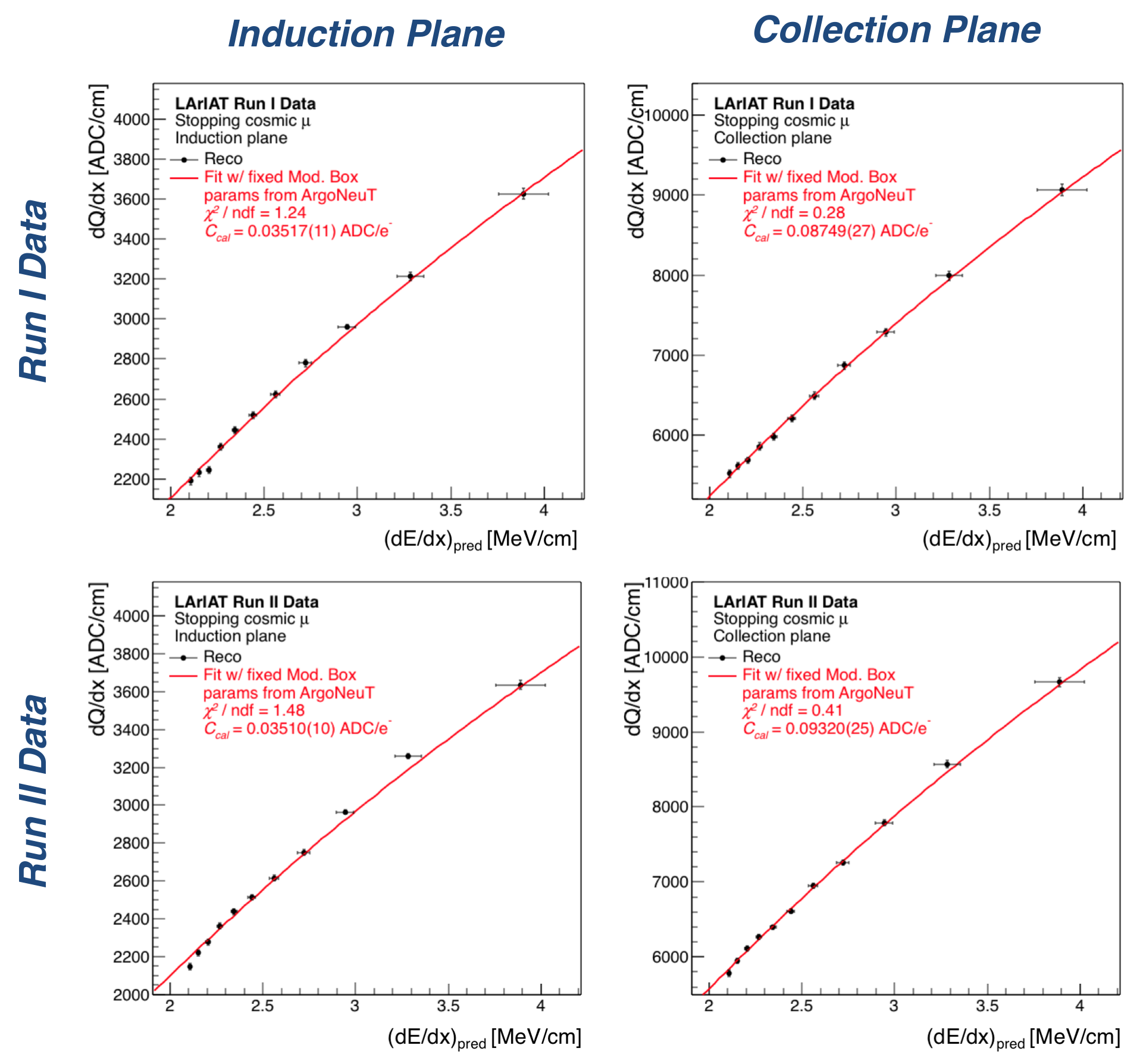}
    \caption{Charge response calibration fits using stopping muon tracks from Run~I and Run~II.  The results from Run~II also apply to data taken in Run~III since the same readout electronics were used.}
    \label{fig:dedx_calibration_fits}
\end{figure}

Equation~\ref{eq:dQdx} is fit for the calibration constant $C_\text{cal}$, using as data points the MPV from the Landau fits to $dQ/dx$ at different residual ranges. The Modified Box recombination model is used to determine $R$, with parameters fixed to those found by ArgoNeuT~\cite{ArgoNeut-recomb}. Uncertainties on the $dQ/dx$ MPV are taken from the fits. The uncertainty in $dE/dx$ arises from uncertainty in the muon endpoint projected along the muon track direction, as estimated from simulations. 

For each wire plane, the best fit value of $C_\text{cal}$ is taken as its final electronics response calibration constant. These fits are displayed in figure~\ref{fig:dedx_calibration_fits} for cosmic muon data taken during Run~I and Run~II, with results summarized in table~\ref{tab:dedx_calibration}. The fits yield charge calibration constants of $C_\text{cal}$~$\approx$~0.035~ADC/e$^-$ on the induction plane (approximately 29 collected electrons per integrated ADC), and $C_\text{cal}$~$\approx$~0.09~ADC/e$^-$ on the collection plane (approximately 11 collected electrons per integrated ADC). Since the charge readout electronics did not change after Run~II, the calibration constants obtained in Run~II were also applied in Run~III. Though the calibration procedure is carried out on both wire planes, typically only the collection plane is used for calorimetry due to the cleaner pulses and stronger charge response observed on the wires of that plane.

\begin{table}[htb]
\centering
\caption{Charge response calibration constants used in the reconstruction of data for each run period. Uncertainties are those returned by the fits in figure~\ref{fig:dedx_calibration_fits}.}
\label{tab:dedx_calibration}
\begin{tabular}{l  c  c }

\hline

&
\multicolumn{2}{c}{\bf Charge Response $C_\text{\bf cal}$ [ADC/e$^-$]}  \\

{\bf Run  Period}
& {\bf Collection Plane}
& \textbf{Induction Plane} \\
\hline

Run I          
& 0.08749(27)  
& 0.03517(11)  \\

Run II \& III
& 0.09320(25) 
& 0.03510(10) \\

\hline
\end{tabular}
\end{table}

\paragraph{Validation with beamline particles}\label{sec:beamline_crosscheck}

\begin{figure}
    \centering
    \includegraphics[width=0.49\textwidth]{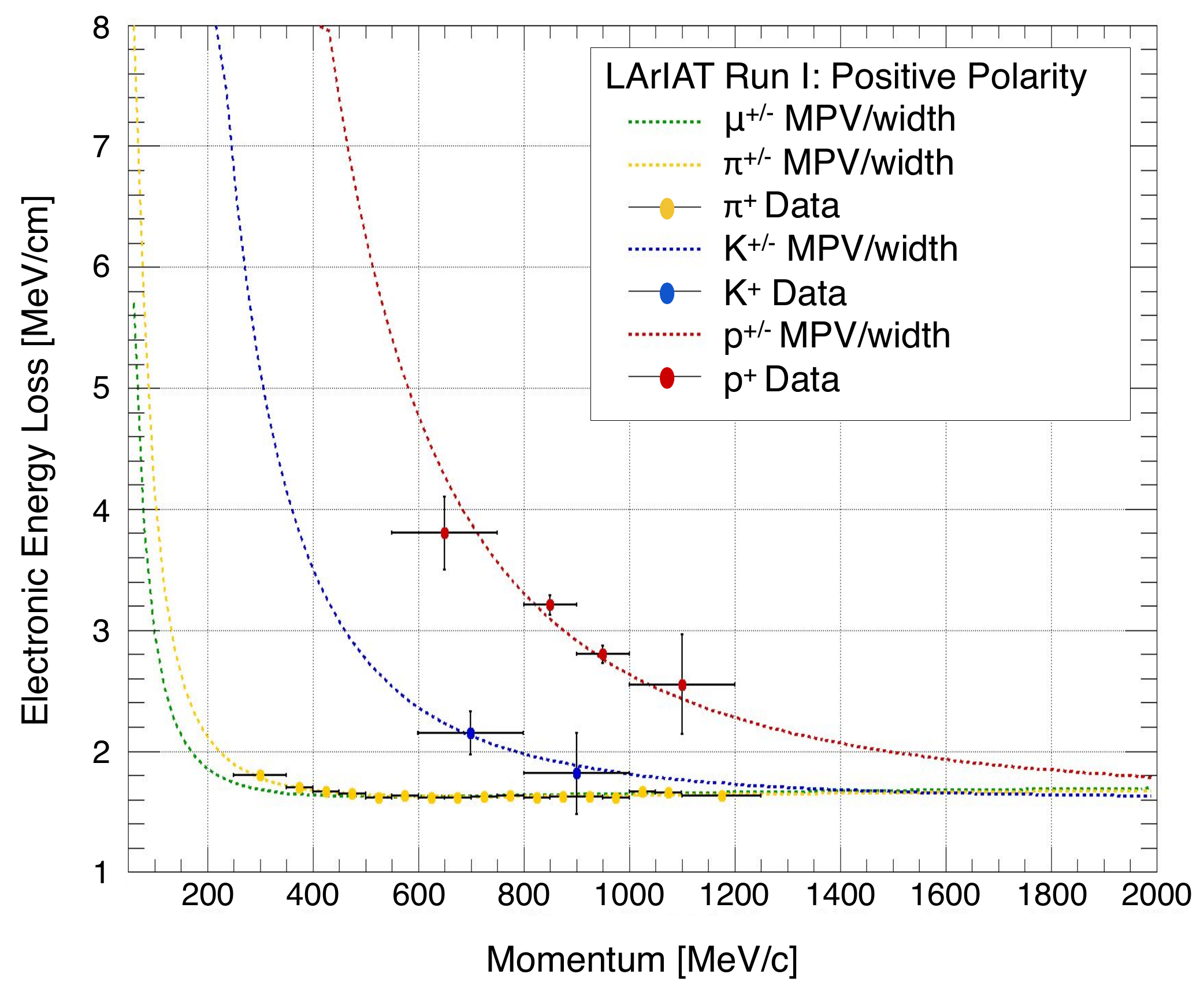}
    \includegraphics[width=0.49\textwidth]{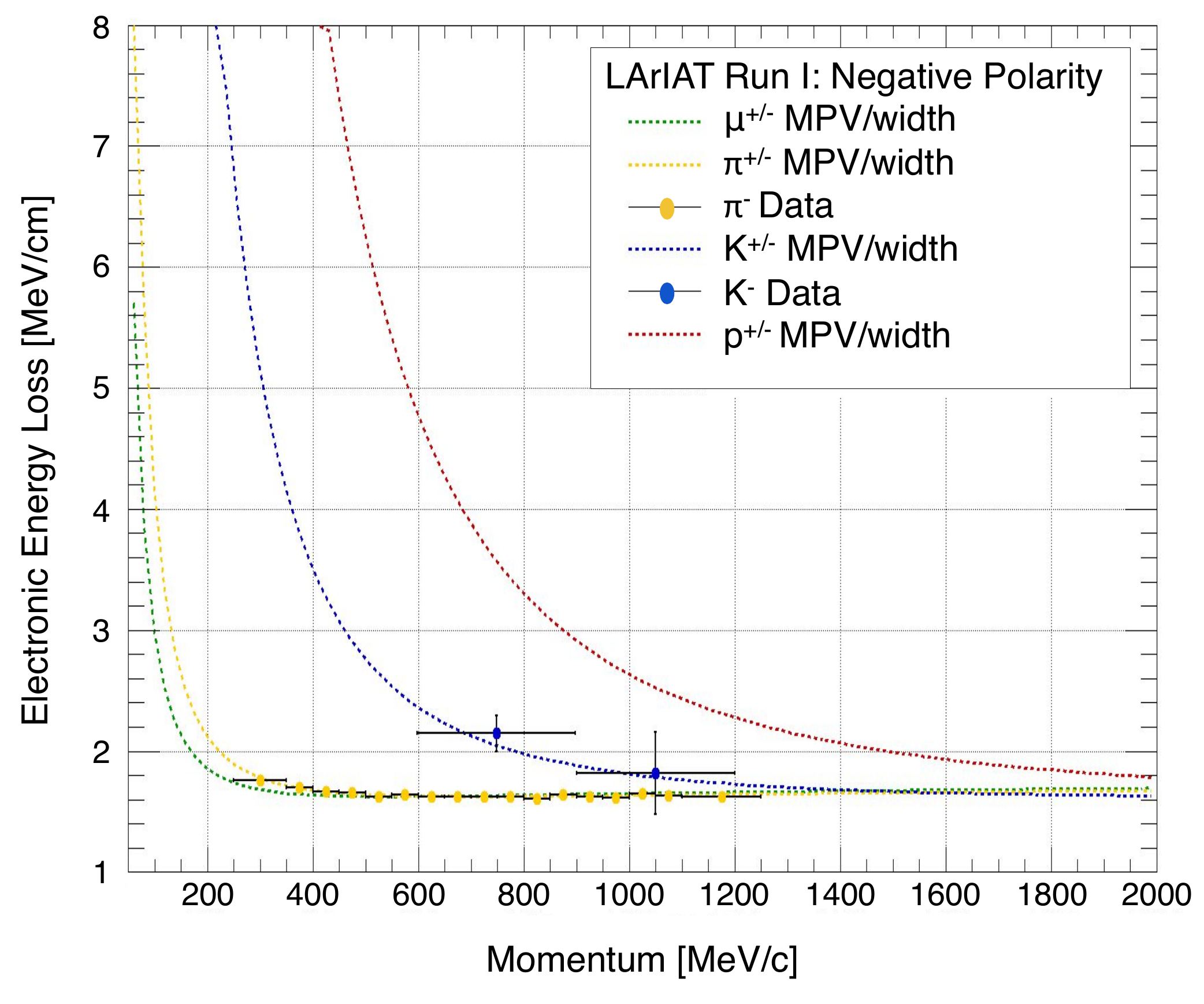}
    \includegraphics[width=0.49\textwidth]{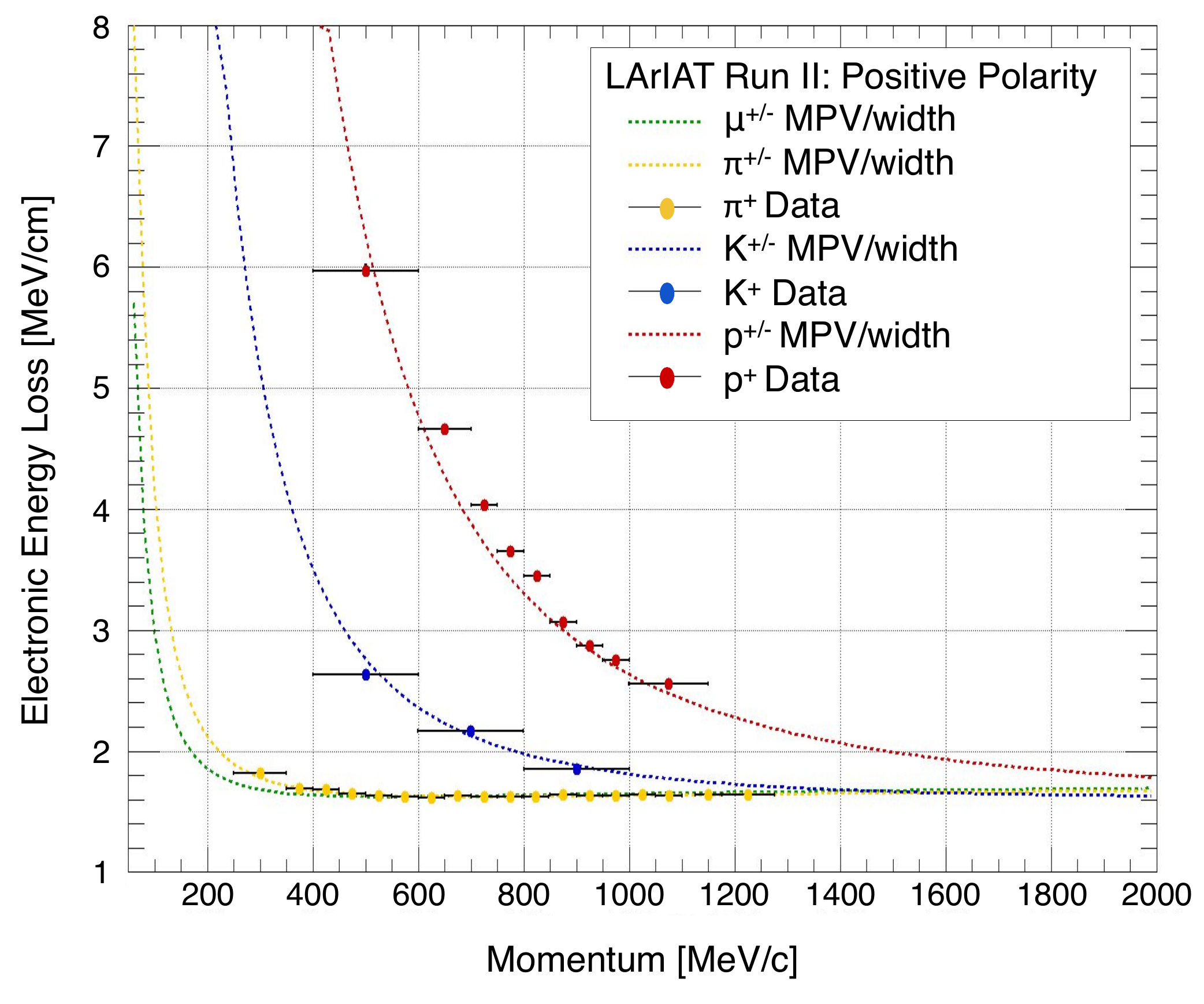}
    \includegraphics[width=0.49\textwidth]{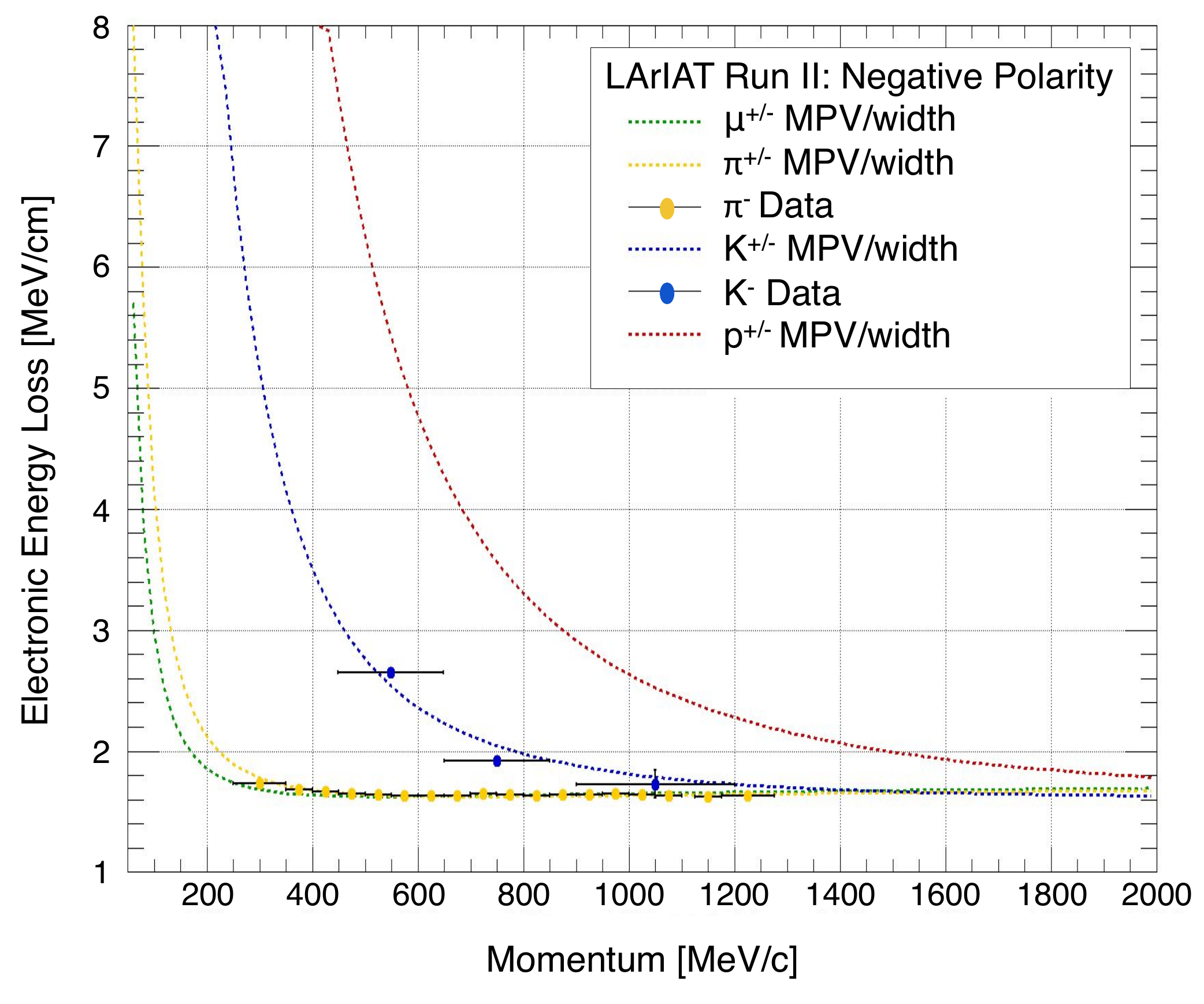}
    \caption{Electronic energy loss, $dE/dx$, plotted as a function of momentum for selected beam particle tracks in Run~I positive polarity data ({\emph{top left}}), in Run~I negative polarity data ({\emph{top right}}), in Run~II positive polarity data ({\emph{bottom left}}), and in Run~II negative polarity data ({\emph{bottom right}}). In all data samples, the calorimetry constants were tuned using the sample of stopping cosmic muons, and those constants were applied with no additional tuning to the beamline samples shown here. }
    \label{fig:dEdxtuning}
\end{figure}

The calorimetry constants determined from the stopping muons are cross-checked using charged particles from the beamline. The momenta of these particles are determined by the beamline spectrometer. The combination of momentum and TOF measurements yields the particle mass, allowing separation of different species. To ensure that momenta and particle ID are measured as well as possible, only golden tracks are selected -- tracks with one and only one hit in each of the four wire chambers (see section~\ref{sec:momentumscale} for details on tracking with the wire chambers). Each track is then extrapolated downstream to determine its entrance point at the front face of the TPC. In order to account for energy loss in material between the last wire chamber and the entrance to the TPC, a simple correction is made to the measured momentum, based on the Monte Carlo simulation.

One final criterion is applied: to ensure that each track has some samples in the minimum-ionizing range before the Bragg peak, the track is required to be at least 10~cm long. The reconstructed $dE/dx$, using the calorimetry constants determined with the stopping muon sample, is calculated for the first twelve space points of the track (typically 5~cm). To avoid edge effects near the field cage, the first space point is excluded.

The samples of pions/muons (which are not distinguishable within the TOF resolution), kaons, and protons are divided into subsamples of momentum, each with a range of 50~MeV/c, from 150~MeV/c to 1100~MeV/c. Then, for each momentum and particle species subsample, the $dE/dx$ histogram is fit to a simple Landau function to extract the MPV and its uncertainty. 
The MPV values and uncertainties for each fit are plotted versus momentum and compared with the Bethe-Bloch predictions for each particle type, shown in figure~\ref{fig:dEdxtuning}. Each of the particle species matches well with its respective Bethe-Bloch prediction over the range of momenta in the beamline.

The successful validation with multiple species of particles from the beamline provides confidence in the accuracy of the calibrations outlined in this section: the channel-specific non-uniformity gain corrections shown in figure~\ref{fig:wirebywire_calibration}, together with the plane-specific ADC-to-electron conversion factors shown in table~\ref{tab:dedx_calibration}.

\section{Conclusions}

The LArIAT detector and supporting beamline instrumentation were successfully operated in the Fermilab Test Beam Facility during three major running periods from 2015 to 2017, collecting data that will allow tests of a variety of hardware configurations, as well as measurements of hadronic interaction cross sections. Hardware performance and cross section measurements will be reported in future papers.  Detailed descriptions of the LArTPC configurations were given in this paper, with the aim of providing the necessary technical background information for future references to LArIAT operation and physics analyses.

\acknowledgments
This document was prepared by the LArIAT collaboration using the resources of the Fermi National Accelerator Laboratory (Fermilab), a U.S. Department of Energy, Office of Science, HEP User Facility. Fermilab is managed by Fermi Research Alliance, LLC (FRA), acting under Contract No. DE-AC02-07CH11359. We also gratefully acknowledge the support of the National Science Foundation, Brazil CNPq grant number 233511/2014-8, Coordena\c{c}\~ao de Aperfei\c{c}oamento de Pessoal de N\'ivel Superior - Brazil (CAPES) - Finance Code 001, S\~ao Paulo Research Foundation - FAPESP (BR) grant number 16/22738-0, Polish National Science Centre grant Dec-2013/09/N/ST2/02793, and the JSPS grant-in-aid (Grant Number 25105008), Japan. The collaboration extends a special thank you to the coordinators and technicians of the Fermilab Test Beam Facility, without whom none of this work would have been possible.

\bibliographystyle{JHEP}
\bibliography{references}

\end{document}